\documentclass[10pt,journal,compsoc]{IEEEtran}
\usepackage{tabularx,booktabs}
\usepackage{caption}
\usepackage{subcaption}
\usepackage{pgfplots}
\usepackage{pgfplotstable}
\usepackage{sfmath}
\usepackage{graphicx}
\usepackage{multirow}
\usepackage{algorithm, algorithmic}
\usepackage{makeidx}
\usepackage{soul}
\usepackage{color}
\usepackage{colortbl}
\usepackage{xcolor}
\usepackage{tikz}
\usepackage{amsthm,amssymb,amsmath}
\usepackage{wrapfig}
\usepackage{tocbibind}
\usepackage[per-mode = fraction]{siunitx}
\usepackage{amsmath}
\usepackage{pifont}

\newcommand{\xmark}{\text{\ding{55}}}
\usepackage{makeidx}
\usepackage{float}
\makeindex
\ifCLASSOPTIONcompsoc
  \usepackage[nocompress]{cite}
\else
  \usepackage{cite}
\fi
\ifCLASSINFOpdf
  
\else
  
\fi
\begin{document}

\title{Gene and RNA Editing: Methods, Enabling Technologies, Applications, and Future Directions}

\author{Mohammed~Aledhari and Mohamed~Rahouti 
\thanks{M. Aledhari is with the Data Science, University of North Texas, Denton, TX 76207 USA. Email: mohammed.aledhari@unt.edu}
\thanks{M. Rahouti is with the Department of Computer \& Information Science, Fordham University, New York, NY, 10023 USA. Email: mrahouti@fordham.edu}
\thanks{Manuscript received xx, 2024; revised xx, 2024.}}

% The paper headers
\markboth{Review Article ,~Vol.~xx, No.~xx, August~2024}%
{Aledhari \MakeLowercase{\textit{et al.}}: Gene and RNA Editing}

%%%%%%%%%%%%%%%
\IEEEtitleabstractindextext{%
\begin{abstract}
Gene and RNA editing methods, technologies, and applications are emerging as innovative forms of therapy and medicine, offering more efficient implementation compared to traditional pharmaceutical treatments. Current trends emphasize the urgent need for advanced methods and technologies to detect public health threats, including diseases and viral agents. Gene and RNA editing techniques enhance the ability to identify, modify, and ameliorate the effects of genetic diseases, disorders, and disabilities. Viral detection and identification methods present numerous opportunities for enabling technologies, such as CRISPR, applicable to both RNA and gene editing through the use of specific Cas proteins. This article explores the distinctions and benefits of RNA and gene editing processes, emphasizing their contributions to the future of medical treatment. CRISPR technology, particularly its adaptation via the Cas13 protein for RNA editing, is a significant advancement in gene editing. The article will delve into RNA and gene editing methodologies, focusing on techniques that alter and modify genetic coding. A-to-I and C-to-U editing are currently the most predominant methods of RNA modification. CRISPR stands out as the most cost-effective and customizable technology for both RNA and gene editing. Unlike permanent changes induced by cutting an individual's DNA genetic code, RNA editing offers temporary modifications by altering nucleoside bases in RNA strands, which can then attach to DNA strands as temporary modifiers.
\end{abstract}

%%%%%%%%%%%%%%%%%
}

\maketitle

\IEEEdisplaynontitleabstractindextext

\IEEEpeerreviewmaketitle

\IEEEraisesectionheading{\section{Introduction}\label{sec:introduction}}

%%%%%%%%%%%%%%%%%%%%%%%%%
%\IEEEPARstart
Genetic mutations contribute to a wide variety of diseases and viruses. They can be inherited from the parents or acquired throughout an individual’s life \cite{cite_016}. Mutations occur when a complementary base is altered or there is a complete loss in gene expression \cite{lodish2008molecular}. Effects of sequence alteration can be harmless, helpful, or hurtful depending on how the mutation modifies the individual’s genetic makeup \cite{cite_004}. Nonetheless, mutations can be more damaging to DNA due to their permanence to a cell’s structure than a mutation is to RNA and protein molecules which have multiple synthesized copies \cite{lodish2008molecular}. Depending on where they occur and whether they alter the function of essential proteins, gene mutations affect human health in various ways \cite{cite_007}.

The scope of this article is to inform readers of the research methods, enabling technologies, and applications of gene and RNA editing to mutations, and the benefits of RNA and gene editing advancements to the future of the medical and technological fields. Current traditional approaches of prescribed pharmaceuticals modify the course of a disease but do not provide a cure \cite{cite-Stanford}. Traditional medicines do not treat or modify the root of disease, which is found in the DNA of a genome. Rather than administering chemically-modified medications to patients, gene and RNA editing provides a new, more holistic pathway due to its ability to edit the naturally-occurring genetic process within the human body. The purpose of gene and RNA editing is to modulate the root of disease by fixing the gene’s DNA mutation \cite{cross2019watch}.

Deoxyribonucleic acid, also known as DNA, is a molecule that stores genetic coding for an entire organism–in this case, the human body. Under a microscope, DNA’s double helix looks like a twisting ladder with linking and repeating steps that consist of the genetic codes, the basis of an individual’s existence. The particular sequence of the links are the building blocks of information that tells us about the individual, like their ancestry and their current state or biological condition, including the possibilities of future biological reactions and functions.

The pattern within the double helix links specific nucleotides, or chemical units of genetic material, to one another. Nucleotides consist of three parts: a 5-carbon sugar molecule, a phosphate group, and a nitrogen-containing base. Nucleotides that virtually bond together within DNA’s double helix are adenosine (A) and thymine; guanine (G) and cytosine (C). These are the base pairs that duplicate in a variety of sequences. The DNA double helix follows the rules of chargeoff making the number of G and A equal to the numbers of C and T \cite{cite_017}.

DNA is a memory bank and an instructor for development, reproduction, and survival \cite{cite_015}. As a memory bank, DNA possesses vital information concerning the organism, or the individual, and it keeps the information similar to how computers store data. As an instructor, DNA conveys stored information to create new cells and proteins for biological functioning.

An important property of DNA is its ability to replicate. To make copies of itself, the double helix unwinds and duplicates the base sequences \cite{cite_016}. New cells are created after replication and division from the parent cell. As DNA continually copies the stored genetic information, it communicates that same information to new daughter cells. New cells possess the same abilities as the parent cell, and can thereby repeat the replication and instruction processes.

DNA does not actively apply information \cite{eisenberg2018rna}. In order for DNA to transfer instructions for biological processes, the double helix needs a messenger. DNA doesn’t leave the safety of the cell’s protective membrane-bound organelles, like the mitochondria and chloroplast, or the cell’s nucleus, also known as the brain of a cell; this is where RNA steps in. RNA is a conduit, a pilot, that drops in. It connects with the split double helix, and obtains instructions from genes and/or DNA to create proteins necessary for biological functions in the body. However, the creation of a protein differs from replication in that the RNA strand does not remain bonded to DNA \cite{wilson2010principles}.

A similar, yet more functional piece of DNA is a gene. “Only a functional piece of DNA is called a gene” \cite{cite_017}.  It makes up one \% of DNA’s sequence \cite{cite_015}. A gene is a specific trait consisting of DNA \cite{cite_017}. Not all genes possess DNA, and not all DNA is considered a gene. Genes only contain inherited information, like if your parents have brown eyes and you have brown eyes, that is an hereditary gene passed down from your parents and/or their parents before them. Within a human individual, there are a two copies of each gene to represent each parent \cite{cite_016}.
 
Genes can consist of either DNA or RNA, and they serve as a trait regulator. These specific sequences code genetic information from parent to child \cite{cite_013}. Genes that contain inherited DNA also have the ability to replicate and call messengers to deliver its stored information to the rest of the body. As a result, both DNA and genes contribute to the creation of proteins for the biological processes.

Since DNA and genes are limited to specific locations in the cell, messengers of RNA come to aid in the process of protein creation. Protein molecules are the workhorses, the engine in the entire body \cite{clancy2008translation}. Gene expression is a function of DNA  “by which cells read out, or express, the genetic instructions in their genes” \cite{wilson2010principles}. Two major steps of the gene expression process are transcription and translation \cite{cite_007}.

RNA is the carrier of DNA information. As a single, short strand consisting of a ribose sugar, RNA is more flexible than DNA \cite{eisenberg2018rna}. It capable of folding into a variety of shapes \cite{wilson2010principles}. From the storage facility of DNA, mRNA (or messenger RNA) carries the blueprints for the protein engine \cite{eisenberg2018rna}. Similar to the structure of DNA, RNA contains complementary bases of adenine (A), cytosine (C), guanine (G), and uracil (U). Uracil (U) substitutes for DNA’s thymine (T) base, because it is resistant to oxidation as RNA leaves and exists outside of the cell’s nucleus.

Transcription is the first step in creating proteins. After unwinding and splitting the double helix, a DNA strand serves as a template for complementary base-pairing between nucleotides \cite{wilson2010principles}. An enzyme connects to and processes the strand forming mRNA, a single-strand copy of DNA, or a gene’s DNA \cite{clancy2008translation}. Once RNA is released from DNA, many more copies of that RNA strand can be created and sent out while still containing the original DNA instructions. “A transcription unit typically carries the information of just one gene, and therefore codes for either a single RNA molecule or a single protein” \cite{wilson2010principles}.

Gene expression is completed with its second step of translation. After moving out of the nucleus, the genetic code, or blueprint, of mRNA is translated and read by tRNA (transfer RNA), and rRNA (ribosomal RNA) forms bonds between the amino acid groups of codons to produce polypeptide chains. Every three nucleotides on the mRNA strand are read as a group of three called codons. These three lettered codes translate into specific amino acids, or instruct the cessation of protein creation. “The mRNA sequence is thus used as a template to assemble—in order—the chain of amino acids that form a protein” \cite{clancy2008translation}.

Human individuals, and all organisms, are born with a genetic code pre-written into the body, strands of information, known as nucleotides, imbedded in DNA and RNA sequences that is translated into proteins, or amino acid sequences, by living cells \cite{cite-GeneticCode}. Mutations cause changes to DNA sequences away from the normal and stable sequence \cite{cite-Genome}. While traditional therapies and medications supply temporary relief, pharmaceuticals impose on the natural system of our bodies, forcing, even to the degree of impeding the body to change, causing numerous debilitating side effects. Medications are currently not capable of solving permanent genetic mutations that cause impediments and diseases without negatively impacting the individual. This shortcoming is most apparent when treating monogenic diseases, diseases caused by inherited mutations to a single gene \cite{cite-Stanford}. Whereas gene and RNA editing provide targeted therapies that lead to a painkiller lasting days or weeks while remaining nonaddictive and reversible \cite{cross2019watch}.

Different to traditional medicines and surgery, gene and RNA editing technologies engineer cells that possess curative properties administered to a range of diseases that currently have no cure \cite{cite-Stanford}. These technologies and therapies are “living drugs” that can heal and modulate damaged tissues or diseased organs \cite{cite-Stanford}. Currently gene and RNA therapies are not so different in delivery from other existing medications \cite{cite-Harries}. Engineered cells are encapsulated, then, they are prescribed and taken regularly, however, their effects do not last indefinitely like that of conventional medicines \cite{cite-Harries}. 
Methods of gene and RNA editing recognizes the differences and similarities of the two techniques. The methods are similar because RNA editing derives from gene editing technologies. Gene editing modifies DNA strands within an individual’s genome with a cut and deletion of the mutation, while RNA editing sends a messenger to DNA and converts the nucleic acid bases. For this reason, researchers have transitioned their focus to RNA editing due to its temporary off-target mutations, making RNA editing a potentially safer alternative to gene editing \cite{cross2019watch}.

Other reasons to limit use of gene editing methods are that it can cause a negative immune response from cells, and thereby cause more pain to a patient or even cause more debilitations, like cancer, if targeting is not efficient. Researchers have evidence that gene editing enzymes, like Cas9, triggers immune responses, or causes accidental and permanent changes to the genome \cite{reardon2020step}. Delivery of gene editing methods are also suspected to make cells cancerous due to interference with the efficiency of gene editing to cut DNA \cite{cite-Clara}. Nonetheless, gene editing is a prospect of treating and even curing previously intractable diseases, especially in cases where the primary cause is a mutation, by targeting specific DNA sequences \cite{ORMOND2017167}.

RNA editing, by contrast, allows clinicians to make temporary fixes that eliminate mutations in proteins, halt their production, or change the way they work in specific organs and tissues \cite{reardon2020step}. These temporary fixes derive from base modification methods such as A-to-I editing and C-to-U editing. Whether these modifications occur within the cell or not, unused RNA strands are degraded by the cell; therefore, any errors introduced by an RNA editing therapy would be washed out, rather than staying with a person forever \cite{reardon2020step}. Researchers have found that RNA editing can have a greater impact with its ability to turn genes on and off, aid chemical reactions, slice and dice other RNAs, and even build proteins by transporting amino acids and linking them together \cite{reardon2020step}.

Enabling technologies provide a platform for gene and RNA editing methods to become a reality. All methods derived from the discovery of DNA’s double helix in the 1950s, and the discovery of restriction enzymes a decade later enabled gene editing technologies that edit genome DNA \cite{cite-Synthego}. The development of various engineered nucleases was realized in 2010 with zinc-finger nucleases, meganucleases, and transcription activator-like effector nucleases \cite{ORMOND2017167}. In 2013, the CRISPR/Cas9 system was adapted from the bacterial species which RNA-guided nucleases that cleave target sequences and edit the mutation \cite{ORMOND2017167}. Major technologies derived from these gene-editing discoveries are ZFNs, TALENs, and CRISPR. Utilizing these technologies paved the way for scientists to discover RNA editing capabilities and technologies with an enzyme derived from Cas9.The enzyme was thereby called Cas13, and other sub-enzymes derive from the discovery.

Biotechnology with editing abilities of mutations is predominantly based on target specificity. The target sequence of editing occurs wherever there is a mutation. Target accuracy of the technology exemplifies the effectiveness the edit. RNA editing technologies do not have the high target specificity that is found with gene editing technologies. However, gene editing is limited to editing mutations within genome DNA.

Currently, it is estimated that there are greater than 6,000 such diseases, or mutations, that affect over 350 million people worldwide \cite{cite-Stanford}. Mutations occur in two cellular areas: non-reproductive cells and gametes or cells that will eventually produce gametes \cite{cite_004}. The categories of mutations that take place in these cells are somatic and germ-line.

Somatic mutations are mutated cells that are passed down to a daughter cell during cell division \cite{cite_006}. These acquired mutations are the most common cause of cancer due to environmental factors. Diseases and cancerous mutations are usually caused by carcinogens like tobacco use, ultraviolet radiation, viruses, chemical exposures, and aging \cite{cite_005}.

Germ-line mutations are inherited genetic alterations passed directly from parent  to child via the sperm or egg cell. A predominantly harmless germ-line mutation is called a polymorphism. This involves a variation at a single base pair \cite{cite_015}. Polymorphism is responsible for many of the normal differences between people such as eye color, hair color, and blood type \cite{cite_016}.

There are different classifications of mutations including hereditary, acquired (or somatic) mutations, and polymorphisms; however, Beyond the cellular areas, the main types of mutations are known as base substitution, insertion, deletion, and frameshift. Base substitutions are the simplest type of gene-level mutation \cite{cite_004}. Base substitution is also known as point mutation. The mutation the nucleotide base of DNA sequences to encode proteins \cite{cite_003}. In the case of substitutions, one chemical base is exchanged for another during DNA replication, such as switching A to a G. Only one codon is affected in this mutation.

Substitutions cause one of three changes:
\begin{itemize}
    \item change a codon to one that encodes a different amino acid and cause a small change in the protein produced. For example, sickle cell anemia is caused by a substitution in the beta-hemoglobin gene, which alters a single amino acid in the protein produced.
    \item change a codon to one that encodes the same amino acid and causes no change in the protein produced. These are called silent mutations.
    \item change an amino-acid-coding codon to a single "stop" codon and cause an incomplete protein. This can have serious effects since the incomplete protein probably won't function.
\end{itemize}

Altogether, there are three different subtypes of point mutations:
\begin{itemize}
    \item Missense mutation: This type of mutation is a change in one DNA base pair that results in the substitution of one amino acid for another in the protein made by a gene. An example of a missense mutation is sickle-cell anemia which mutates a codon in the gene and negatively impacts hemoglobin, or the amount of oxygen-carrier proteins in the blood \cite{lodish2008molecular}.
    \item Nonsense mutation: A nonsense mutation is also a change in one DNA base pair. Instead of substituting one amino acid for another, however, the altered DNA sequence prematurely signals the cell to stop building a protein. This type of mutation results in a shortened protein that may function improperly or not at all.
    \item Silent mutation: Some mutations that change DNA bases do not have any effect on the sequence of amino acids in the protein. These mutations are called silent mutations and they do not affect the structure or function of the protein because there is no effect on the amino acid sequence \cite{cite_007}.
\end{itemize}

Regarding methods and technology for point mutations, there are currently no established therapeutic approaches \cite{Bhakta}.

Insertions and deletions can result in frameshift mutations, because the entire frame of the base shifts to make room for an additional base or the subtraction of a base. “If one or two bases are deleted the translational frame is altered resulting in a garbled message and nonfunctional product. A deletion of three or more bases leave the reading frame intact. A deletion of one or more codons results in a protein missing one or more amino acids. This may be deleterious or not” \cite{cite_003}.  Insertion and deletion occur when the replicating strand "slips," or wrinkles, which allows the extra nucleotide to be incorporated, or a nucleotide is omitted from the replicated strand \cite{cite_004}.

Frameshift mutations are the results of insertions and/or deletions from the DNA template during the replication process. The alteration occurs during translation because ribosomes read the mRNA strand in terms of codons, or groups of three nucleotides, also known as “reading the frame” \cite{cite_004}. When the ribosome encounters the mutation, it will read the mRNA sequence differently, which can result in the production of an entirely different sequence of amino acids in the growing polypeptide chain \cite{cite_004}. For example, consider the sentence with each word representing a codon: "The fat cat sat”; if we delete the first letter and analyze the sentence in the same way, there’s a lack of communication and information \cite{cite_002}.  The result generates truncated, or shortened, proteins that are as useless as "hef atc ats at" is uninformative \cite{cite_002}. This error changes the relationship of all nucleotides to each codon, and effectively changes every single codon in the sequence \cite{cite_004}.

A couple of lesser known types of mutations are duplication and repeat expansion. A duplication consists of a piece of DNA that is abnormally copied one or more times, which thereby alters the function of the resulting protein \cite{cite_007}. Repeat expansion, on the other hand, is a mutation that increases the number of times that a short DNA sequence, like a trinucleotide, is repeated. These mutations are “unstable (dynamic) mutations that often change size in successive generations” \cite{Paulson:2018aa}. Currently, at least 30 genetic diseases, such as Fragile X, Spinal and Bulbar Muscular Atrophy, Myotonic Dystrophy, and Huntington’s disease, are believed to be caused by repeat expansion mutations \cite{cite_012}.

Many mutations can be harmless, however, there are the rare mutations that cause debilitating side effects, diseases, and even death. A change in a gene’s instructions for making a protein can cause the protein to malfunction or to be missing entirely. When a mutation alters a protein that plays a critical role in the body, it can disrupt normal development or cause a medical condition, a condition called a genetic disorder \cite{cite_016}.

Diseases, viruses, and disabilities are linked to mutations. In the development of disease, the resistance, immunity, age, and nutritional state of the person exposed, as well as virulence or toxicity of the agent and the level of exposure, all play a role in determining whether disease develops \cite{cite_006}. Many common human diseases are due to mutations in single genes, arising by spontaneous germ-line mutations \cite{lodish2008molecular}. "Spontaneous" refers to the fact that the changes occur in the absence of chemical, radiation, or other environmental damage \cite{pray2008dna}. A few causes of mutations and their effects are listed below:

\textbf{Errors in DNA Replication:}
On very, very rare occasions the DNA enzyme will incorporate a noncomplementary base into the daughter strand. During the next round of replication the misincorporated base would lead to a mutation \cite{cite_003}. This type of mispairing is known as wobble. It occurs because the DNA double helix is flexible and able to accommodate slightly misshaped pairings \cite{cite_003}. Errors in DNA replication can be germ-line or somatic mutations that can cause cancer \cite{cite_003}. 

\textbf{Errors in DNA Recombination:}
DNA often rearranges itself by a process called recombination which proceeds via a variety of mechanisms. Occasionally DNA is lost during replication leading to a mutation \cite{cite_003}.

\textbf{Chemical Damage to DNA:}
Many chemical mutagens, some exogenous, some man-made, some environmental, are capable of damaging DNA. Many chemotherapeutic drugs and intercalating agent drugs function by damaging DNA \cite{cite_003}.

\textbf{Radiation:}
Gamma rays, X-rays, even UV light can interact with compounds in the cell generating free radicals which cause chemical damage to DNA \cite{cite_003}.

People who survive high doses of radiation, like the survivors of the atomic bomb blasts in Japan in 1945, display definite chromosomal abnormalities in certain types of their circulating white blood cells \cite{cite_006}. Radiation can lead to an increase in cancers such as leukemia (a white-blood cell cancer) \cite{cite_003}.

Different diseases and disorders occur due to genetic alteration and other factors that contribute to mutation. Maternal age plays an important role in predisposing toward genetic injury, such as congenital malformations and down syndrome \cite{cite_006}. Down syndrome is a mutation that involves an error during cell division where the total number of chromosomes are 47 instead of the typical baby’s 46 chromosome count, a chromosome being a DNA molecule with genetic material \cite{cite_006}. Therefore, these babies are born with extra genetic material.

Germ-line mutations don’t need environmental factors, they are virtually genetic in nature. Since many germ-line mutations have occurred for a long time, they have more treatment options available, though treatments do not guarantee results. Diseases such as Turner’s syndrome can be treated with hormones to increase linear growth and their maturation during pubic years \cite{cite_006}.

Some mutations are spontaneous and can result from either environmental factors or germ-line mutations. Retinal tumors in children are a hereditary form of retinoblastoma. However, there’s a spontaneous form that occurs on rare occasions due to a somatic mutation of the chromosome. In this case, whether it’s a somatic or germ-line mutation, the treatment for the disease will remain the same.

Medicinal treatments of DNA mutations might worsen the disease or cause a more debilitating mutation. Applying drugs and chemicals to an individual “involves metabolism of the drug by detoxification enzymes into reactive intermediates that damage DNA. The mutations that remain are those not removed by DNA repair enzymes. In contrast to viruses, drugs and chemicals have been shown to cause mutations not only in human cells in culture but also in a living host” \cite{cite_006}. Therefore, it is difficult to determine the best pathway to combat these DNA mutations.

Current therapies and treatments for DNA mutations include medications, vaccines, and holistic care, such as therapy. Gene and RNA editing are newer forms of treatment that are not as convenient or affordable as conventional, medicinal care. However, gene and RNA editing is becoming more affordable with newer technologies and companies that are investing in the future for medical advancement.

Start-ups and academic labs are creating enabling technologies that implement RNA editing methods. This is accomplished by engineering molecules that commandeer our own enzymes to precisely edit RNA \cite{cross2019watch}. Several RNA therapies have already been licensed for use in the clinic according to Harries \cite{cite-Harries}. 

Medicines and therapies from gene and RNA editing provide essential evidences of applications within the medical field. Gene editing therapies are currently useful for patients with anemia and eye degeneration. However, gene editing can cause more suffering to a patient with longterm diseases and disorders. RNA editing has improved drug delivery with less side effects to the patient, particularly those suffering from blood disorders. Though, therapies for RNA editing is limited at this time.

Future directions for gene and RNA editing are curing and the prevention of diseases and viruses, like cancer and HIV. An increase in RNA editing will expand therapeutic applications. With companies investing in editing research and technologies, and creating their own platforms, there is a greater potential for editing in the medical field, rather than using pharmaceuticals to solve genetic disorders derived from in vivo mutations. Some editing technologies and companies are implementing deep learning environments to observe and mimic the transcription process of DNA and RNA for further research into mutation correction.

%%%%%%%%%%%%%%%%%%%%%%%%%
\subsection{Research Problem}
	RNA and gene editing research are limited due to the delivery systems and thereby the applications to viruses and diseases among the human population. Recent technologies have been developed to diagnose and identify specific RNA-based diseases and viruses similar to COVID and HIV. However, gene editing is more complex and concrete. Research in the highly-adaptable RNA editing process remains prevalent within methodologies due to the predominant focus on gene editing applications; however, the same technologies are being adapted and adjusted for RNA editing. Nonetheless, the problem of delivery and efficient on-target technologies limits clinical trials without chemical modification of the drug therapy.

\subsection {Purpose of the Study}
The purpose of this study is to inform researchers and students to the benefits of RNA and gene editing. The applications of gene and RNA editing technologies to population diseases, viruses, and genetic debilitations is virtually provided through therapies to alleviate symptoms and pain. Understanding of these methods and technologies optimizes future studies and applications to the editing research field. Modification of the methods is a goal to enhance the abilities of the technologies. Confidence on the material and methods will inspire students to initiate their own solutions to the research problems. Reading the literature review will save time and instill the best research practices for the collection of data and rewriting problems and solutions based on their own understanding of the material. The review of relevant literature enables students to complete the editing process individually by potentially combining the editing methods.

\subsection{Audience} 

Students, researchers, biologists, programmers, and medical field individuals are those with the most interest in RNA and gene editing; therefore, this literature will address methodologies, technologies, and applications with relevance to the editing research field. The topic will motivate students and researchers to build on and improve the methods within the article. The audience will discover that the review is easier to read and understand from a layman’s point of view which expands the defined audience to researchers not privy to the information. 

\subsection{Contribution} 
The impact of this review to the gene and RNA editing research field is significant to future discoveries and applications. Knowledge of the methods and technologies will influence the development of skills in gene and RNA editing. Kennesaw State University contributes to students’ studies by providing the means and ability to expand the editing research field. Creating a review of gene and RNA editing research aids students and researchers to progress technologies and applications. Companies and start-ups subsidize funds for research to further clinical applications. While, contributions from individual researchers allows for the material to relate to students who are not as savvy on the topic of investigation. 

Further, this paper's scope falls within biotechnology as it explores innovative gene and RNA editing methods and technologies, such as CRISPR, which are pivotal for developing advanced therapeutic applications and improving the detection and treatment of genetic diseases and public health threats.

\subsection{Motivation}
The motivation of this literature review derives from expanding education and studies in biotechnology to improve human conditions. Gene and RNA editing provides a solution to hereditary debilitations that limit individuals. Inspiring individuals to edit mutations enables furtherment of methods and technologies. Knowledge of the editing applications and their potential to aid in disease rehabilitation motivates start-ups to finance research into the research field.

\subsection{Paper Organization}
\textcolor{black}{This review is structured to progressively build upon a foundational understanding of the gene and RNA editing field. The introduction provides essential definitions and terminology, establishing a baseline for readers. As the review advances, terminology will expand, deepening the initial knowledge. The section on related works elucidates the contributions and methodologies of various authors in the field. The architecture section delves into the methods of gene and RNA editing, highlighting the hardware and software utilized to develop these systems. The enabling technologies section integrates hardware and software to detail specific gene and RNA editing devices and their potential applications. The applications section covers both therapeutic uses and medications in clinical trials, including initiatives by companies and start-ups invested in gene and RNA editing research. Finally, the future directions section explores potential advancements and new arenas for applying these technologies, providing insight into the trajectory of gene and RNA editing research.}

%\begin{figure*}
%\centering
%\includegraphics[width=5in]{Figures/GeneRNAedit_roadmap.jpg}
%\caption{Paper roadmap.}
%\label{fig:roadmap}
%\end{figure*}

%%%%%%%%%%%%%%%%%%%%%%%%%

\begin{table*}[!ht]
\scriptsize
\caption{\textcolor{black}{Summary of existing surveys and reviews related to gene and RNA editing.}}
\label{tab:survey}
\begin{tabular}{|>{\arraybackslash}p{2.5cm}|>{\arraybackslash}p{0.9cm}|>{\centering\arraybackslash}p{0.8cm}|>{\centering\arraybackslash}p{0.7cm}|>{\centering\arraybackslash}p{0.6cm}|>{\centering\arraybackslash}p{1cm}|>{\centering\arraybackslash}p{0.9cm}|>{\arraybackslash}p{7cm}|}
\hline
\textbf{Refs. (Author)}
& \textbf{\tiny Year of \newline Publication}
& \textbf{\tiny Gene Editing}
& \textbf{\tiny RNA Editing}
& \textbf{\tiny Methods}
& \textbf{\tiny Enabling \newline Tech}
& \textbf{\tiny Applications}
& \textbf{\tiny Overview}
\\ \hline \hline

% ########################################################################\

Adli \cite{adli2018crispr} & 2018 & \checkmark{} & \xmark{} & \checkmark{} & \checkmark{} & \xmark{} & History of gene editing and development of CRISPR systems \\ \hline

% ########################################################################\

Beisel \cite{beisel2018crispr} & 2018 & \xmark{} & \xmark{} & \xmark{} & \checkmark{} & \xmark{} & Gene expression profile capture and assessment of viral cell behavior to develop and improve editing technologies \\ \hline

% ########################################################################\

Eisenberg et al. \cite{Ishinoe00580-17} & 2018 & \xmark{} & \checkmark{} & \checkmark{} & \checkmark{} & \xmark{} & A-to-I RNA editing method and engineered technology via organism bacteria strains \\ \hline

% ########################################################################\

Ishino et al. \cite{Ishinoe00580-17} & 2018 & \checkmark{} & \xmark{} & \checkmark{} & \checkmark{} & \checkmark{} & Sequencing methodologies for technologies and applications throughout the gene editing history \\ \hline

% ########################################################################\

Pal \cite{cite_042} & 2018 & \checkmark{} & \xmark{} & \xmark{} & \checkmark{} & \checkmark{} & Gene therapies and treatments in the healthcare system as approved by the FDA \\ \hline

% ########################################################################\

Bergman \cite{cite-Perspectives}  & 2019 & \checkmark{} & \xmark{} & \checkmark{} & \checkmark{} & \xmark{} & Perspectives on gene technologies and methods to evaluate bioethics and types of applications  \\ \hline

% ########################################################################\

Roth et al. \cite{roth2019genome} & 2019 & \xmark{} & \checkmark{} & \checkmark{} & \xmark{} & \checkmark{} & RNA editing activity and methods that are applicable through certain indexes and systems to tissues of the body \\ \hline

% ########################################################################\

Harries \cite{cite-Harries}  & 2019 & \xmark{} & \checkmark{} & \xmark{} & \xmark{} & \checkmark{} & RNA editing delivery and applications with consideration to the benefits and limitations as compared to gene editing \\ \hline

% ########################################################################\

DeWeerdt \cite{cite_046} & 2019 & \xmark{} & \checkmark{} & \xmark{} & \xmark{} & \checkmark{} & Therapeutic applications and history of RNA editing and their three therapeutic categories \\ \hline

% ########################################################################\

Samuel \cite{SAMUEL2019}  & 2019 & \xmark{} & \checkmark{} & \checkmark{} & \xmark{} & \checkmark{} & Understanding basics of RNA editing techniques and methods with enzymes utilized for the processes \\ \hline

% ########################################################################\

Sturm \cite{cite_003}  & 2019 & \checkmark{} & \xmark{} & \checkmark{} & \xmark{} & \checkmark{} & Genetic mutations with gene editing methods and applications to repair damage to DNA  \\ \hline

% ########################################################################\

Yan et al. \cite{yan2019crispr} & 2019 & \checkmark{} & \checkmark{} & \xmark{} & \checkmark{} & \xmark{} & CRISPR-Cas systems with explanation of the RNA editing and detection technologies \\ \hline

% ########################################################################

 Abudayyeh et. al \cite{cite_033}  & 2020 & \xmark{} & \checkmark{} & \xmark{} & \checkmark{} & \xmark{} & Overview of CRISPR/13 technologies and the enzymes it utilizes \\ \hline

% ########################################################################

Aquino-Jarquin \cite{aquino2020novel} & 2020 & \xmark{} & \checkmark{} & \checkmark{} & \checkmark{} & \checkmark{} & Focuses on RNA editing and different technologies and applications \\ \hline

% ########################################################################

Gearing \cite{cite_036} & 2020 & \xmark{} & \checkmark{} & \xmark{} & \checkmark{} & \xmark{} & RNA editing technologies and how they operate in organisms \\ \hline

% ########################################################################

Li et al. \cite{li2020applications} & 2020 & \checkmark{} & \xmark{} & \xmark{} & \checkmark{} & \checkmark{} & Gene editing technology for human mutations as applied to targeted therapy  \\ \hline

% ########################################################################

Reardon \cite{reardon2020step} & 2020 & \xmark{} & \checkmark{} & \xmark{} & \checkmark{} & \checkmark{} & Comparison of editing technologies, and applications of RNA editing therapies \\ \hline

% ########################################################################

Storch \cite{storch2020crispr} & 2020 & \xmark{} & \xmark{} & \xmark{} & \checkmark{} & \xmark{} & Detection platforms designed by Cas proteins to enable exact target sequencing for viruses \\ \hline

% ########################################################################

Tian et al. \cite{tian2020dna} & 2020 & \xmark{} & \xmark{} & \xmark{} & \checkmark{} & \xmark{} & Detection platforms designed by Cas proteins to enable exact target sequencing for viruses \\ \hline

% ########################################################################

Mah \cite{cite_001}  & 2021 & \checkmark{} & \xmark{} & \checkmark{} & \checkmark{} & \checkmark{} & Gene editing methods, technologies, and applications from a CRISPR standpoint \\ \hline

% ########################################################################\

Scarpelli et al \cite{cite_006}  & 2021 & \xmark{} & \xmark{} & \xmark{} & \xmark{} & \xmark{} & Provides background and definition to human disease and mutations \\ \hline

% ########################################################################

%\textcolor{black}{Gaj et al. \cite{gaj2021next}} & 2021 & - & - & - & - & - & Advancements in CRISPR technologies and their transformative applications in gene and cell therapy. \\ \hline

\textcolor{black}{Gaj et al. \cite{gaj2021next}} & 2021 & \checkmark{} & \xmark{} & \checkmark{} & \checkmark{} & \checkmark{} & Advancements in CRISPR technologies and their transformative applications in gene and cell therapy. \\ \hline

% ########################################################################

%\textcolor{black}{Bharathkumar et al. \cite{bharathkumar2022crispr}} & 2022 & - & - & - & - & - & Overview of therapeutic applications of CRISPR/Cas-based modifications. \\ \hline

\textcolor{black}{Bharathkumar et al. \cite{bharathkumar2022crispr}} & 2022 & \checkmark{} & \xmark{} & \checkmark{} & \xmark{} & \checkmark{} & Overview of therapeutic applications of CRISPR/Cas-based modifications. \\ \hline

% ########################################################################

%\textcolor{black}{Liu et al. \cite{liu2022crispr}} & 2022 & - & - & - & - & - & Review of diverse CRISPR-Cas tools available for gene editing. \\ \hline

\textcolor{black}{Liu et al. \cite{liu2022crispr}} & 2022 & \checkmark{} & \xmark{} & \xmark{} & \checkmark{} & \checkmark{} & Review of diverse CRISPR-Cas tools available for gene editing. \\ \hline

% ########################################################################

%\textcolor{black}{Wang et al. \cite{wang2023crispr}} & 2023 & - & - & - & - & - & Review of a decade of CRISPR technology, outlining its monumental impact on genome editing and future possibilitie. \\ \hline

\textcolor{black}{Wang et al. \cite{wang2023crispr}} & 2023 & \checkmark{} & \xmark{} & \checkmark{} & \checkmark{} & \xmark{} & Review of a decade of CRISPR technology, outlining its monumental impact on genome editing and future possibilities. \\ \hline

% ########################################################################

%\textcolor{black}{Saber et al. \cite{saber2023review}} & 2023 & - & - & - & - & - & Latest CRISPR-based genome-editing tools, including base editing and prime editing. \\ \hline

\textcolor{black}{Saber et al. \cite{saber2023review}} & 2023 & \checkmark{} & \xmark{} & \checkmark{} & \checkmark{} & \xmark{} & Latest CRISPR-based genome-editing tools, including base editing and prime editing. \\ \hline

% ########################################################################

%\textcolor{black}{Villiger et al. \cite{villiger2024crispr}} & 2024 & - & - & - & - & - & The use of CRISPR technologies for editing the genome, epigenome, and transcriptome. \\ \hline

\textcolor{black}{Villiger et al. \cite{villiger2024crispr}} & 2024 & \checkmark{} & \xmark{} & \checkmark{} & \checkmark{} & \xmark{} & The use of CRISPR technologies for editing the genome, epigenome, and transcriptome. \\ \hline

% ########################################################################

%\textcolor{black}{E. Deneault \cite{deneault2024recent}} & 2024 & - & - & - & - & - & Advances in therapeutic gene editing applications, particularly focusing on their potential to treat various genetic disorders. \\ \hline

\textcolor{black}{E. Deneault \cite{deneault2024recent}} & 2024 & \checkmark{} & \xmark{} & \checkmark{} & \xmark{} & \checkmark{} & Advances in therapeutic gene editing applications, particularly focusing on their potential to treat various genetic disorders. \\ \hline

% ########################################################################

%\textcolor{black}{Chehelgerdi et al. \cite{chehelgerdi2024comprehensive}} & 2024 & - & - & - & - & - & Mechanisms, challenges, and applications of CRISPR-based gene editing, with a special focus on its use in cancer therapy. \\ \hline

\textcolor{black}{Chehelgerdi et al. \cite{chehelgerdi2024comprehensive}} & 2024 & \checkmark{} & \xmark{} & \checkmark{} & \xmark{} & \checkmark{} & Mechanisms, challenges, and applications of CRISPR-based gene editing, with a special focus on its use in cancer therapy. \\ \hline

% ########################################################################

This article & 2024 & \checkmark{} & \checkmark{} & \checkmark{} & \checkmark{} & \checkmark{} & Discusses the state-of-art security frameworks, challenges, and techniques for securing both inter- and intra-vehicular communication along with related applications and use-cases \\ \hline

% ########################################################################

\end{tabular}
\end{table*}

%---------------------------------------------------------------------------------

\section{Related Works}
Multiple researchers and authors have work associated with the research for this article. In order to fully understand the basis of gene and RNA editing, references are cited throughout the work. These authors’ works are related to research and results involving the previous and present modulations to genes and RNA. The research remains as a standing point for the current discoveries and technologies, while also propelling and inspiring future endeavors and discovers in biotechnological editing. Contributions of fifteen authors are regarded and detailed below, along with a table (Table \ref{tab:survey}) that compares thirty authors’ works relevant to gene and RNA editing research.

The engineering of programmable RNA editing methods via ADAR enzymes contributes to RNA technologies \cite{aquino2020novel}. The article highlights A-to-I base modification of RNA editing and their technologies. Descriptions of editing platforms are CIRTS, RESCUE, RESTORE, and LEAPER. The author assesses the efficiencies and limitations of each editing technique. To understand the programmable nature of RNA editing, Aquino-Jarquin highlights point mutations and the role of ADAR enzymes in the editing process. Future improvements to these technologies are discovered by comparing and contrasting each editing specificity of nucleotide sequence and off-targeting possibilities.

Basic understanding of gene editing methods and technologies expands on gene editing techniques and tools to incorporate applications and future developments in the field \cite{cite_001}. Mah expands on the DSB (or double-strand break) that occurs with genome editing. Definitions and detailed images of gene editing contribute to research and different types of techniques. There are four primary gene editing technologies that have developed and been utilized throughout the history of the genome. Mah describes each technology, limitations, and explains how they built on one another. Restriction enzymes, ZFNs, TALENs, and CRISPR are the four techniques that companies implement for gene editing, so that individuals can determine the best method for application.

Focusing on RNA editing, particular with the enzyme Cas13 that was founded via CRISPR technology, contributes to the efficient value and future directions of RNA editing methods and technologies as compared to gene editing systems \cite{cite_036}. Gearing contributes to the RNA editing enabling technologies with mention of editing methods A-to-I editing and C-to-U editing. With these methods, the author notes the technologies REPAIR and RESCUE that is applicable to the contrasting methods. Advantages of RNA editing systems compared to gene editing systems is highlighted in the article and applied to this literature review in order to contrast the two technologies.

Evidence of RNA editing therapies in present and future endeavors contributes to the application of our research with barriers and basics about the RNA therapy methodology applied today \cite{cite_046}. The author examines three predominant categories of RNA treatment. Introducing future directions of RNA editing provides a baseline for applications. Therapeutic approaches deliver a sum of the three RNA therapy categories, and knowing each category is essential to expand on the elements of the editing therapies today. The findings are supported with other researchers from RNA editing fields. The author reiterates the editing process and the limitations of RNA therapies by explaining the methods and medicines such as RNAi.

Research regarding RNA editing allows for a comparison of gene editing methods and technologies \cite{reardon2020step}. Gene editing methods and architectures are MNs, ZFNs, TALENs, and CRISPR. The author notes applications of RNA editing therapies including cancer, pain, and genetic disorders. Historical findings of the first RNA editing methods and technologies are identified. According to Reardon, Thorsten Stafforst discovered the mRNA protein significance to editing, however, gene editing had a breakthrough, and RNA editing was placed in the background. Reardon explains that gene editing can cause permanent damages, while RNA editing is a more beneficial alternative because it mimics genetic variations rather than cutting them from the gene. ADAR proteins are highly effective in the RNA editing process, which adds to future research and clinical trials.

Analyzing RNA editing technology with the implementation of different types of enzymes iterates evidence that supports applications of therapies and correction of human mutations with RNA editing technologies \cite{Cox1019}. The engineered transcriptome technology of Cas13 is highlighted, because there are different types of Cas13 proteins with specific characteristics that affect RNA editing activity. Cox defines the sequence parameters and flexibility of RNA editing and therapeutic AAV vectors. REPAIR systems are contrasted with other editing systems, and Cox notes the RNA editing technological advantages, particularly that the additional residues lends to technological specificity improvement.

Informing students and researchers about therapies with the application gene editing technologies allows for improvement to the technologies and expansion of defined therapies \cite{LIU201717}. The article contributes to therapeutic applications of Cas9 technology and the current results of clinical trials. The author defines the different applications of CRISPR by defining in vivo and in vitro deliveries. The article explains Cas9’s role in gene editing and how guide RNA directs the enzyme to efficiently target sites in order to correct genetic mutations. Liu provides evidence of clinical trials involving applications to cancer and immunotherapy via engineered Cas9’s enzymes.

Therapeutic applications contribute to gene editing influenced therapies \cite{shim2017therapeutic}. Deliveries that are currently being adapted for clinical trials gives researchers an understanding of system implementation benefits and limitations. The author addresses current research trends and delivery strategies for gene editing-based therapeutics in non-clinical and clinical settings. One of the main delivery strategies focused on is in vivo with CRISPR/Cas9. He surmises that non-viral and viral vectors are applicable to gene therapies. Based on evidence non-viral delivery is better than viral due to the non-viral vector’s flexibility for deliverable components.  Research and clinical trials from external sources support the data the author provides in order to inform the audience of the best delivery systems available for an individual/patient.

Expansion on RNA editing strategizes its ability as an immune protector which contributes to the RNA editing methods and technologies of this article \cite{eisenberg2018rna}. One RNA method that is highlighted is A-to-I editing. It is examined as a beneficial alternative to gene editing due to its ability to edit non-coding areas of the genome. Evidence shows that this RNA editing method and implementation diminishes the effect of an immune reaction by other cells, which occurs in other editing methods. Knowing the RNA editing methods aids in preparing and creating architectures that will fulfill the precise base modification for future technologies and applications.

Architecture are applied through certain preferable technologies, such as CRISPR; however some of these technologies are improved with software algorithms and detection devices like CARMEN and SHERLOCK \cite{cite_033}. Emphasizing RNA editing methods and technologies is essential to improving the site specific sequencing devices. Adjustable and customizable technologies align the targeted sequencing area with the optimal base modification for utilization. In one article, the author details the usage of Cas13 technology REPAIR, which is an A-to-I base modification. More research and lab work of A-to-I editing expanded the capability of RNA systems to convert C-to-U base conversion. With this new method, Abudayyah notes another technology that was developed called RESCUE. Cytosine deamination is defined and ADAR evolution is displayed via residual mutations. Abudayyeh's many research articles contribute a sufficient amount of RNA editing evidence, including a comparison of the editing efficiencies in RNA methods and technologies.

Several gene therapy applications in the healthcare industry are mentioned to understand the efficiency and applicability of editing to each individual \cite{cite_042}. Compartmentalization of the predominate gene technologies that currently residing in the gene therapy application domain is informative for necessary and desirable medical advances. In 2017, the FDA approved cell-reprogramming therapies to treat leukemia and other sensory conditions due to the proven successes of the therapies. According to the article, cells have been engineered to treat HIV, lymphoma, and Crohn’s disease. Pal utilizes university lab research to support the explanation of each treatment, and the specific cells that are being engineered for their relative conditions. Microsoft and other tech giants are investing into gene editing research to fine tune the technologies for healthcare applications.

Identifying viral infections is vital to accurately and efficiently target RNA sequences for both gene and RNA editing \cite{storch2020crispr}. Detection devices provide a targeted approach to therapy applications. Illustrations and laboratory research are displayed with florescent markers detailing the significance and efficiency of detection technologies possessing viral sequences of RNA. The author explains the basic methods and technologies of CRISPR and how it contributed to the discovery and application of detection platforms. Comparing detection technologies emphasizes the most efficient editing technologies currently available. The best editing technologies are capable of precise sequence cutting or exact base modification; however, each detection and editing technology has both strengths and weaknesses. Therefore, the chosen technology is based on the researcher's and/or the patient’s preference, particularly in regard to the desired application.

The definition and usage of gene expression in gene and RNA systems is a significant concept for researchers and students to understand in order to proceed in editing applications \cite{beisel2018crispr}. Capturing gene expression profiles of viral RNA transcript sequences can provide information regarding the behavior of viral cells. Studying the characteristics of viral cells can improve gene and RNA editing systems.  By utilizing other RNA bacteria sequences to engineer Cas proteins that will track and identify gene expression profiles with abnormalities. Evidence within the article introduces record-seq technology that will improve treatments and reduce the risk of an immune response within the body to the engineered cells. Particular emphasis on CRISPR systems implementing RNA editing provides insight into how devices utilize proteins for the RNA process.

An introduction to different types of gene and RNA editing technology contributes to the architectures and delivery conduits available for editing systems \cite{tian2020dna}. The article explains the editing, in one scenario, is guided by hairpin DNA probes to reduce sequence limitations. The method is effective and specifically covers in vivo and in vitro delivery applications as compared to other gene editing technologies. Most gene editing technologies are large in size, and thereby harder to pack and deliver. However, this newly advancing method and technology proves its effectiveness as a guiding tool. The article provides testing results and materials used in the process that verifies the performance of FEN1 cleavage with HpDNA in vitro on both DNA and RNA cells. The system does not have strict requirements for target site sequences which makes it an ideal alternative technology for gene and RNA editing in biotechnology.

Expansion on RNA approaches supports the methods and technologies of RNA editing \cite{yablonovitch2017evolution}. A-to-I editing is viewed as the predominant practice for RNA editing. The post-transcriptional modification optimizes cell protein function, diversity, and flexibility so the cell adapts to varying environments. The author discusses results that pertain to RNA editing evolution including the dynamic editing process with various results affected by a number of factors. The article identifies advances of technologies such as ADAR protein evolution and high-throughput sequencing, and their evident role in RNA editing and adaptation. Organism locations of ADAR enzymes and other applicable bacteria for RNA editing is identified and examined to determine the effectiveness of A-to-I editing, particularly noting the adaptability of the system.

%\newpage
%%%%%%%%%%%%%%%%%%%%%%%%%
\section{Architectures and Platforms}
To distinguish methods concerning RNA and gene editing, the basic editing process and relationship between RNA and DNA will be defined. The effects of an individual’s DNA blueprint identify characteristics and behaviors that editing can overcome and/or erase entirely. Adaptation via RNA editing and erasure via gene editing.

\begin{figure*}
\centering
\includegraphics[width=7in]{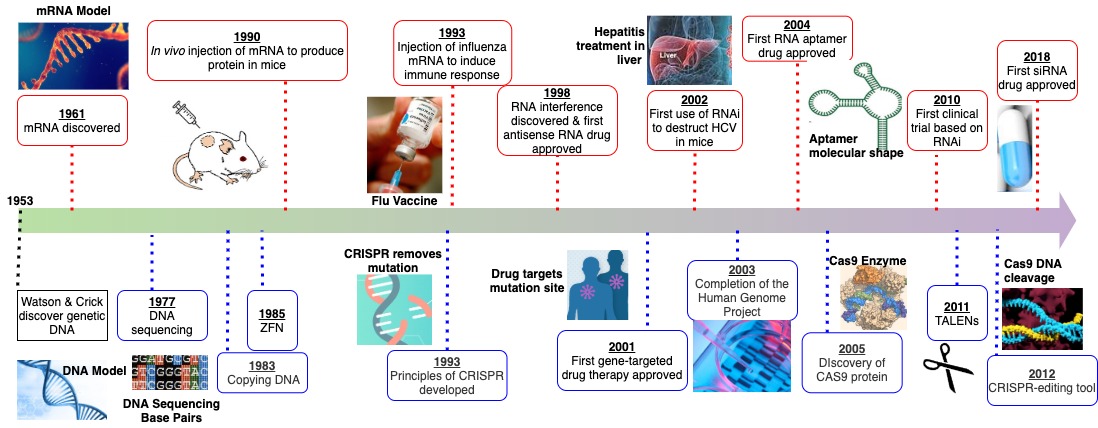}
\caption{Timeline of Gene and RNA editing discoveries and developments. Blue represents gene editing developments. Red represents RNA editing developments. }
\label{marker:timeline}
\end{figure*}

%%%%%%%%%%%%%%%%%%%%%%%%%
\subsection{Gene Editing Methods}
From the moment that DNA’s double helix was discovered in the 1950s, the idea of using gene editing to treat disease or alter traits became a holy grail for researchers in molecular genetics \cite{cite_006}. According to Li, the development of genetic engineering (manipulation of DNA or RNA) established a novel frontier in genome editing. The concepts of gene therapy arose initially during the 1960s and early 1970s \cite{friedmann1992brief}. Gene and RNA editing began (Fig. \ref{marker:timeline}) with the discovery of restriction enzymes that protect beneficial bacteria against abnormal bacteria. The discovery fueled the era of recombinant DNA technology, or gene editing, which then contributed to discovers in RNA editing \cite{adli2018crispr}.

\begin{figure}[h!]
\centerline{
\includegraphics[width=0.50\textwidth]{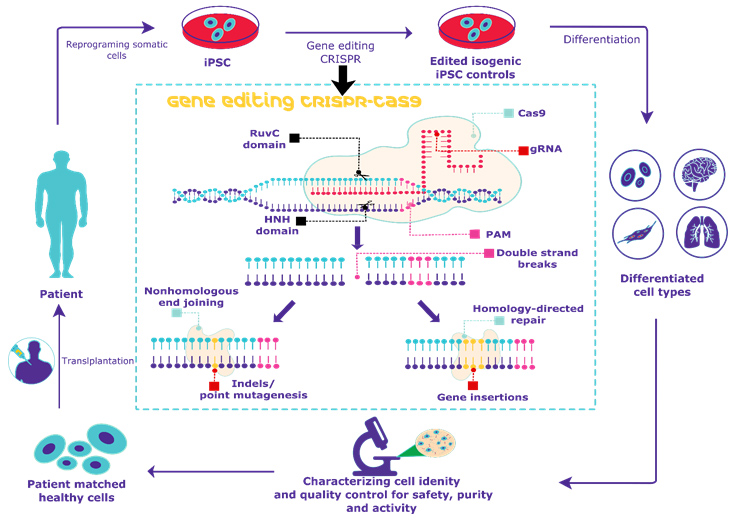}}
\caption{CRISPR gene editing process in the human body \cite{cite-imagegeneediting}.}
\label{marker:human}
\end{figure}

The most frequented definition of gene editing is "the manipulation of the genetic material of a living organism by deleting, replacing, or inserting a DNA sequence, typically with the aim of improving or correcting a genetic disorder" \cite{cite_genomedef}. Gene editing is used to make highly specific changes in the DNA sequence of a living organism, essentially customizing its genetic makeup. In the process of gene editing, enzymes are engineered to target a specific DNA sequence, where they introduce cuts into the DNA strands, enabling the removal of existing DNA and the insertion of replacement DNA \cite{cite_006}. The most well-known tool for gene-editing is CRISPR-Cas9 (Fig. \ref{marker:human}), Cas9 being the modified enzyme utilized in the editing process. In recent years, CRISPR has also been modified via a different enzyme of Cas13 to proceed in the RNA editing process.
	
\begin{figure}[h!]
\centerline{
\includegraphics[width=0.50\textwidth]{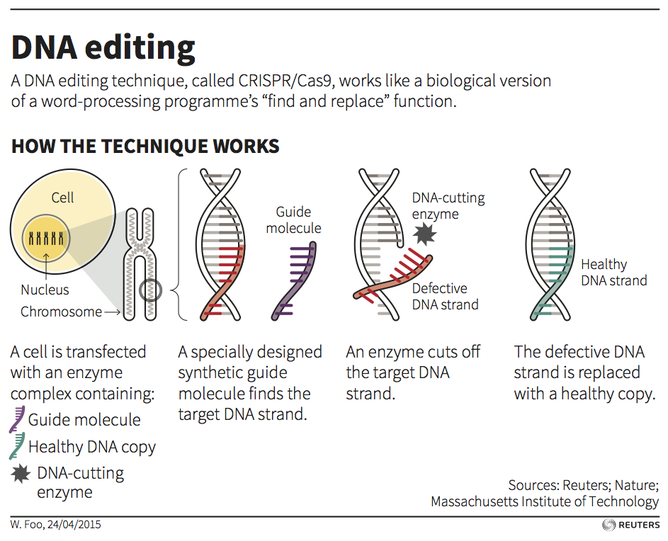}}
\caption{DNA editing \cite{cite-geneeditingpic}.}
\label{marker:DNAediting}
\end{figure}

	In order to truly correct genetic mistakes, researchers needed to be able to create a double-stranded break in DNA at precisely the desired location in the more than three billion base pairs that constitute the human genome (Fig. \ref{marker:DNAediting}). Once created, the double-stranded break could be efficiently repaired by the cell using a template that directed replacement of the “bad” sequence with the “good” sequence. However, making the initial break at precisely the desired location—and nowhere else—within the genome was not easy \cite{cite_006}. In order to accomplish the goal of gene editing a few approaches were used. \par

	Restriction Enzymes were the original genome editor, because it was found that these particular enzymes can be used to manipulate fragments of DNA, like those in genes. A restriction enzyme is a protein produced by bacteria that cleaves DNA at specific sites along the molecule \cite{cite_006}. Since these enzymes can cleave to sites in a cell, then they are easily used to cleave to bacteria within a molecule to eliminate infected organisms. In this methodology, the DNA code is cut and new material, such as a new sequence of nucleotides, is inserted. This method is not utilized today beyond that of molecular cloning, because restriction enzymes are limited by the nucleotide patterns they recognize. However, restriction enzymes play key roles in other processes such as DNA mapping, epigenome mapping, and constructing DNA libraries \cite{cite_001}. \par
	The next method discovered in the 1980s is ZFNs (Fig. \ref{marker:Types}a), or Zinc Finger Nucleases, which increased the genome recognition potential for editing processes. In order to edit DNA, scientists needed a more precise tool that lessened deleterious off-target effects. The idea was to have an engineered enzyme that cleaves to DNA similar to that of a restriction enzyme. However, scientists wanted to target genes to edit the designated areas. For this reason, ZFNs needed separate DNA-binding and DNA-cleavage domains \cite{Carroll:2011aa}. The purpose of this method is to restore the integrity of a newly edited gene.\par
	The third gene editing discovery in 2011 was transcription activator-like effector nucleases, also known by the acronym TALENs (Fig. \ref{marker:Types}b). It is structurally and methodically similar to ZFNs in that it uses the Fokl enzyme/nuclease to cut DNA. However, the DNA binding domains differ because TALENs use transcription activator-like effectors (TALEs), which enables single nucleotide resolution \cite{cite_001} This ability increases targeting capabilities and specificity compared to ZFNs. Despite its high labor and monetary cost, which hindered TALENs widespread adoption, the method is still used in the medical field to correct diseases characterized by loss of skin integrity leading to potentially fatal skin blisters and increased risk for skin cancer. Agriculturally, scientists found a way to create pathogen-resistant rice with TALENs. \par
 
 \begin{figure}[h!]
\centerline{
\includegraphics[width=.50\textwidth]{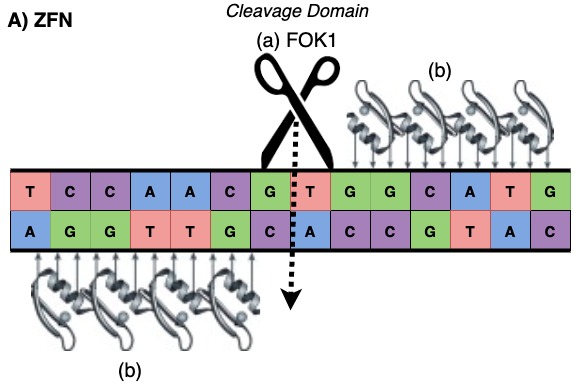}}
\caption{ZFN.}
\label{marker:ZFN}
\end{figure}

\begin{figure}[h!]
\centerline{
\includegraphics[width=.50\textwidth]{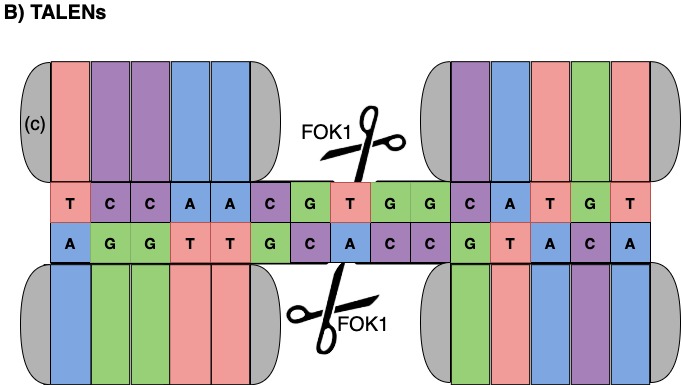}}
\caption{TALENs.}
\label{marker:TALENs}
\end{figure}

\begin{figure}[h!]
\centerline{
\includegraphics[width=.50\textwidth]{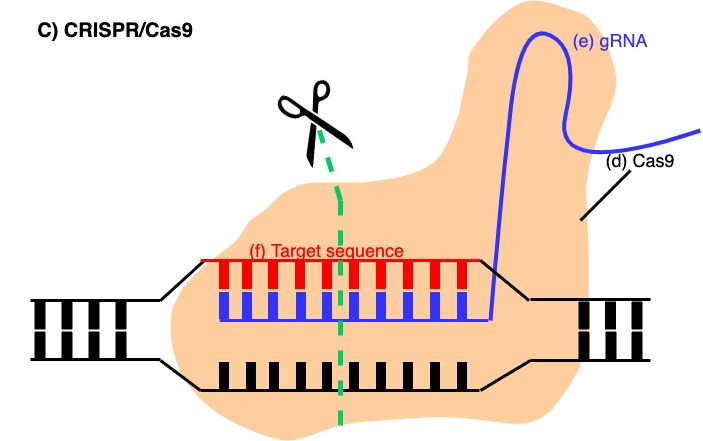}}
\caption{Types of gene editing procedures. A) ZFNs, B) TALENs, C) CRISPR-Cas9.}
\label{marker:Types}
\end{figure}

	In 2012, scientists discovered a new method derived from CRISPR-Cas9 (Fig. \ref{marker:Types}c), a system that has long existed in bacteria to help them fight off invading viruses. CRISPR is one of the most prestigious gene editing technology today due to its usage as a bacterial-defense system \cite{beisel2018crispr}. Supposedly, CRISPR could be used to edit genetic material in DNA. It is a two-component system consisting of a guide RNA and a Cas9 nuclease/enzyme that cuts the DNA within the ~20 nucleotide region defined by the guide RNA that scientists are able to customize. Algorithms have been developed to assess chances of off-target effects. The reason CRISPR is preferable in comparison to the previous methods is due to its customizable-nature and its cost-effectiveness. Despite the multitude of companies involved in the future of medicine via gene editing, “DNA editing can cause unwanted mutations in other parts of the genome — ‘off-target effects’ — which might create new problems” \cite{reardon2020step}.
 \par

\begin{figure}[h!]
\centerline{
 \includegraphics[width=.50\textwidth]{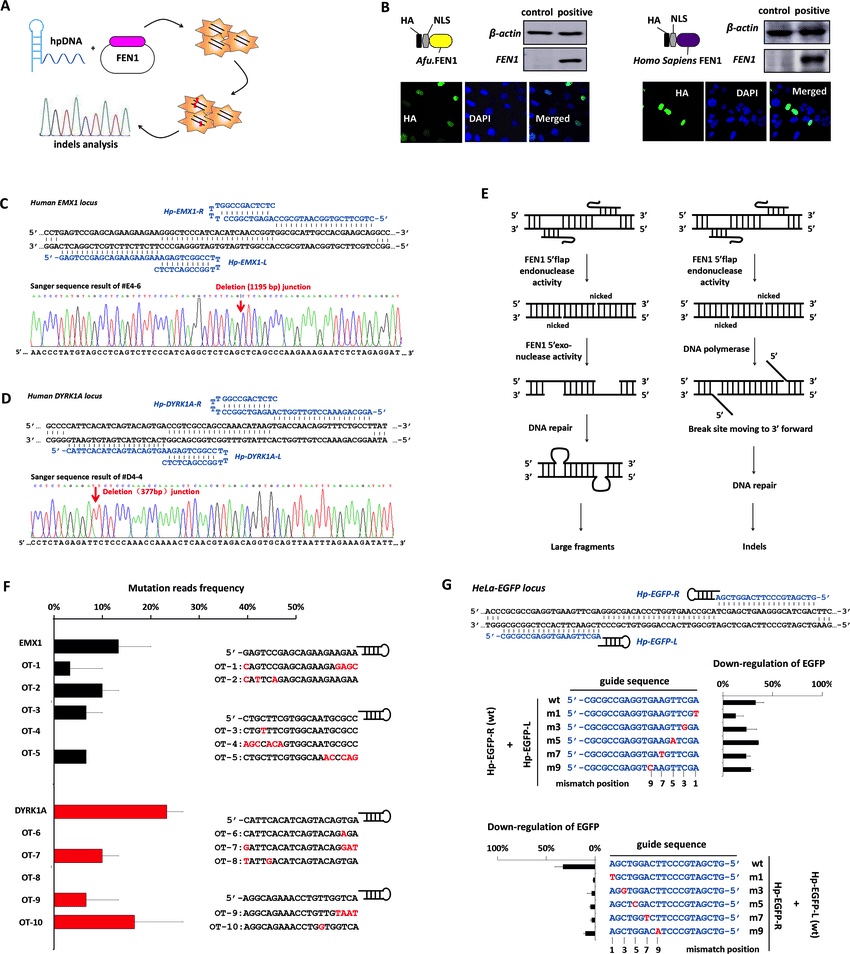}}
\caption{The HpSGN system can be harnessed to facilitate genome editing in human cells. (A) Schematic diagram of hpDNA-guided disruption of genomic DNA in HEK293A cells heterologously expressing FEN1. (B) Engineering of A. fulgidus/Homo sapiens FEN1 with NLS enables the import of FEN1 into mammalian nucleus. (C) The location of hpDNAs for the EMX1 gene and a typical Sanger sequencing result of the HpSGN-edited product with large fragment deletions. (D) The location of hpDNAs for the DYRK1A gene and a typical Sanger sequencing result of the HpSGN-edited product with large fragment deletions. (E) The hypothesis of large fragment deletions by the HpSGN system. (F) The frequency of mutation reads of the EMX1 gene and the DYRK1A gene edited by the HpSGN system. NGS was employed to analyze HpSGN-edited products. n = 2. (G) The efficiency of downregulation of EGFP in groups transfected with NLS-FEN1 plus wild/mutated probe of Hp-EGFP-L/R. n = 2 \cite{tian2020dna}.}
\label{marker:Tiansystem}
\end{figure}

	However, other methods of gene editing have stemmed from this research to improve the targeting specificity of CRISPR. One that is slightly low for practical applications is a structure-guided nuclease (SGN) DNA editing tool without sequence limitation. It was constructed based on fusing flap endonuclease 1 (FEN1) \cite{tian2020dna}. SGN functions like scissors. However, FEN1 is effective at capturing and detecting specific single base mutations. Utilizing FEN1 as a targeting system and combining it with another system that affects DNA might contribute to the gene editing process. The method includes a hairpin DNA probe-SGN system (hereafter designated the HpSGN system) (Fig. \ref{marker:Tiansystem}). The HpSGN system is composed of FEN1 nuclease (gray) and a hairpin DNA probe \cite{tian2020dna}. This system reaches the goal without no need to change proteins. The dual-functional feature of FEN1 in our HpSGN system makes it possible to develop a broad-spectrum antiviral kit to manage infections caused by either RNA or DNA viruses or both \cite{tian2020dna}. The versatile functions of our HpSGN system may also find significant applications in regulating protein expression by cleaving coding genomic DNA and mRNA. However, modifications to both RNA and DNA simultaneously may compromise specificity. For example, the HpSGN system cannot specifically target RNA molecules without affecting the integrity of its cohabitating genomic DNA in the nucleus \cite{tian2020dna}.\par
 
\subsection{RNA Editing Methods}
Editing, like splicing, represents a form of processing that has the capacity to amplify genetic diversity and alter gene product function by modifying the information transfer process at the posttranscriptional level \cite{SAMUEL2019}. RNA editing has become a generic term applied to a bewildering array of post-transcriptional processes whose effect is to change the nucleotide sequence of a mature RNA species relative to the encoding DNA sequence \cite{gray2012evolutionary}. RNA editing occurs when an enzyme selectively changes a nucleotide in the messenger RNA so that the meaning of its codon changes from one amino acid to another \cite{KNAPP2003602}.  Nucleotide changes are introduced into an RNA sequences which contributes to sequence variation(s). RNA editing “amplifies genetic plasticity by allowing the production of alternative protein products from a single gene” \cite{CHO2007417}. To correct single point mutations in RNA, researchers are designing antisense oligonucleotides that bind to an enzyme in our cells called ADAR (Fig. \ref{marker:RNAcorrections}). By also binding a complementary strand of RNA, these oligonucleotides coax ADAR into changing a mutation in an adenosine (A) base of the RNA into an inosine (I), which the cell reads as guanosine (G) \cite{cross2017crispr}. \par
	The reason that RNA editing is more beneficial than gene editing is due to the flexibility of the RNA strand as compared to DNA. It encompasses a wide variety of mechanistically and phylogenetically unrelated processes that change the nucleotide sequence of an RNA species relative to that of the encoding DNA \cite{gray2012evolutionary}. \par
	Another reason is due to the fact that “dysregulated RNA editing has been found in different types of cancers,” therefore, RNA editing has a greater ability in therapeutic benefits \cite{BOEGEL201925}. Modifying RNA is also easier because there’s a higher variety of them available to edit than DNA. RNA editing has been observed in mRNAs, tRNAs, and rRNAs, in mitochondrial and chloroplast encoded RNAs, as well as in nuclear encoded RNAs \cite{OHMAN2001125}. For these reasons, RNA editing possesses more methods than gene editing, “without the need to make hard-wired mutations at the DNA level” \cite{yablonovitch2017evolution}. \par
	There are two general methods of RNA editing via base modification and insertion/deletion that occur after the transcription process of DNA, but before it is translated into a protein, is A to I editing, and the other is C to U editing. RNA editing of either type leads to the formation of transcripts whose sequence differs from that of the genome template \cite{SAMUEL2019}. Subsets of base modifications are Cytidine and Adenosine Deaminase, and a subset of the insertion/deletion method is via Trypanosomatid Protozoa \cite{OHMAN2001125}. The only issue that is encountered with any editing activity is that “the roles of editing events in genes are directly associated with an immune response by signaling groups of proteins that respond in the presence of a virus” \cite{chan2020rna}. Nevertheless, Compared to gene editing, RNA editing can safely be approached because engineering RNA lowers the risk for permanent genomic changes, and even though off-site RNA editing can occur, it could be reversible \cite{aquino2020novel}. \par

\subsubsection{A-to-I Editing}
Adenosine-to-inosine (A-to-I) RNA editing removes amino group via chemical conversion of adenosine, which results in codon changes that alter the function of a chosen protein \cite{Zinshteyn:2009aa}. It is performed by adenosine deaminases acting on RNA (ADARs), proteins that bind to RNAs and alter their sequence by changing a familiar base of adenosine into a molecule called inosine \cite{reardon2020step}. Inosine is translated, or read, as guanosine, so A-to-I editing can alter the amino acid sequences of proteins, as well as affect other transcriptional processes like alternative splicing and microRNA (miRNA) binding \cite{yablonovitch2017evolution}. It is speculated that the ADAR proteins evolved as a defense against viruses, but many viruses with double-stranded RNA are unaffected by the enzymes \cite{reardon2020step}. Therefore, the idea is to evolve ADAR in the body so it can modify bases on its own. Human genome is “catalyzed by enzymes of the ADAR protein family1–3 that bind to double-stranded RNA (dsRNA) structures and alter the RNA molecule by modifying adenosines to inosines” \cite{roth2019genome}. \par
	Since diseases and disorders can mutate DNA and RNA, like how tumors change G to A, ADARs are similar to an overwrite function that changes letters chemically, without breaking the RNA molecule’s ‘backbone’ \cite{reardon2020step}. Two functional human ADAR orthologs, ADAR1 and ADAR2, which consist of N-terminal double-stranded RNA–binding domains and a C-terminal catalytic deamination domain…ADAR1 has been found to target mainly repetitive regions, whereas ADAR2 mainly targets nonrepetitive coding regions \cite{Cox1019}. \par
	First attempts to correct a mutated RNA with A-to-I editing was with muscular dystrophy disorder.  A protein is missing in this mutation due to a codon in the sequence; however, no definitive solution was discovered. Despite the method’s lack of effects on muscular dystrophy, A-to-I substitution can have a significant physiological or clinical impact \cite{aquino2020novel}. According to research, this method “has a plethora of biological functions, but its detection in large-scale transcriptome datasets is still an unsolved computational task” \cite{giudice2020investigating}. \par
	Though this form of editing works in plants and animals well, the human body is “largely nonadaptive in the case of human A-to-I editing” \cite{10.1093/molbev/msy011}. Mammalian studies reveal that most A-to-I editing sites are in non-coding parts of the genome, these are areas of the DNA that do not provide instructions for creating proteins. Editing of these non-coding sites is thought to have a critical role in protecting against activation of innate immunity that can occur during RNA editing attempts \cite{eisenberg2018rna}. For this reason, there are different methods implemented into the RNA editing process to prevent an immune reaction. \par
	The physiological function of A-to-I editing is largely unknown, with evidence for functionality present for only a few incidences \cite{10.1093/molbev/msy011} \cite{nishikura2016editing}. A-to-I editing events in vertebrate coding regions are neutral or slightly deleterious, likely resulting from promiscuous activities of editing enzymes \cite{xu2014human}. \par

\subsubsection{C-to-U Editing}
C-to-U editing, like A-to-I editing, is another substitutional RNA editing method changing is cytidine to uridine. It modifies the base and it takes place in apoprotein B (apoB) \cite{cite-typesRNAediting}. The process affects the metabolism of lipoproteins and necessitates a single strand template with clear-cut characteristics in the vicinity of the edited base \cite{cite-typesRNAediting}. Results in the modification of a glutamine codon to a stop codon. Plays a role in adaptive immune response which utilizes antibodies in order to neutralize invading pathogens \cite{cite-typesRNAediting}. As a result, C-to-U editing has a complexity of site-selectivity, and it creates stop codons that in turn reduce protein abundance \cite{blanc2003c}.
 \par
	The molecular mechanism of C-to-U editing involves the hydrolytic deamination of a cytosine to a uracil base. C-to-U editing is mediated by RNA-specific cytidine deaminases and several complementation factors, which have not been completely identified\cite{blanc2003c}. Here, we review recent findings related to the regulation and enzymatic basis of C-to-U RNA editing. More importantly, when C-to-U editing occurs in coding regions, it has the power to reprogram genetic information on the RNA level, therefore it has great potential for applications in transcript repair. If it is possible to manipulate or mimic C-to-U editing, T$>$C or A$>$G genetic mutation-related diseases could be treated. Enzymatic and non-enzymatic site-directed RNA editing are two different approaches \cite{Vu:2017ab}. The best characterized example of C-to-U RNA editing involves the nuclear transcript encoding intestinal apolipoprotein B, or apoB \cite{blanc2003c}. changes a CAA to aUAA stop codon, generating a truncated protein, apoB48. ApoB RNA editing has important effects on lipoprotein metabolism, and its emergence defines distinct pathways for intestinal and hepatic lipid transport in mammals. C-to-U editing of apoB RNA requires a single-strand template \cite{blanc2003c}. \par
	C-to-U editing can alter the protein sequence when occurring in coding regions, it provides a means for enhancing proteomic diversity without altering the genome, which could be advantageous when different protein functions are needed in different tissues or developmental times \cite{10.1093/molbev/msy011}. We consider a C-to-U editing in coding regions nonsynonymous if it creates a missense or nonsense change; otherwise it is regarded as synonymous. The adaptive hypothesis of C-to-U editing asserts that nonsynonymous editing is beneficial and hence predicts a higher frequency of nonsynonymous editing than synonymous editing, which is expected to be neutral \cite{10.1093/molbev/msy011}. Hundreds to thousands of coding sites were recently found to be C-to-U edited or editable in humans \cite{10.1093/molbev/msy011}, which increases the productivity and effectiveness of the C-to-U editing method. \par

\begin{figure}[h!]
\centerline{
\includegraphics[width=.50\textwidth]{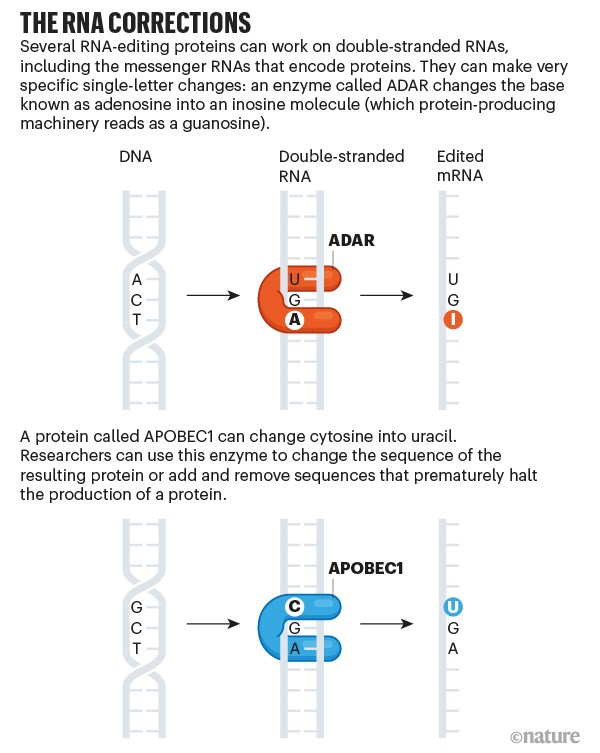}}
\caption{RNA corrections: A-to-I editing (top) usage of ADAR. C-to-U editing (bottom) use of APOBEC \cite{reardon2020step}.}
\label{marker:RNAcorrections}
\end{figure}

 	The importance of C-to-U editing is due to the many genetic diseases that are caused by T-to-C point mutations \cite{bhakta2020rna} Base conversions such as C-to-U and A-to-I, are not only changed by ADAR, there is also a family of enzymes called APOBEC (Fig. \ref{marker:RNAcorrections}) that is primarily used in C-to-U editing. Tsukahara admits that the researchers in Japan have succeeded for the first time in the world in artificial C-to-U conversion of mutated RNA using APOBEC1 \cite{bhakta2020rna}. The studies found are hopeful in therapies that treat genetic diseases by restoring wild-type sequences at the mRNA level \cite{bhakta2020rna}. \par
\par

\subsection{Comparison of Editing Methods}
Differences between the two editing methods is distinguished by their approaches. The absolute requirement for a double-stranded RNA template distinguishes A-to-I and C-to-U RNA editing because the former requires a pre-mRNA template containing intronic regions and is thus biochemically confined to unspliced transcripts. A further distinction biochemically is that ADAR enzymes do not require additional cofactors \cite{blanc2003c}. Reasons for emerging interest in A-I editing are that this type of editing prevents an innate immune response due to endogenous dsRNAs, the critical role of ADAR1 in overcoming resistance to an immune checkpoint, the realization that ADAR can serve as a potential therapeutic target in a subset of cancer types, and the prospects of recruiting ADAR to reprogram RNA nucleic acids have vast technological and therapeutic applications \cite{roth2019genome}. Future development requires extensive monitoring of ADAR activity to ensure acceptable level of off-targets \cite{roth2019genome}. \par

	There are benefits to each method that affect life sustainment and quality of life. Liu finds that both posttranscriptional modification methods are slightly deleterious or neutral, probably resulting from off-target activities of editing enzymes. Nonsynonymous editing level at a site is negatively correlated with the evolutionary conservation of the site \cite{10.1093/molbev/msy011}. Together, these findings refute the adaptive hypothesis; instead findings indicate that, at least in humans, most events of each type, whether C-to-U or A-to-I editing, RNA modification likely manifests cellular errors rather than adaptations, demanding a paradigm shift in [its] research \cite{10.1093/molbev/msy011}. Even though Stafforst published a strategy for designing chemically modified guide RNAs with built-in structures that effectively lasso the ADAR proteins found naturally in our own cells \cite{cross2017crispr} due to the challenges like drug-delivery headaches that occur with gene editing \cite{merkle2019precise}. 
\par

%%%%%%%%%%%%%%%%%%%%%%%%%
\subsection{Hardware}
\begin{figure}[h!]
\centerline{
\includegraphics[width=.5\textwidth]{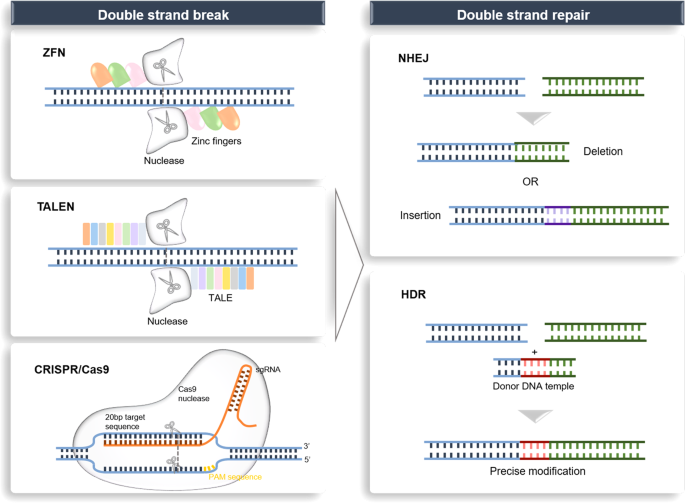}}
\caption{Gene editing methods applicable pathways.}
\label{marker:Geneediting}
\end{figure}

As with many techniques in modern molecular biology, gene and RNA editing hardware is dominated by microscopes to view the molecular process of the enzymes utilized, pipetting, liquid handling, and the complexing and delivery of key reagents\cite{cite-CriSprArchitecture} (see Fig. \ref{marker:Pipettingtools}).
Currently, the most popular gene engineering techniques apply DNA-cutting enzymes by RNA-guided nucleases to achieve genome editing through DSB cut and repair \cite{Danner:2017te}. Gene editing hardware induces targeted DNA double-strand breaks (DSBs), which are then repaired by one of two major pathways\cite{Danner:2017te}(see Fig. \ref{marker:Geneediting}): the error-prone, non-homologous end joining repair system (NHEJ), or the homologous recombination-based double-strand break repair pathway (HDR) \cite{GUHA2017146}. Precise gene editing relies on HDR due to the high rates of precise sequence modifications \cite{Danner:2017te}. \\
 
\begin{figure}[h!]
\centerline{
\includegraphics[width=.5\textwidth]{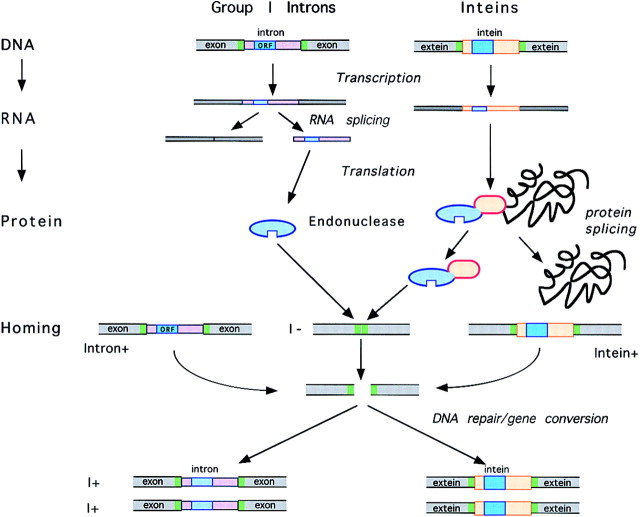}}
\caption{Meganuclease architecture. Proteins recognize and cleave long DNA target or homing sites of 15–40 bp in a homologous allele that lacks the intron or intein sequence, and they promote the lateral transfer of their encoding intron or intein to these sites by a targeted transposition mechanism termed “homing ” \cite{belfort1995prokaryotic}. Intron mobility or homing is targeted and initiated by an endonuclease making a site-specific DNA double-strand break at a homing site (green region) in an intronless allele (I-) of the host gene. Double-strand break repair transfers the mobile intron from an intron-containing allele (I+), which includes the open reading frame (ORF) encoding the endonuclease activity. In the case of group I introns, the invading intron reconstitutes the intact host gene at the RNA level by self-splicing out the intron sequence. In contrast, inteins work at the protein level to splice out the invading protein and ligate the host protein components together \cite{jurica1998dna}.}
\label{marker:MNarchitecture}
\end{figure}

\begin{figure}[h!]
\centerline{
\includegraphics[width=.5\textwidth]{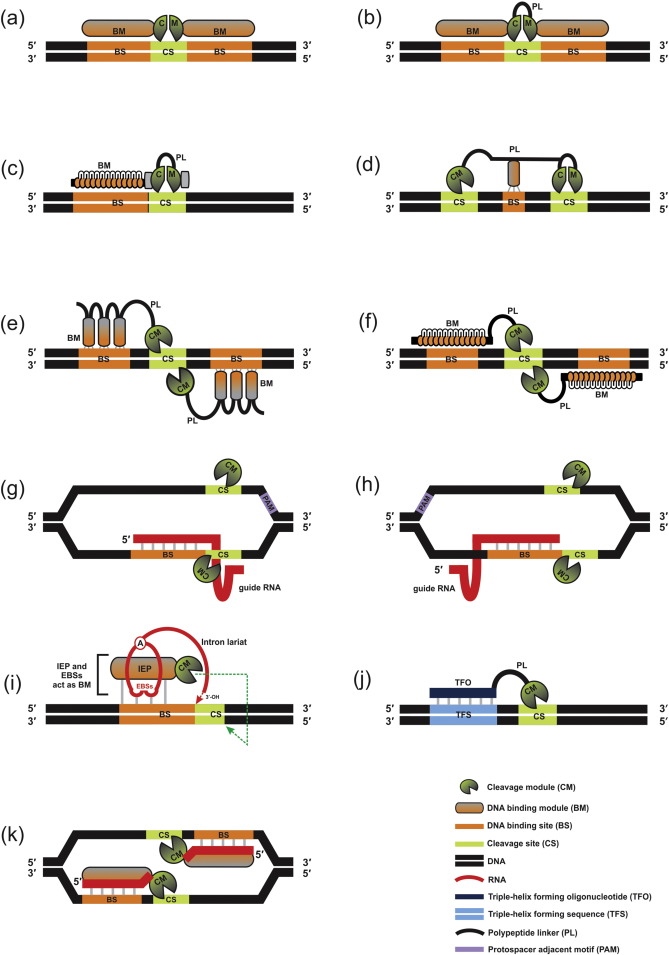}}
\caption{Programmable gene editing architectures \cite{GUHA2017146}. (a) Single-motif LAGLIDADG homing endonucleases, (b) double-motif LAGLIDADG homing endonucleases, (c) megaTAL, (d) MegaTev, (e) zinc-finger nucleases (ZFN), (f) transcription activator-like effector nucleases (TALENs), clustered regularly interspaced short palindromic repeats (CRISPR)/CRISPR-associated proteins (Cas) systems using (g) Cas9 or (h) Cpf1, (i) targetrons, (j) triplex-forming oligonucleotide (TFO) nucleases, and (k) structure-guided nucleases (SGNs). EBS = exon-binding site; IEP = intron-encoded protein. The nuclease domain of FokI is used to engineer ZNFs, TALENs, and SGNs. Elements of this figure have been adapted from Hafez and Hausner et al. NRC Research Press License number: 3981970186164.}
\label{marker:Programmable}
\end{figure}

\begin{figure}[h!]
\centerline{
\includegraphics[width=.5\textwidth]{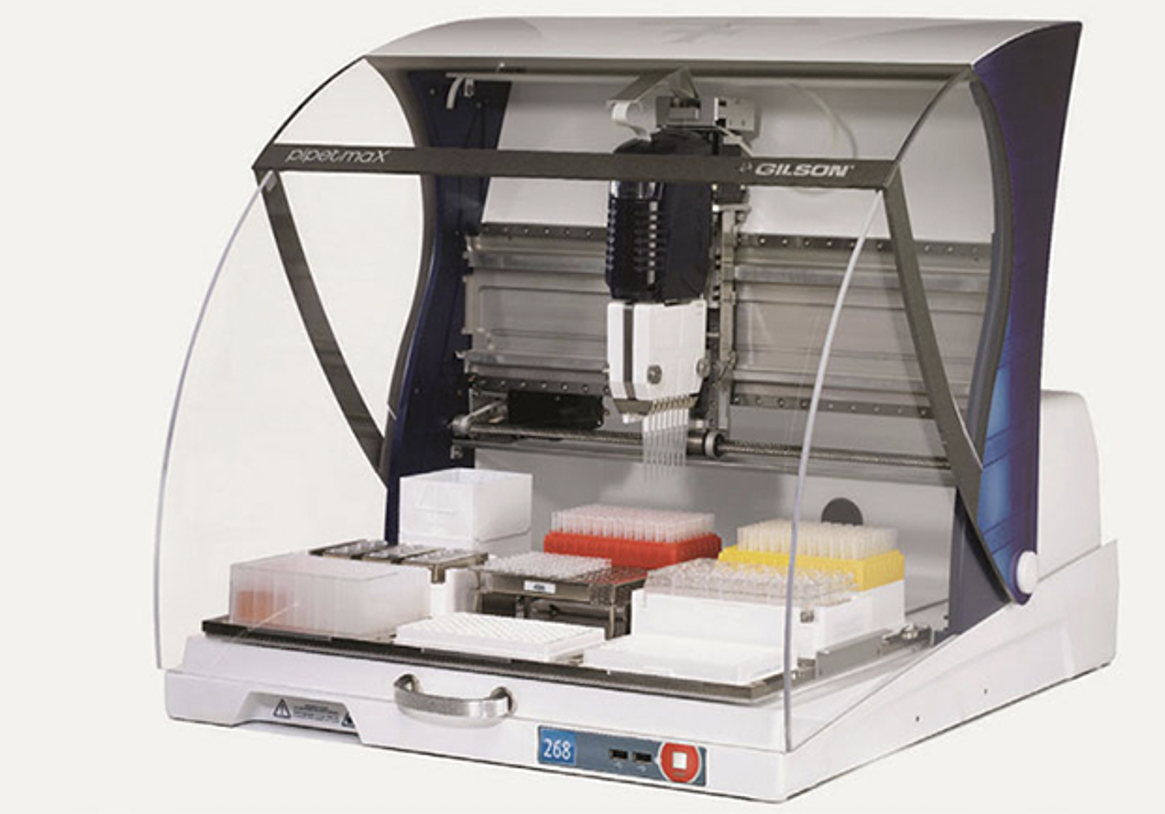}}
\caption{Pipetting tools for gene and RNA editing \cite{cite-CriSprArchitecture}.}
\label{marker:Pipettingtools}
\end{figure}

	Gene editing hardware can be summarized into the following tools: meganucleases (MNs), zinc finger nucleases (ZFNs), transcription activator-like effector nucleases (TALENs), clustered regularly interspaced short palindromic repeat (CRISPR)-associated nuclease Cas9, and targetrons \cite{GUHA2017146}. These devices rely on different hardware associations to achieve editing in double-stranded DNA viral vectors of differentiated human cells with RNA interference (RNAi) to mediate gene knockdown approaches \cite{GUHA2017146}. The first group consists of MNs, ZFNs and TALENs, which achieve sequence-specific DNA-binding via protein-DNA interactions\cite{GUHA2017146}. The second group is comprised of two sub-groups: (i) CRISPR/Cas9 and targetrons, which are RNA-guided systems, and (ii) peptide nucleic acids (PNAs), triplex-forming oligonucleotides (TFOs), and structure-guided endonucleases (SGNs), which are DNA-based-guided systems\cite{GUHA2017146}.\\
	Meganuclease hardware, or homing endonucleases, (Fig. \ref{marker:MNarchitecture}) are highly site-specific dsDNA endonucleases that can be reengineered to expand their target site repertoires using various strategies, such as computational structure-based design, domain swapping, combined with yeast surface display for efficient detection of MN\cite{GUHA2017146}. Three types of MN architectures are single (Fig  \ref{marker:Programmable}a) and double (Fig. \ref{marker:Programmable}b) motif LAGLIDADG homing nucleases, megaTAL(Fig.  \ref{marker:Programmable}c), and megaTEV (Fig.  \ref{marker:Programmable}d). MNs are also in demand as components of vector/cloning systems (e.g. HomeRun vector assembly system) and synthetic biology applications (e.g. iBrick) that require rare-cutting enzymes\cite{GUHA2017146}. \\
The non-modular structure of the LAGLIDADG has an intron-encoded homing endonuclease called I-Cre I that binds to the homing site of DNA and cleaves the site \cite{jurica1998dna}. The structure of I-Crel provides the first view of a protein encoded by a gene within an intron \cite{heath1997structure}. LAGLIDADG motif, which is found in many mobile intron endonucleases, forms a novel helical interface and contributes essential residues to the active site \cite{heath1997structure}. The LAGLIDADG interface is formed by an extended, concave beta-sheet from each enzyme monomer that contacts each DNA half-site, resulting in direct side-chain contacts to 18 of the 24 base pairs across the full-length homing site\cite{jurica1998dna}. The beta-ribbon architecture facilitates the recognition of extended DNA sequence motifs such as homing sites by maintaining a space between each accessible side chain on the MN protein surface, which parallels the spacing of DNA\cite{jurica1998dna}. Flexibility of the beta-strands also allows contacts with nucleotide bases over a longer distance; however the MN structure is limited due to its lack of modular configuration \cite{jurica1998dna}. \\

MegaTAL and megaTEV are improved versions of homing nucleases that possess modular configuration as single-chain nucleases. With the architecture of MegaTAL, the DNA-binding region of a transcription activator-like (TAL) effector is appended to a site-specific MN for cleaving a desired genomic target site \cite{GUHA2017146}. Unlike MegaTAL, MegaTEV has a version designed with a dual nuclease. MegaTEV architecture fuses the DNA-binding and cutting domain from a MN, known as Mega, I-OnuI, with another nuclease domain derived from the GIY-YIG homing nuclease (Tev, I-TevI) \cite{GUHA2017146}. This protein was designed to position the two cutting domains ~ 30 bp apart on the DNA substrate and generate two DSBs with non-compatible single-stranded overhangs for more efficient gene disruption \cite{GUHA2017146}.

ZFN hardware has a modular assembly (MA) method of generating engineered zinc finger proteins (ZFPs) \cite{bhakta2010generation}. It was the first practical method for creating custom DNA-binding proteins \cite{bhakta2010generation}. The first zinc finger nuclease to cleave an endogenous site was created using MA, as was the first artificial transcription factor to enter phase II clinical trials \cite{bhakta2010generation}. In addition to this “modular assembly” architecture, there are interactions between adjacent zinc-finger domains, such as oligomerized pool engineering (OPEN) \cite{Gaj:2016aa}. It is a rapid, publicly available strategy for constructing multi-finger arrays to induce targeted alterations with high efficiencies of 1 to 50\% at 11 different target sites located within three endogenous human genes \cite{maeder2008rapid}. Another ZFN hardware is used with phage display (Fig. 9). A large number of zinc-finger domains are thereby able to recognize and identify distinct DNA triplets \cite{Gaj:2016aa}. These domains can be fused together in tandem using a canonical linker peptide to generate multiple zinc-finger proteins that can target a wide range of possible DNA sequences \cite{Gaj:2016aa}.
 
%\begin{figure}[h!]
    %\centerline{
    %\includegraphics[width=.5\textwidth]{Figures/arap2005.png}}
    %\caption{Phage display hardware is the cylindrical structure within the image that attaches to the mouse epithelial cells and binds to determine peptide sequences \cite{arap2005phage}.}
    %\label{marker:Phagedisplay}
%\end{figure}

TALENs hardware consists of TALE proteins which are bacterial effectors \cite{Gaj:2016aa}. Assembly of a custom TALEN or TAL effector construct is accomplished in 5 days and involves two steps: (i) assembly of repeat modules into intermediary arrays of 1–10 repeats and (ii) joining of the intermediary arrays into a backbone to make the final construct \cite{cermak2011efficient}. After the basic custom TALEN is the most straightforward approach of Golden Gate assembly which customizes TAL effector proteins and TALEN constructs using module, array, last repeat and backbone plasmids \cite{cermak2011efficient}. However, high-throughput TALE assembly methods have also been developed, including FLASH assembly (Reyon et al. 2012), iterative capped assembly (Briggs et al. 2012), and ligation independent cloning (Schmid-Burgk et al. 2013), among others \cite{Gaj:2016aa}. More recent advances in TALEN assembly, though, have focused on the development of methods that can enhance their performance, including specificity profiling to uncover nonconventional RVDs that improve TALEN activity, directed evolution as means to refine TALE specificity (Hubbard et al. 2015), and even fusing TALE domains to homing endonuclease variants to generate chimeric nucleases with extended targeting specificity (Boissel et al. 2014) \cite{Gaj:2016aa}. Since TALENs has a highly repetitive structure, its efficiency through the use of lentivirus (Holkers et al. 2013) or a single adeno-associated virus (AAV) particle makes cell delivery challenging \cite{Gaj:2016aa}.

CRISPR system consists of two to three parts: CRISPR RNA (crRNA) and trans-activating crRNA (tracrRNA) or a single guide RNA (sgRNA), and a Cas protein \cite{cite-Kramer}. A synthetic sgRNA, which can be constructed with E-CRISP (Heigwer et al. 2014), and Cas proteins, such as Cas9 or Cpf1, contain crRNA and tracrRNA elements \cite{Gaj:2016aa}. Short segments of foreign DNA are integrated within the CRISPR locus and transcribed into crRNA, which then anneal to tracrRNA to direct sequence-specific degradation of pathogenic DNA by the Cas9 protein \cite{Gaj:2016aa}. For this reason, crRNA is used to target specific sequences, and tracrRNA serves as a bridge connecting the crRNA and the Cas9 protein \cite{cite-Kramer}. To increase ease, scientists designed sgRNAs to combine both crRNA and tracrRNA together into one longer RNA \cite{cite-Kramer}. Therefore, crRNA and tracrRNA are expressed as a single construct, sgRNA \cite{GUHA2017146}.

The predominant protein that CRISPR utilizes to cleave DNA is Cas9, particularly because it does not require any engineering for retargeting \cite{GUHA2017146}. Cas protein genes are transcribed with the linker sequences and act as an adaptive defense mechanism \cite{cite-Kramer}. Cas9 introduces a double-strand break (DSB), which is then repaired via one of two major pathways, nonhomologous end-joining (NHEJ) or homology-directed repair (HDR), to produce the desired genomic change \cite{cite-Toell}. HDR uses an undamaged DNA template to repair the DSB, allowing new sequences to be introduced into the gene of interest \cite{cite-Toell}. As such, gene insertions or corrections can be enabled by HDR \cite{cite-Toell}. Target recognition by the Cas9 protein only requires a seed sequence within the crRNA and a conserved protospacer-adjacent motif (PAM) upstream of the crRNA binding site \cite{Gaj:2016aa}. PAM is an essential DNA cleavage landmark for Cas9, and PAM easy to find in the genome—it’s like looking for the word “the” in a book, and any complementary 20-base segment adjacent to a PAM can work as the target site\cite{cite_Waltz}. After PAM motif is recognized by the Cas9 nuclease, it is then mediated by an arginine-rich motif present in Cas9 and located immediately downstream of the gRNA target site \cite{Gaj:2016aa}.

Usage of genetic circuits to enable spatiotemporal control of induced Cas9, such as small molecule-regulated approaches for temporal control and tissue-specific promoters for spatial control, can help to balance gRNA on-target activity with off-target effects \cite{cite-Toell}. By designing various gRNAs, this system can be utilized for targeted mutagenesis by inducing the NHEJ pathway or it can be applied to repair or replace alleles by utilizing the cellular HDR repair mechanism with the presence of a user-provided DNA corrective template. a nuclease-deficient version of Cas9 protein has been generated by mutating positions in different domains \cite{GUHA2017146}. This variant is commonly known as “dead” Cas9 or dCas9 that is applicable to CRISPR architecture \cite{GUHA2017146}.

CRISPR architecture can be introduced viral or non-viral systems depending on the application \cite{cite-Toell}, and the tool can cost less than US \$50 to assemble \cite{cite_Waltz}. Traditionally, viral systems deliver gene-editing components as nucleic acids (DNA and mRNA) \cite{cite-Toell}. Non-viral systems allow more flexibility, because they deliver the gene-editing materials either as plasmid DNA, mRNA, or as a RNP complex of Cas9 protein and sgRNA \cite{cite-Toell}. Lentivirus, adenovirus, and adeno-associated virus are also conduits for CRISPR/Cas9 delivery \cite{cite-Toell}. Lentivirus and adenovirus have a large packaging capacity and thus, a high efficiency for in vivo gene transfer, while adeno-associated viruses have a lower immunogenicity, they have a smaller packaging limit \cite{cite-Toell}. Adding to these delivery system options, there’s electroporation which delivers gene-editing materials through nanometer-sized pores in the cell membrane while utilizing an electrical current \cite{cite-Toell}. An example of electroporation hardware is Lonza’s Nucleofector™ which is becoming the standard for non-viral delivery \cite{cite-Toell}.

Hardware that accurately optimizes sample preparations for gene and RNA editing includes motorized pipettes, microplate pipetting, and fully automated. Gilson produces PLATEMASTER®, which streamlines workflow, and PIPETMAN® L Multichannel or new electronic Bluetooth®-enabled PIPETMAN® M Connected Multichannel ranges to make setting up multiple 96-well plates for clonal cell culture straightforward and fast, giving researchers the maximum probability of detecting mutations, or modifiable sequences (Fig. \ref{marker:Pipettingtools}) \cite{cite-CriSprArchitecture}. Gilson PIPETMAX® is an automated liquid handler that removes the risk of manual error and automatically calculates the volumes required to be transferred to create a normalized plate for the editing process \cite{cite-CriSprArchitecture}.

	Another hardware device for gene and RNA delivery is the Light Plate Apparatus (LPA). It can deliver independent \numrange{310}{1550} nm light signals to each well of a 24 well plate with intensity control over three orders of magnitude and millisecond resolution \cite{Gerhardt_2016}. All components can be purchased for under \$400, and the device can be assembled and calibrated by a non-expert in one day \cite{Gerhardt_2016}. The LPA precisely controls gene expression from blue, green, and red light responsive optogenetic tools in bacteria, yeast, and mammalian cells and simplify the entrainment of cyanobacterial circadian rhythm \cite{Gerhardt_2016}. The LPA dramatically reduces the entry barrier to optogenetics and photobiology experiments \cite{Gerhardt_2016}. The core of the LPA is a printed circuit board (PCB) outfitted with a Secure Digital (SD) card reader, an Atmel ATMega328a microcontroller, 3 LED drivers, 48 solder-free LED sockets, a power regulating circuit, and other standard electronics components, and the only external connection is to a 5V power supply \cite{Gerhardt_2016}.
 
Since CRISPR is utilized for both gene and RNA editing, the hardware can be used for both processes; however, different enzymes are utilized along with complexes, such as MRB1 and RECC. The characterization of constituent subunits of the mitochondrial RNA-binding complex 1 (MRB1) implies its importance to the editing process \cite{hashimi2013dual}. Architecture of the trypanosome RNA editing accessory complex, MRB1, is essential in understanding RNA sequence functions, spatial and temporal organization \cite{ammerman2012architecture}. Two functions for MRB1: mediating multi-round kRNA editing by coordinating the exchange of multiple gRNAs required by RECC to edit lengthy regions of mRNAs, and then linking kinetoplastid RNA (kRNA) editing with other RNA processing events \cite{hashimi2013dual}. This type of editing centers on a paradigm that guide RNAs (gRNAs) provide a blueprint for uridine insertion/deletion into mitochondrial mRNAs by RECC \cite{hashimi2013dual}. Research suggests that MRB1 core complex participates in multifaceted dynamic and RNA-dependent associations that coordinate the numerous roles of the MRB1 complex in mitochondrial RNA biogenesis and editing \cite{ammerman2012architecture}. For this reason, MRB1 complex has been ascribed multiple functions and overlapping protein compositions \cite{ammerman2012architecture}.

%%%%%%%%%%%%%%%%%%%%%%%%%
\subsection{Software}
Gene and RNA editing need software algorithms to hasten the programming process, and thereby the editing applications \cite{cite_Waltz}. Each gene editing hardware utilizes a different type of software that works for that specific editing process. RNA editing software is limited due to the fewer methods and hardware available for the editing process as compared to gene editing. However, many algorithms used to target editing sequence sites for gene editing are also utilized in RNA editing, because algorithms can rapidly search with few inputs from the user \cite{cite_Waltz}.

Software for ZFNs and TALE is very similar limited since there’s only one type of editing method and hardware utilized in each process. ZFN software uses 3D protein modeling and energy calculations through computer-based softwares \cite{GUHA2017146}. Researchers have identified potential residues within the FokI dimer interface that are responsible for ZFN dimerization \cite{GUHA2017146}. While TALE software is the code used by TALE proteins to recognize DNA (Boch et al. 2009; Moscou and Bogdanove 2009). In a matter of months, this software discovery enabled the creation of custom TALENs capable of modifying nearly any gene \cite{Gaj:2016aa}. However, these editing architectures must be customized for every use and require far more expertise and resources to assemble \cite{cite_Waltz}. Constraints for these editing devices are that the target has to be located near a landmark that the molecular scissors can recognize, and it must be unlike any other 20-base segment anywhere else in the genome \cite{cite_Waltz}.

Traditional gene modification techniques involve shuttling DNA into cells without knowing where in the genome it will stick, while editing with CRISPR is like placing a cursor between two letters in a word processing document and hitting “delete” or clicking “paste” \cite{cite_Waltz}. The repeats of CRISPR is a genetic phenomenon found in microbes that scientists adapted to disable a gene or add DNA at precise locations in the genetic code\cite{cite_Waltz}. Therefore, CRISPR is used to knock out genes in animal models to study their function, give crops new agronomic traits, synthesize microbes that produce drugs, create gene therapies to treat disease, and potentially to genetically correct heritable diseases in human embryos \cite{cite_Waltz}.
 
A benefit to the CRISPR softwares have the advantage of low-cost and user-friendliness. LPA hardware signals are programmed using an intuitive web tool named Iris with source code and detailed descriptions of the staggered start algorithm, waveform handling, file specifications, randomization and de-randomization procedure, and LPF creation using Python \cite{Gerhardt_2016}. Gilson hardware applies a software wizard called qPCR Assistant that helps to set up gene and RNA editing experimentation and keep track of numerous samples with ease \cite{cite-CriSprArchitecture}. Its flexibility allows the researchers to create and store multiple experiments and use Gilson’s preferred labware, while creating an output file that can be imported directly to the thermocycler \cite{cite-CriSprArchitecture}.

Other softwares utilized by gene and RNA editing hardwares include Protospacer Workbench, E-CRISP, and CHOPCHOP. Protospacer Workbench is an offline software for rapid, flexible design of Cas9 sgRNA \cite{macpherson2015flexible}. It helps to design, analyze, and share CRISPR target-sites for any organism or set of FASTA formatted sequences, and it has the option of using sequence mapping softwares such as Bowtie or BLAST \cite{cite-protospacer}. Another software is E-CRISP. It uses a fast indexing approach to find binding sites and a binary interval tree for rapid annotation of putative gRNA target sites \cite{heigwer2014crisp}. Harvard University’s software CHOPCHOP asks the user to enter the organism, the gene, and some optional advanced parameters \cite{cite_Waltz}. Within seconds, the algorithm finds within the target gene all the possible 20-base segments located near a PAM motif, ranks them based on their uniqueness in the genome and other parameters, and generates a list of matching gRNAs \cite{cite_Waltz}. Nonetheless, RNAs can get distracted by decoy segments, called off-target sites, and may end up mutating the wrong gene \cite{Gaj:2016aa}.

For sequence alignment and to avoid off-target effects, there are a few software tools that aid in predicting cleavage efficacy at the mismatched locus by designing off-target scores (for example, CFD score, MIT score, etc.) \cite{chuai2018deepcrispr}. Four most recent off-target prediction algorithms including CFD score, MIT score, CROP-IT score and CCTop score) \cite{lin2018off}. The prediction results of these algorithms were compared with data generated by whole sgRNA off-target cleavage detection techniques like GUIDE-seq, Digenome-seq, high-throughput genome-wide translocation sequencing (HTGTS), direct in situ breaks labeling sequencing (BLESS), and integration-deficient lentiviral vector capture (IDLV) \cite{chuai2018deepcrispr}. These are essentially hypothesis-based methods that use empirically defined off-target criteria to identify off-target sites \cite{chuai2018deepcrispr}. Effective learning-based prediction of the whole genome off-target profile is needed \cite{chuai2018deepcrispr}. However, GUIDE-seq is the most rigorous framework for gene-wide identification and delineation of off-target effects to date \cite{lin2018off}.

Deep learning applications include alternative splicing predictions, binding target predictions for regulatory proteins, protein secondary structure and biomedical image analysis \cite{lin2018off}. Two algorithms that use deep neural networks to predict off-target mutations in gene editing (i.e. deep convolutional neural network and deep feedforward neural network) are preferred to the state-of-the-art off-target prediction methods (i.e. CFD, MIT, CROP-IT, and CCTop) and the three traditional machine learning models (i.e. random forest, gradient boosting trees, and logistic regression) due to the latter being limited methodologies rather than fully operational \cite{lin2018off}. For example, CCTop score (Stemmer et al., 2015) and MIT score (Hsu et al., 2013) only consider the positions and counts of the mismatched sites of sgRNA-DNA as the features to score the potential off-targets \cite{lin2018off}. On the other hand, recurrent neural networks (RNN), convolutional neural networks (CNN) and long short-term memory (LSTM) have been demonstrated successful, though none are applicable to CRISPR as of yet \cite{lin2018off}.

Recently a comprehensive computational platform to unify sgRNA on-target and off-target site prediction into one framework with deep learning, surpassing available state-of-the-art in silico tools was discovered\cite{chuai2018deepcrispr}. The software is DeepCRISPR, and it fully automates the identification of sequence and epigenetic features that may affect sgRNA knockout efficacy in a data-driven manner \cite{chuai2018deepcrispr}. DeepCRISPR uses a deep convolutionary denosing neural network (DCDNN)-based autoencoder to learn the underlying representation of sgRNA regions, which contain both the DNA nucleotide sequence and epigenetic information \cite{chuai2018deepcrispr}.

\begin{figure}[h!]
\centerline{
\includegraphics[width=.5\textwidth]{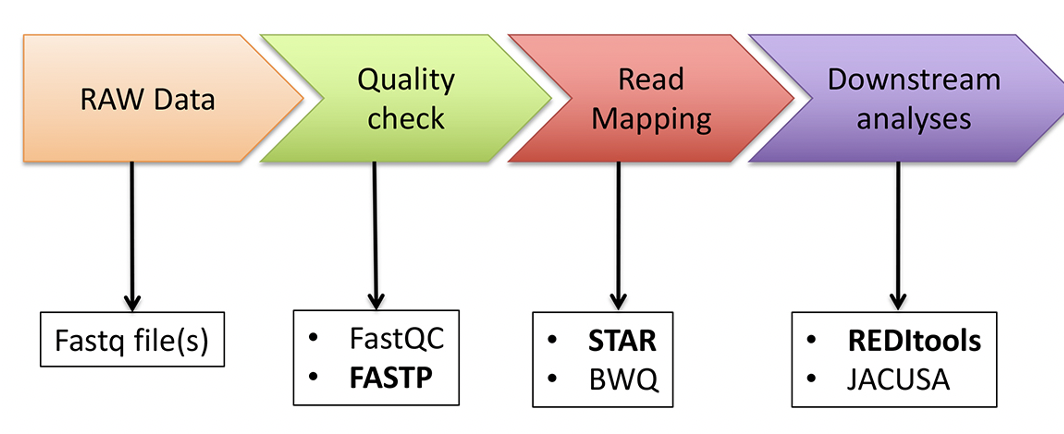}}
\caption{RNA editing bioinformatics pipeline \cite{cite-RNAsoftware}.}
\label{marker:RNAeditingBio}
\end{figure}

When it comes to RNA editing software specific to RNA site sequences, there's REDItools and QEdit. Quantifying RNA editing, or QEdit, is relevant to compare independent samples and study its potential role in different experimental conditions \cite{cite-RNAsoftware}. QEdit is an RNA editing quantification in deep transcriptome data \cite{Lo-Giudice:2020aa}. A bioinformatics pipeline can be used to investigate RNA editing in RNAseq datasets (Fig. \ref{marker:RNAeditingBio}), and RNA editing candidates can be detected using REDItools \cite{cite-RNAsoftware}. A web interface of REDItools is REDIportal, which inherits its layout from all web pages developed in Bootstrap, CSS and JavaScript \cite{10.1093/nar/gkaa916}.

%\begin{figure}[h!]
    %\centerline{
    %\includegraphics[width=.5\textwidth]{Figures/REDItoolsprocess.png}}
    %\caption{REDItools process \cite{10.1093/nar/gkw767}.}
    %\label{marker:REDItools}
%\end{figure}

REDIportal allows two main searches, at position level and sample level \cite{10.1093/nar/gkaa916}. The software, available online at http://srv00.recas.ba.infn.it/atlas/, is the largest and comprehensive collection of RNA editing in humans including more than 4.5 millions of A-to-I events detected in 55 body sites from thousands of RNAseq experiments \cite{10.1093/nar/gkw767}. Users can interrogate the database in the conventional way by providing a genomic region or query the collection of samples by providing one or more run accessions and limiting results according to tunable AEI or ADAR expression values \cite{10.1093/nar/gkaa916}. The search dropdown menu now includes the Gene View page by which users can visualize RNA editing events in their genic context, zooming on specific gene regions if needed \cite{10.1093/nar/gkaa916}. Retrieved sites and samples are shown in dynamic and sortable tables automatically generated by DataTables in server-side mode to easily handle millions of rows \cite{10.1093/nar/gkaa916}. REDIportal embeds RADAR database and represents the first editing resource designed to answer functional questions, enabling the inspection and browsing of editing levels in a variety of human samples, tissues and body sites \cite{10.1093/nar/gkw767}.

\begin{figure}[h!]
\centerline{
\includegraphics[width=.5\textwidth]{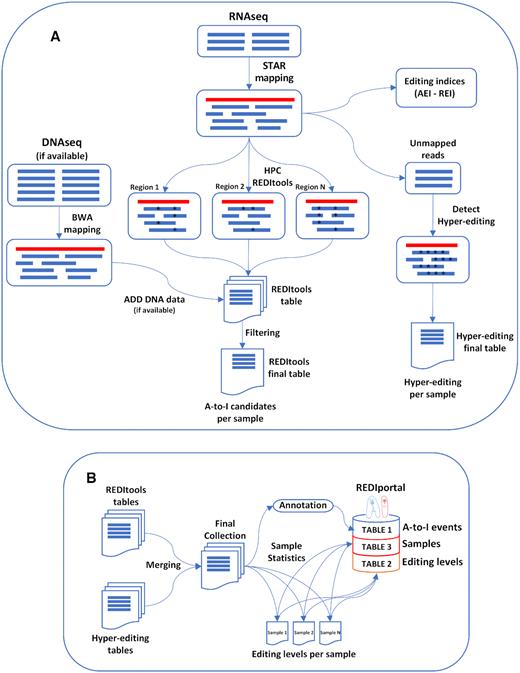}}
\caption{Data processing and database construction \cite{10.1093/nar/gkaa916}. (A) RNAseq data in fastq format are aligned on the human genome by STAR and converted in BAM files. In parallel, if DNAseq reads are available, are aligned on the same genome version by BWA. RNAseq BAM files are analyzed by HPC REDItools and the editing calling is distribute across different computing nodes, each working on a given genomic region. Resulting REDItools tables undergo to further filtering steps before the generation of the final table of A-to-I candidates. RNAseq unmapped reads are re-analyzed to detect hyper-edited reads and provide a list of hyper-editing sites per sample. RNAseq BAM files are further used to calculate the AEI and REI indices. (B) REDItools table and hyper-editing tables are merged in the final REDIportal collection. All events are annotated and stored in the MySQL TABLE1. They are also used to interrogate all RNAseq data to recover RNA editing levels and populate the MySQL TABLE2. Main RNA editing statistics per sample are also computed and collected in the MySQL TABLE3. Blue rectangles are reads, red rectangles are genomic regions, while black stars are A-to-I candidates.}
\label{marker:REDIport}
\end{figure}

For each aligned RNAseq experiment, FeatureCounts is a software for counting reads to genomic features such as genes, exons, promoters and genomic bins \cite{10.1093/nar/gkaa916}. It was applied to count the number of reads per gene, and a custom script was used to normalized these values in TPM (transcripts per million) \cite{10.1093/nar/gkaa916}. Same aligned reads were also used as inputs to calculate the indices of AEI (a weighted average of editing events occurring in all adenosines within Alu elements) and REI (a weighted average over all known recoding sites like those residing in coding protein genes) that are relevant metrics to measure the RNA editing activity, globally or at recoding sites, respectively (Fig. \ref{marker:REDIport}) \cite{10.1093/nar/gkaa916}. The AEI was calculated using the RNAEditingIndexer program \cite{10.1093/nar/gkaa916}.

%%%%%%%%%%%%%%%%%%%%%%%%%
\section{Enabling Technologies of RNA and Gene Editing}
Understanding technology that enables RNA and gene editing in a pharmaceutically-driven world is essential to field and experimental studies. CRISPR may be one of the most prestigious forms of editing technologies currently available due to its cost effectiveness and customizable nature, however, many other companies have jumped into the race to improve and customize the editing process.

The difficulty of any editing technology is target accuracy and specificity. Nonetheless, there are a few types of editing technologies available with greater accuracy and with the ability to lessen the body’s immune response during the editing process. Targeting specificity can be ensured via databases like REDItools. Previously, the foundational technologies of ZFNs, TALENs, and CRISPR to contributed to the gene and RNA editing methods that many companies and scientists use today.

%%%%%%%%%%%%%%%%%%%%%%%%%
\subsection{Gene Editing Technologies}
The technologies for gene editing began as an idea in the 1940s, and by the 1970s the first gene editing technology was available. The approach for targeted genome editing is essentially the modification of any sequence of interest in living cells or organisms \cite{Joung:2013aa}.

For DNA editing, the most well-known technology is CRISPR. However, technologies like TALENs and ZFNs contributed to the discovery and creation of CRISPR. Even though CRISPR is less costly and more customizable, it is not the only technology used for editing today. Each technology has its advantages and benefits that affect the results of therapies and medicinal purposes, as will be addressed in the applications section of this review.

When comparing the different types of editing technologies available today, it is important to identify the strengths of their different platforms (see Table \ref{tab:comparisongene}). Constructs that affect and/or limit the platform are the recognition site, modification pattern, target sequence size, specificity, target limitations, difficulties of engineering, and difficulties of delivering. The recognition site is a region in DNA involving a restriction enzyme that recognizes different nucleotide sequences. If the scientist desires to edit a particular site, then they use the restriction enzyme to target that sequence. A modification pattern is determined by the type of nuclease, or enzyme that cuts the DNA. Target sequence size depends on the amount of base pair (also known by the initialism bp) that the particular technology can target at a time. Targeting size is very important because it determines the efficiency and effectiveness of the technology. Specificity is very similar to target sequence size, however, specificity also determines the tolerances of the technology, such as the number of positional matches or mismatches of the targeted sequence \cite{li2020applications}.

\subsubsection{ZFNs}
Zinc Finger Nucleases (ZFNs) were the first programmable nucleases, an enzyme used to cleave chains of nucleotides, with the ability to target and replace genetic custom sites. To accomplish this task, ZFNs are composed of two parts: an engineered nuclease, also known as Fok1, fused to zinc-finger DNA-binding domains \cite{cite_001}. The binding domain consists of two finger modules that overlap to create a Zinc Finger Protein. ZFNs thereby increases the length of the DNA recognition site and consequently increases specificity \cite{cite_001}.

Fok1 is in the DNA-cleaving domain of ZFNs. It derives from a type of restriction enzyme that has the ability to recognize a specific DNA sequence and cleaves it to make room for the newly edited sequence. In fact, the discovery of Fok1 has led to the development of artificial enzymes with new specificities \cite{Wah:1998vj}.
 
ZFNs integrate insertions or deletions at targeted gene sites. The process of ZFN begins when the zinc-finger binds with the DNA and recognizes a 3-base pair site, or a codon, and combines to develop longer sequences. When the two domains of DNA-binding and DNA-cleaving fuse together, a pair of scissors is created. ZFNs make double strands break and dissociate from DNA (Fig. \ref{marker:ZFN}). They contribute to gene deletion (non-homologous) and gene integration at the target site (homologous recombination).
 
Benefits from ZFNs include: The medical field highly benefited from ZFNs ability to target the main receptor for HIV and edit tumor-infiltrating lymphocytes–a treatment for metastatic melanoma. 
\begin{itemize}
    \item Creation of cell lines that produce higher yields of proteins or antibodies
    \item Rapid disruption of, or integration into, any genomic loci
    \item Mutations made are permanent and heritable
    \item Works in a variety of mammalian somatic cell types
    \item Edits induced through a single transfection experiment
    \item Knockout or knock-in cell lines in as little as two months
    \item Single or biallelic edits occur in 1–20 \% of clone population
    \item No antibiotic selection is required for screening\cite{cite-Merck}
\end{itemize}

\subsubsection{TALENs}
Transcription activator-like effector nucleases, or TALENs, has the same modification pattern as ZFNs with a different recognition site. The recognition site is a repeat-variable diresidue (RVD) tandem repeat region of the TALE protein, a natural protein secreted by plant pathogenic bacteria \cite{li2020applications}. TALE proteins assist infections by binding to specific DNA sequences and activating the expression of host genes \cite{Moore:2014aa}.

The platform recognizes target sites that consist of two TALE protein DNA-binding sites (DBD) that flank a 12- to 20-bp (abbreviation for base pair) sequence recognized by the Fok1 cleavage domain \cite{Gaj:2016aa}. The DBD consists of repetitive sequences of residues. Two amino acids positioned and 12 and 13 remain in the binding domain after TALEs protein binds. The pair of residues determine the nucleotide specificity for TALENs and are referred to as RVD, or repeat variable diresidue \cite{Moore:2014aa}. Combining the repeating regions directs a TALE synthesized protein to the nucleus and thereby modulates transcription \cite{Moore:2014aa}. The nuclease Fok1 is thereby created and fused to DNA-binding domains to create site-specific DSBs, or double-strand breaks. Fok1 stimulates DNA recombination to achieve TALEN-induced targeted genomic modification \cite{li2020applications}.

\subsubsection{CRISPR}
The acronym CRISPR stands for “Clustered Regularly Interspaced Short Palindromic Repeats of genetic information that some bacterial species use as part of an antiviral mechanism” cite035). CRISPR technology can simultaneously detect nucleic acids of many viruses and pinpoint specific ones \cite{storch2020crispr}. CRISPR can store corresponding DNA snippets to encountered viral RNA sequences. For this reason, CRISPR is also harnessed to record gene expression \cite{beisel2018crispr}.

In 1987, clustered regularly interspersed short palindromic repeats (CRISPRs) were discovered in E. coli and in many other bacteria species \cite{li2020applications}. CRISPR describes the unique continuously repeating DNA sequences within bacterial genomes. For this reason, CRISPR was adapted as an acquired immunity defense against bacteria, like that in viruses and diseases. However, in the 1980s, it was difficult to predict the biological function of these unusual repeated sequences due to the lack of sufficient DNA sequence data, especially for mobile genetic elements \cite{Ishinoe00580-17}. Nonetheless, CRISPRs were easier to detect due to their discovery in E.coli. With the first CRISPR discovery, scientists knew what to look for in other genome bacteria.

In the 2000s, the initial stages of CRISPR contributed to a span of a new technology that could edit DNA. The function of the short repeat sequences remained a mystery until 2005 when similarities were discovered to bacterial DNA. The spacer regions of CRISPRs were comparable to the sequences of bacteria and viral reproductions in DNA \cite{Ishinoe00580-17}. Spacers derive from sequences in parasitic elements that are inserted into CRISPR arrays and then transcribed and employed as guides to identify and inactivate parasitic genomes \cite{Shmakove01397-17}. Experiments revealed that an immune defense was adapted from these sequences to defend against offending foreign DNA by instigating guide-RNA to cleave DNA \cite{li2020applications}. It wasn’t until 2007 that CRISPR was able to function as an immunity safeguard against bacteria and viral reproductions via  CRISPR-associated genes, also known as Cas enzymes. Cas was discovered when several genes previously proposed to encode DNA repair proteins specific for an archaea, or a multitude of bacteria and eurykarotes, were identified as being strictly associated with CRISPR \cite{Ishinoe00580-17}. CRISPR and Cas proteins work together to form an immunity system to protect against invading bacteria and viruses, and thereby contribute to a mode of action for scientists to manipulate DNA and RNA. Therefore, CRISPR–Cas systems are able to acquire foreign genetic material by directly capturing DNA from an invader \cite{beisel2018crispr}.
 
CRISPR‐Cas systems gradually developed classes and types by the time 2012 when scientists discovered an array of new Cas proteins. These CRISPR-Cas systems are divided into two classes, known respectively as 1 and 2. The classes are based on the structural variation of the Cas genes, or the encoded effector proteins. Class 1 CRISPR–Cas systems consist of multiprotein effector complexes, whereas class 2 systems comprise only a single effector protein \cite{li2020applications}. Six CRISPR-Cas types and at least 29 subtypes have been reported \cite{Makarova:2011aa}. Each class currently includes three types: types I, III, and IV in class 1; and types II, V, and VI in class 2. Types I, II, and III possess the following proteins: Cas3 for type I, Cas9 for type II, and Cas10 for type III \cite{Ishinoe00580-17}. Type I and type III systems are referred to in the CRISPR-associated complex for antiviral defense (also known as Cascade system) \cite{Makarova:2011aa}. The most frequently used subtype of CRISPR systems is the type II CRISPR/Cas9 system, which depends on a single Cas protein from the a Strep bacteria, or SpCas9 \cite{li2020applications}.

Mechanistically, the CRISPR/Cas9 system comprises two components: a single-stranded guide RNA (sgRNA) and a Cas9 nuclease \cite{li2020applications}. The protein targets particular DNA sequences complementary to the targeting sequence within the sgRNA located immediately upstream of a compatible PAM, also known as a protospacer adjacent motif  \cite{Gaj:2016aa}. The target sequence site for CRISPR sgRNA contains a unique 20 bp sequence along with a compatible short DNA sequence (usually 2-6 base pairs in length) that is targeted for cleavage by the CRISPR system. The PAM is required for a Cas nuclease to cut and is generally found 3-4 nucleotides downstream from the cut site \cite{li2020applications}. The sgRNA binds to the target sequence via base pairing and Cas9 cleaves the DNA to generate a DSB. Then, genome repair is initiated. The CRISPR/Cas9 system introduces small insertions and deletions.
 
Three main categories of genetic edits that the CRISPR company heralds can be performed with the CRISPR/Cas9 system: disruption, deletion, and insertion. If a single cut is made in the genome, a process called non-homologous end joining can result in the addition or deletion of base pairs, disrupting the original DNA sequence and causing gene inactivation. Deletion occurs when a larger fragment of DNA can be deleted by using two guide RNAs that target separate sites. After cleavage at each site, non-homologous end joining unites the separate ends, deleting the intervening sequence. Insertion, or correction, is no different to insertion during DNA transcription, however, the CRISPR/Cas9 machinery adds a DNA template to the cell to make a correction to the gene. The process of repairing double-strand DNA lesions is also known as homology directed repair (HDR), and it can only be performed when the same piece of DNA is present in the nucleus.
 
CRISPR technology is constantly making improvements and discoveries in therapeutics and in the medical field, even with its additions of deep learning to the biochemical system. Miri exhorts that CRISPR technology can be used to produce universal T-cells, equipped with recombinant T cell receptor (TCR) or chimeric antigen receptor (CAR), through multiplex genome engineering using Cas nucleases \cite{14575691720200915}. However, gene editing technologies can result in the irreversible permanent change of genome information, and this process is also facing inevitable security risks and ethical problems \cite{li2020applications}. Nonetheless, scientists are beginning to optimize CRISPR gRNA to accurately predict the sgRNA on-target knockout efficacy and off-target profile with deep learning, also known as DeepCRISPR \cite{chuai2018deepcrispr}. CRISPR has limitations concerning the cell types that it can edit, such as neurons which are difficult to modify for nervous system diseases \cite{li2020applications}.
	
\begin{table*}[!ht]
\small
\caption{Comparison of Gene editing technologies.}
\label{tab:comparisongene}	 
\begin{tabular}{ | m{3cm} | m{3cm}| m{3cm} | m{3.4cm} | m{3.4cm} | } 
\hline
& Site-specific Recombinase Technology & ZFN & TALEN & CRISPR-Cas9 \\ 
\hline
Modification pattern &  & Fok1 nuclease & Fok1 nuclease & Cas9 nuclease \\ 
\hline
Target sequence size in base pairs (bp) & Needs specific sites indicated & Typically 9-18 bp per ZFN monomer; 18-36 bp per ZFN pair & Typically 14-20 bp per TALEN monomer; 28-40 bp per TALEN pair & Typically 20 bp guide sequence and PAM sequence \\ 
\hline
Specificity &  & Tolerating a small number of position mismatches & Tolerating a small number of position mismatches & Tolerating positional/multiple consecutive mismatches \\
\hline
Efficiency percentage & 0.03\%-0.3\% & 0-12\% & 0-76\% & 0-81\% \\
\hline
\end{tabular}

\end{table*}

\subsection{RNA Editing Technologies}
Contrary to the limitations of gene editing tech, RNA technologies are not limited to the specificity of the nucleus and RNA can record DNA sequences for capturing gene expression profiles \cite{beisel2018crispr}. RNA molecules freely move throughout the cell and can deliver genetic material to many different genes. RNA is a safer conduit of editing compared to gene editing, because the alteration of RNA bases offers a more temporary alternative to DNA editing \cite{cross2017crispr}. It wasn’t until CRISPR/Cas9 was discovered and utilized to manipulate genomes that scientists realized the capabilities of RNA editing via the gRNA that Cas utilizes to cut the DNA. Therefore, gene editing technologies paved the way for RNA editing technologies. \par
	RNA technologies stem from the methods A-to-I editing and C-to-U editing. Through the usage of these RNA technologies, disease-relevant sequences can be rescued to yield functional protein products \cite{Cox1019}. When it comes to RNA editing, the process needs only two proteins: one protein makes a DNA version, or template, of the RNA sequence, and thereby generates a new DNA strand from RNA \cite{beisel2018crispr}. The base conversion of RNA technologies contributes to therapeutics more than it does to diseases in present day. As technology is improving and more companies join the pursuits of RNA editing, the methods are becoming a scientific reality. \par
	Seven RNA-editing systems currently constitute the precision and specificity of bp conversions, including the ability to edit point mutations that are known to cause cancers and diseases. CRISPR inspired newer RNA editing machines and technologies with the discovery of the Cas13 protein. Spurred by the research, scientists have engineered programmable RNA-guided machines for RNA editing purposes. RNA base-editing platforms that have recently been engineered to perform programmable base conversions on target RNAs mediated by ADAR enzymes are known by the acronyms CIRTS, REPAIR, RESCUE, RESTORE, CURE, and LEAPER \cite{aquino2020novel}. Each system utilizes a different pathway for the editing process, whether its A/I or C/U (see Table \ref{tab:comparisonRNA}).\par

\subsubsection{CRISPR/Cas13}
CRISPR-Cas9 encountered limitations with targeting single-stranded RNA genomes most viruses that infect humans \cite{cite_034}. Nonetheless, CRISPR stores the memory of past encounters with foreign DNA in its unique spacers that reveals further study into the utilization of different enzymes and proteins by CRISPR technology \cite{Shmakove01397-17}. Thereby, the discovery of the Cas13 enzyme emerged from RNA interference \cite{reardon2020step}. RNAi, or RNA interference, is a biological response to double-stranded RNA that mediates resistance to both parasitic and pathogenic nucleic acids, and regulates the expression of protein-coding genes \cite{hannon2002rna}. With RNAi, Cas13 is capable of gene silencing by targeting mRNA, or knockdown, and base editing \cite{cite_033}. \par
	Cas13 enzyme was found to cut RNA instead of DNA \cite{reardon2020step}. The enzyme digests only RNA and not DNA, and exerts its RNA-cleaving activity on any nearby RNA that it encounters by generating a signal that indicates the presence of a sequence of interest \cite{storch2020crispr}. Since Cas13 can be directed to target specific sequences in RNA, it is also relatively easy to deliver into cells, and it is adapted to naturally target viral RNA in bacteria \cite{cite_034}. \par
	CRISPR-Cas13 consists of four subtypes, Cas13a, Cas13b, Cas13c and Cas13d and differentiates itself from Cas9 by lacking a DNA nuclease domain \cite{yan2019crispr}. Programmed Cas13a can bind and cleave RNA, protecting bacteria from RNA viral production sites and serving as a powerful platform for RNA manipulation \cite{abudayyeh2016c2c2}. The target specificity of all the Cas13s is determined by a 28–30 nucleotide spacer and requires a guide RNA of approximately 64 nucleotides \cite{yan2019crispr}. Experiments of the Cas13s revealed the non-specific RNA breakdown activity, which required the design of sensitive and specific diagnostic detection technologies, like SHERLOCK \cite{Gootenberg:2017aa}. \par
	Types of Cas13 strains along with their efficiencies include: \\
\begin{itemize}
\item {LwaCas13a} 
\begin{itemize}
\item{Strain: Leptotrichia wadeii} 
\item{with ~50 \% knockdown on luciferase and endogenous transcripts }
\end{itemize}
\item{PspCas13b}
\begin{itemize}
\item{Prevotella sp. P5-125} 
\item{with 90-95 \% knockdown on luciferase reporter}
\end{itemize}
\item{RfxCas13d}
\begin{itemize}
\item{Ruminococcus flavefaciens} 
\item{with 80-95 \% on mCherry reporter and endogenous transcripts}
\end{itemize}
\end{itemize}
Note: Wanted to make this a table, but it didn't work due to the length of sentences that overlapped the second page. Don't know how to adjust. 

\subsubsection{REPAIR}
The degradable character of the RNA transforms technologies like REPAIR into a tool with reversible consequences \cite{doi:10.1080/14737159.2019.1568242}. REPAIR is a system, referred to as RNA Editing for Programmable A to I Replacement (REPAIR), which has no strict sequence constraints, can be used to edit full-length transcripts containing pathogenic mutations \cite{Cox1019}. Using Cas13, we developed REPAIR, a programmable RNA editing system that allows for temporary modifications to base pairs of genetic variants in transcripts \cite{cite_033}.The system has high-specificity and minimizes facilitation of viral delivery or negative immune responses. REPAIR presents an RNA-editing platform with broad applicability for research, therapeutics, and biotechnology \cite{Cox1019}.

REPAIR fulfills precise A-to-I edits. It works by fusing Cas13b to the adenosine deaminase domain of ADAR2, a system termed as REPAIR version 1, or REPAIRv1 \cite{Cox1019}. REPAIRv2 was engineered to improve the first version, and, as a result, it has a dramatically higher specificity than version 1. From genesis of the system via Cas13, A-to-I conversion in internal transcripts has led to the correction of two disease-relevant mutations: a form of diabetes and anemia \cite{aquino2020novel}. Other advantages to REPAIR system are that the exact target site can be encoded which creates a favorable A-C mismatch ideal for ADAR-editing activity; there are no targeting sequence constraints, and no motif preference surrounding the target adenosine \cite{Cox1019}. According to Gearing, REPAIRv2 has “successfully edited 33/34 sites with a maximum of 28 \% editing efficiency,” the highest on-target specificity and lowest amount of off-target editing \cite{cite_036}. However, REPAIR is an offspring of CRISPR-Cas13, and thereby operates in a similar fashion.

\subsubsection{CIRTS}
CRISPR-Cas-Inspired RNA Targeting System, or CIRTS, involves functional chemical modifications to all mRNAs expressed in genes \cite{aquino2020novel}. CIRTS is a protein engineering strategy for constructing programmable RNA control elements, and it provides a platform to probe fundamental RNA regulatory processes \cite{Rauch:2019aa}. The synthetic RNA system is built entirely from human-derived proteins, and it mimics the Cas13 functions of gRNA binding, target sequence site RNA recognition and effector functions \cite{Rauch:2019aa}.

A benefit to CIRTS is that it is smaller than CRISPR-Cas programmable protein systems which is due to the way it is engineered. The engineering of CIRTS consists of: (1) an RNA hairpin-binding protein that provides the core of this system because it is a selective, high-affinity protein binder on an engineered gRNA; (2) a gRNA with a specific RNA structure that interacts with the engineered hairpin binding protein, as well as a complementary sequence to the target RNA of interest; (3) a protein that could bind to the gRNA to stabilize and protect it prior to target interaction; and (4) an epitranscriptomic regulator that acts on the targeted RNA \cite{Rauch:2019aa}. An epitranscriptomic regulator allows the functional modifications to RNA transcripts in a cell without affecting the nucleotide sequence \cite{cite_044}. With this ability, CIRTS provides a platform to avoid immune issues when applied to epitranscriptome-modulating therapies \cite{Rauch:2019aa}.

Due to the versatility of CIRTS, it is capable of combining multiple protein domains, where a specific domain of enzymes might work as “writer,” “reader,” and/or “eraser” to target virtually any transcript of interest \cite{aquino2020novel}. 

\subsubsection{RESCUE and RESCUE-S}
RESCUE, known as RNA Editing for Specific C to U Exchange, is also a base-editing tool that is capable of precise cytidine-to-uridine conversion in RNA with increased cytidine deamination activity \cite{Abudayyeh382}. While RESCUE-S is programmable RNA cytidine deamination that uses a bifunctional editor and affects A and C. Researchers discovered that adenosine deaminases can accept other bases, resulting in a novel cytidine deamination mechanism that edits dsRNA \cite{aquino2020novel}. When applying a specific transcript at residue spots consisting of phosphorylation, editing levels occurred between 5 and 28 \%, resulting in increased cell growth \cite{Abudayyeh382}. However, the RESCUE system has not been applicable in therapeutics due to its off-target effects \cite{aquino2020novel}. 

\subsubsection{RESTORE}
Recruiting Endogenous ADAR to Specific Transcripts for Oligonucleotide-mediated RNA Editing, or RESTORE, is a programmable system also used for site-directed RNA editing. The system is composed of an antisense oligonucleotide (ASO) chemically modified and engineered in two segments: (1) a programmable specificity domain that determines target mRNA binding, and (2) an unchanging ADAR recruiting domain to guide internal human ADARs to the ASO:mRNA hybrid to edit transcripts \cite{merkle2019precise}.

Editing for RESTORE occurs most efficiently at the transitional level. Results of editing ranges from 75 to 85 \%, which allows scientists to repair clinically relevant mutations, even point mutations disease-relevant transcripts \cite{aquino2020novel}. For this reason, the RESTORE system minimizes off-target editing effect caused by overexpression of ADAR enzymes \cite{aquino2020novel}. \par

\begin{table*}[!ht]
\small
\caption{\textcolor{black}{Comparison of RNA editing technologies.}}
\label{tab:comparisonRNA}	 
\begin{tabular}{ | m{2.7cm} | m{3.8cm}| m{3cm} | m{3.4cm} | m{3cm} | } 
\hline
 & CIRTS & RESCUE & RESTORE & LEAPER \\ 
\hline
 Targeted bases on transcripts evaluated & G-to-A mutation in the coding region of firefly luciferase gene & Simultaneous targeting of an A and a C in transcripts & 5'-UAG-3' triplet in the 3'UTRs; 5'-UAU-3' and CAA motifs in the ORF regions & 5'-UAG-3', 5'-UAC-3', 5'-AAG-3', 5'-CAG-3' motifs \\ 
\hline
gRNA (measured by amount of nucleotides) & 20-40 nt & 30 nt & short single-stranded sequence (63-95 nt) & approx. 111-151 nt long for high editing efficiency \\ 
\hline
 Deaminase & ADAR2  & ADAR2 & endogenous ADARs & endogenous ADARs \\
\hline
RNA editing strategy & A gRNA with an engineered hairpin interacts with the hairpin RNA binding domain to drive a protein complex & Cas13b-ADAR2 fusion & synthetic ASO with 2'-O-methyl, phosphorothioate modifications & short engineered ADAR-recruiting RNAs (arRNAs) \\
\hline
Efficiency & higher gRNA-dependent editing efficiency  & RNA-editing rates up to 42\% & editing efficiency up to 75-85\% (ADAR1) & editing efficiencies of up to 80\% (arRNA)  \\
\hline
Delivery of editing system & viral delivery (AAV)  & plasmid transfection & ASOs transfection & plasmid or viral vector, or as a synthetic oligonucleotide  \\
\hline
Clinically Relevant Applications &   & Clinically Relevant Applications &  PiZZ mutation causing alpha1-antitrypsin deficiency & Hurler syndrome \\
\hline
\end{tabular}
\end{table*}

\subsubsection{CURE}
Programmable C to U RNA, or CURE, is probably one of the newest forms of RNA editing. CURE is the first cytidine-specific C-to-U RNA Editor that comprises the cytidine deaminase enzyme APOBEC fused to Cas13 and acts in conjunction with gRNAs designed to induce loops at the target sites \cite{Huang:2020aa}. Cytosine base editors (CBEs) like CURE can enable efficient, programmable reversion of T•A to C•G point mutations in the human genome \cite{yu2020cytosine}. The uniqueness of this system is that it does not deaminate, or remove an amino group, from adenosine \cite{Huang:2020aa}.

Compared to RESCUE-S, CURE creates fewer missense mutations, yet both editing techniques overlap in motif preferences of UCC and AC \cite{Huang:2020aa}. Unlike RESCUE, CURE has a version that can edit RNAs in the nucleus, while RESCUE is limited to RNAs outside of the nucleus \cite{Huang:2020aa}.

\subsubsection{LEAPER}
Leveraging Endogenous ADAR for Programmable Editing of RNA is also known by the acronym LEAPER. Instead of utilizing gRNAs, ADAR-recruiting RNAs (or arRNAs) are used. The system works by employing short engineered arRNAs to engage endogenous ADAR1 enzymes to change a specific A-to-I bases \cite{aquino2020novel}. This makes the system capable of achieving editing efficiencies of up to 80 \%, with minimal global off-target effects and limited editing of non-target adenosines in the target site \cite{qu2019programmable}.

The study demonstrated that the targeted RNA editing by ADAR proteins could be carried out in the presence or absence of the catalyst Cas13 protein \cite{aquino2020novel}. The utilization of arRNAs enable effective editing on endogenous, or internal, transcripts with high editing rates \cite{qu2019programmable}. The uniqueness of arRNA usage allows for LEAPER to repair cancer-relevant mutations via a premature stop codon, and thereby and restores the cell’s function \cite{qu2019programmable}. The system ensures high-efficiency RNA editing without causing an innate immune response form the cells \cite{aquino2020novel}.

\subsection{Detection Technology}
To implement a solution for a malady, the first step is to identify and then target the source of the patient’s symptoms and/or pain. Ways to accomplish this include technologies that have been adapted from RNA and gene editing devices. However, effective on-target editing is difficult without identifying the mutated source of illness. Methods for detection and treatment of mutations have been discovered using Cas enzymes \cite{cite-IDTSpencer}. Cas13 enzymes are most useful in detection technology because they are well suited for nucleic acid detection \cite{cite_033}.

Detection devices rely on the discovery of high-throughput screening (HTS) defined as the use of automated equipment to rapidly test thousands to millions of samples for biological activity at the model organism, cellular, pathway, or molecular level \cite{ATTENERAMOS2014916}. HTS is a high-tech way to hasten the drug discovery process through the use of robotics, data processing, and/or control software that was discovered and engineered from and to assist the RNA editing processes \cite{TOMAR2019741}. With the advent of high-throughput detection coupled with the development of sequencing-based techniques for all RNA identification, new types of RNA chemical modifications can be detected and discovered \cite{aquino2020novel}.
 
Testing evidence shows that detection technologies can scale to test many DNA and RNA samples while simultaneously testing for multiple pathogens \cite{ackerman2020massively}. Two of the predominant detection technologies are CARMEN and SHERLOCK. From these two platforms, other methods have been discovered and are being developed for clinical trials. A couple of acronyms that cover current DNA methods are derived from the SHERLOCK’s platform are DETECTR and HOLMES \cite{cite-IDTSpencer}. The difference compared to SHERLOCK is that both DETECTR and HOLMES are more effective in DNA diagnostics due to their utilization of Cas12 \cite{chen2018crispr}. “DNA endonuclease-targeted CRISPR transreporter,” or DETECTR, is similar to the SHERLOCK method, except it utilizes Cas12a and makes the system sensitive to DNA detection \cite{chen2018crispr}. HOLMES, or “one-HOur Low-cost Multipurpose highly Efficient System,” uses a purified Cas12a enzyme \cite{li2018crispr}. After testing the ability of HOLMES, it is found that the platform is most effective in detecting DNA, rather than RNA, which probably makes it a better detection system for gene editing \cite{li2018crispr}.
 
Two methods that use Cas13 enzymes are PAC-MAN and CARVER \cite{cite-IDTSpencer}. “Cas13-assisted restriction of viral expression and readout”, or CARVER, utilizes Cas13a instead of PAC-MAN’s Cas13d \cite{cite-IDTSpencer}. CARVER uses Cas13 to detect and destroy viral RNA, especially those of single-strand RNA viruses \cite{freije2019programmable}. The method combines Cas13-mediated cleavage of viral RNA with a Cas13-based readout using the SHERLOCK platform to measure viral RNA levels which contributes to potential rapid diagnostic and antiviral drug development \cite{freije2019programmable}.

PAC-MAN, also known as Prophylactic Antiviral CRISPR in huMAN cells, is applied for other viral diseases simply by changing the crRNA sequence used \cite{cite-IDTSpencer}. The method was developed to target SARS-CoV-2, IAV, and potentially all sequenced coronaviruses, and according to studies, it can effectively inhibit SARS-CoV-2 reporters \cite{abbott2020development}. By using Casd, the method is suspected to have strong target cleavage activity with viral mRNA, which is where coronaviruses reside \cite{cite-IDTSpencer}. However, the delivery systems for these methods will need to be investigated further before it can be used in patients \cite{abbott2020development}.

\subsubsection{SHERLOCK}
SHERLOCK is a nucleic acid detection system that was established in 2017 with the employment of the “RNA collateral effect” of Cas13a and an isothermal amplification method \cite{li2018crispr}. The Specific High-Sensitivity Enzymatic Reporter unLOCKing (SHERLOCK) detection system uses Cas13s non-specific RNase activity to cleave fluorescent reporters upon target recognition, allowing for the design of sensitive and specific diagnostics \cite{Gootenberg:2017aa}. The system can be tuned for single-nucleotide distinction at any position on the target \cite{cite_033} to provide for more specific targeting for gene and RNA editing.

SHERLOCKv2, the predominant version of detection, combines multiple enzymes/nucleases to allow simultaneous detection of many nucleic acid sequences, which makes it clinically useful for determining whether a patient-derived specimen is infected with any of several viruses \cite{cite-IDTSpencer}. Purification of Cas13 is the essential first step for the SHERLOCK detection process \cite{cite_033}. By then increasing the single-base distinction capabilities of Cas13, a synthetic mismatch is added in the guide sequence and placed in the third base of the spacer \cite{cite_033}. Optimal CRISPR RNA (crRNA) can then be determined by SHERLOCK’s maximal distinction ability \cite{cite_033}.
 
Since SHERLOCK allows multiplexed, portable, and ultra-sensitive detection of RNA or DNA from clinically relevant samples, applications to the medical field have been established by the FDA \cite{kellner2019sherlock}. Presently, the main application for SHERLOCK’s application is for detection of viral RNA, particularly COVID-19 diagnosis \cite{cite-IDTSpencer}. It is known as a CRISPR-based diagnostic test \cite{cite-IDTSpencer}. It is available from IDT* as the Sherlock™ CRISPR SARS-CoV-2 kit, which was granted as an Emergency Use Authorization kit by the FDA \cite{cite-IDTSpencer}.

\subsubsection{CARMEN}
Combinatorial Arrayed Reactions for Multiplexed Evaluation of Nucleic acids, or CARMEN, is a platform for scalable, multiplexed pathogen detection \cite{ackerman2020massively}. CARMEN contains CRISPR-based nucleic acid detection reagents that are used to simultaneously differentiate 169 human-associated viruses \cite{ackerman2020massively}. The method extracts viral RNA from samples and make copies of the genetic material which is then added to a detection mixture consisting of Cas13 \cite{cite_029}. The detection process of these two technologies, CARMEN and CRISPR-Cas13, is completed in under eight hours \cite{cite_029}.  Evidence of CARMEN’s ability to rapidly identify infected samples informs the design of improved CRISPR-based diagnostics \cite{ackerman2020massively}.

Combining the platforms of SHERLOCK and CARMEN improves each system in diagnostic testing. CARMEN offers impressive throughput and flexibility \cite{cite_029} while simultaneously benefitting from the specificity of SHERLOCK \cite{ackerman2020massively}. CARMEN enables CRISPR-based diagnostics, including SHERLOCK, in comprehensive disease surveillance to subsequently improve patient care and public health \cite{ackerman2020massively}.

%%%%%%%%%%%%%%%%%%%%%%%%%

\section{Applications of RNA and Gene Editing}
Currently, the best fields for gene and RNA editing are medical treatments and therapeutic applications. Previous enabling technologies contribute to applications in the editing field with proven solutions to deter adverse and painful reactions. For genetic diseases and viruses predominately caused by point mutations, base conversion technologies have influenced the medical and therapeutic arenas. Many diseases have a genetic cause, including more than 10,000 monogenic diseases caused by mutations in individual genes\cite{cite_035}. Therefore, applied RNA and gene editing technologies affect different types of cancers; it assists in HIV prevention and decreases pain for patients with debilitating conditions.

\begin{figure}[h!]
\centerline{
 \includegraphics[width=0.50\textwidth]{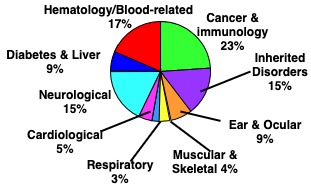}}
 \caption{Percentages of applications based on the number of gene editing start-up companies.}
\label{marker:GenePercent}
\end{figure}

\begin{figure}[h!]
\centerline{
\includegraphics[width=0.50\textwidth]{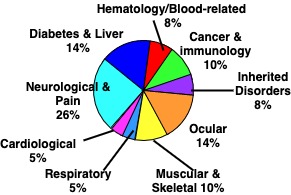}}
\caption{RNA editing company application percentages.}
\label{marker:RNAPercent}
\end{figure}

To understand the applications of gene and RNA editing, there are many companies that provide avenues of various treatments. For this reason, it is important to review companies that have developed platforms within the editing field. Companies that are involved in gene and RNA editing possess a number of established applications based on the previous enabling technologies that convert base pairs. Some companies overlap in gene and RNA editing applications due to their investment in both processes (Fig. \ref{marker:GenePercent} and Fig. \ref{marker:RNAPercent}).
  
Therapeutic applications (Fig. \ref{marker:therapyapps}) are treatments that assist in recovery of an illness or injury.  According to Law Insider, treatment includes any use of a Licensed Product for the prevention or treatment of disease or injury\cite{cite_043}. The intention of therapeutic applications is to have a beneficial effect on the body or mind, and/or produce a useful or favorable result or effect (Webster). Therapeutic applications are a viable alternative to traditional medicines, like chemotherapy \cite{cite_042}. \par
	Gene and RNA editing are effective contributors to therapy. Applying editing technologies has proven useful in treatment of genetic diseases and viruses. Editing therapies can also prevent cell degradation \cite{reardon2020step}. \par
	Gene editing companies that provide therapy for a number of maladies include CRISPR, LocanaBio, Ultragenyx, intellia Therapeutics (NTLA), bluebird bio (BLUE), Horizon Discovery Group, Inscripta, Precision BioSciences, and many more. \par
	RNA editing therapeutic companies are Korro Bio, BEAM Therapeutics, Shape Therapeutics, ProQR therapeutics, Ribozyme, Tessera Therapeutics, Triplet Therapeutics, Neubase Therapeutics, and Arrowhead. Each company targets a different disorder or disease to improve treatment variability. RNA editing techniques are linked to muscular dystrophy, acute pain, PCSK9 which is a regulator of cholesterol, tumors, hemophilia, Alzheimer’s, cystic fibrosis, the eye disease of Hurler’s syndrome, and so much more. The RNA editing therapies to treat these conditions depend on the base conversion process that the company utilizes. \par
\begin{figure}[h!]
\centerline{
\includegraphics[width=0.50\textwidth]{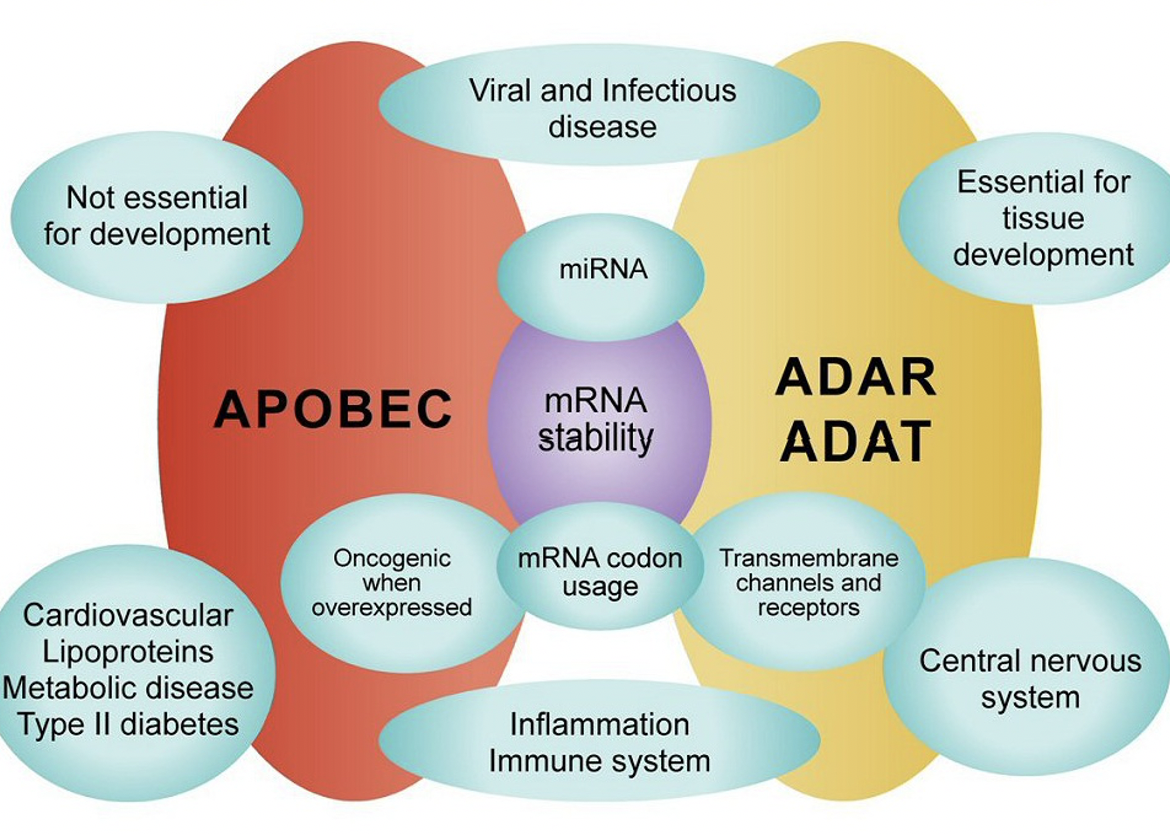}}
\caption{RNA Drug Delivery \cite{cite-drugdeliveryRNA}.}
\label{marker:RNADrugs}
\end{figure}
\begin{figure}[h!]
\centerline{
\includegraphics[width=0.50\textwidth]{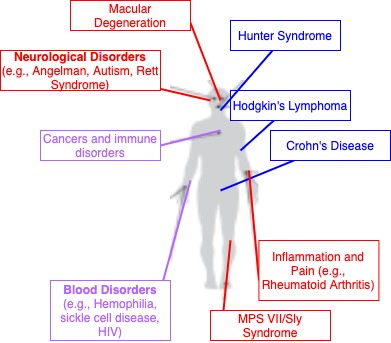}}
\caption{Areas for therapy applications. Red is RNA editing applications. Blue is gene editing applications. Purple contributes to both systems of delivery.}
\label{marker:therapyapps}
\end{figure}

	Concerning medical applications, the question that companies are answering is what diseases and disorders can be cured or treated via medicinal means via gene and RNA editing applications. Medical applications can include devices such as systems like REPAIR or CRISPR applied to a treat a particular disease or editing of a pharmaceutical for usage. RNA editing pathway enzymes, ADAR and APOBEC, target the biomedical field medications, because they are implemented in high throughput screening for small molecule drug candidates (Fig. \ref{marker:RNADrugs}) \cite{cite-drugdeliveryRNA}. \par

\subsection{Gene Editing Therapeutic Applications}
In 2017, researchers demonstrated human applications of gene therapy to treat conditions via in vivo (inside) and ex vivo (outside) editing.  Therapeutic applications based on gene-editing technology can exert permanent and elaborate proofreading effects at the genome level \cite{shim2017therapeutic}. The remarkable progress in nuclease engineering has enabled accurate modification of human genomes for therapeutic purposes \cite{shim2017therapeutic}. In fact, the FDA went on to approve therapies involving cell-reprogramming for treating B-cell leukemia and inherited vision and hearing loss \cite{cite_042}. Current preclinical research on genome editing primarily concentrates on viral infections, cardiovascular diseases (CVDs), metabolic disorders, primary defects of the immune system, hemophilia, muscular dystrophy, and development of T cell-based anticancer immunotherapies \cite{li2020applications}.

Therapeutic application studies, primarily based in the United States, provide statistics on the efficiencies that gene editing has been able to treat. The major diseases treated via gene-editing therapeutics are viral infection (33 \%), blood disease (19 \%), and neoplasms (13 \%) \cite{shim2017therapeutic}. Delivery strategies for gene editing are divided into viral and non-viral vector systems, non-viral ex vivo being the predominant application of the two by 70 \% \cite{shim2017therapeutic}. In ex vivo editing therapy, cells are isolated from a patient to be treated, edited, and then re-administered to the patient \cite{li2020applications}. For in vivo editing therapy, engineered enzymes are delivered by viral or non-viral approaches and directly injected into the patient for systemic or targeted tissue (such as the eye, brain, or muscle) effect \cite{li2020applications}.

CRISPR/Cas9-based disease therapeutics can be achieved both in vivo and ex vivo. In ex vivo therapy, cells are isolated and edited outside of the body using engineered nucleases, after which they are transplanted back into the body (e.g., in cancer immunotherapy; see Figure 1). In in vivo therapy, genetic materials are directly injected into the body (e.g., in genetic disease therapy). Ex vivo editing makes it easier to control the delivery of CRISPR/Cas9 components (such as for variables such as the dose), and more delivery modes are available using this approach.
Inhibition of a viral infection and cancer immunotherapy are the main ex vivo applications of CRISPR/Cas9. Cells of the hematopoietic system, such as T cells, are particularly suitable for ex vivo modification, as they can be easily isolated from the blood, expanded ex vivo, and transplanted back into a patient with limited immune response.2 For example, Schumann et al. disrupted PD-1 and CXCR4 immune checkpoint genes in human T cells by delivering CRISPR/Cas9 RNPs via electroporation, which successfully prevented the inhibition of T cells from killing cancer cells.17 Moreover, resistance to HIV-1 infection has been obtained through disrupting coreceptors of HIV-1 using CRISPR/Cas9-encoding plasmids, with minimal off-target effects.18
\cite{cite-Toell}
	
\subsubsection{Neurodegenerative Disorder Therapies}
LocanaBio company highlights its application by creating a new class of gene therapies  to modify dysfunctional RNA associated with a range of severe disorders including neuromuscular, neurodegenerative and retinal diseases to restore or improve health and make a meaningful difference in patients’ lives (41). Therapeutic applications for neurodegenerative diseases correct defective genes by delivering functional genetic material to cells with repeated administration \cite{10.3389/fnins.2020.580179}. \par
ZFNs showed impressive results in modifying the HIV CCR5 coreceptor surface protein in the autologous CD4 T lymphocytes of persons infected with HIV \cite{GUHA2017146}

\subsubsection{Blood Disorder Therapies}
Blood pathologies that occur in the genes are treatable via gene therapy applications. In HIV treatments, scientists at the at the Fred Hutchinson Cancer Research Center Lab at the University of California, Los Angeles and Washington have cultured and reprogrammed blood cells from bone marrow to kill HIV-infected cells \cite{cite_042}. CRISPR gene-editing applications treat hemoglobin pathologies and sickle cell disease. These diseases are acquired from mutations in a gene that encodes a key component of hemoglobin, the oxygen-carrying molecule in blood, therefore, in order to treat the disease, fetal hemoglobin is increased (035). However, gene editing therapies do not reduce pain and require multiple years of treatment.

Bluebird bio (BLUE) utilizes gene editing, gene addition, and gene-based immunotherapy. With these techniques, BLUE applies therapy to cerebral dystrophy and a cancer of blood cells, also known as multiple myeloma (023). Similar to CRISPR company, BLUE also treats sickle cell disease and anemia with their technologies. In gene addition, BLUE adds a functional gene that can occur either outside (ex vivo) or inside (in vivo) the body (023). Gene addition works in tandem with immunotherapy to utilize a delivery system called a vector to insert new genes directly into cells without causing an individual to be affected by the original disease (023). A vector is much like a delivery truck full of packages. The truck is used to make sure the packages (in this case genes) are delivered to the right address (the cells) (023). Viruses are used as a carrier because they have a natural ability to deliver genetic material into cells \cite{cite_037}. For now BLUE has a medicine called Zynteglo to temporarily treat a blood genetic disorder (023).

\subsubsection{Cancer and Immunotherapies}
Cancer is the focal point of many gene therapies, if only it is to alleviate the pain that chemotherapy causes. Stem cell transplantation combined with the CRISPR/Cas9 system is another approach for the therapy of genetic mutations, like cancer, and it is found to be a useful cell source for cell replacement therapy without immune rejection problems \cite{li2020applications}. One of the principal approaches of cancer immunotherapy is transfer of natural or engineered tumor-specific T-cells into patients, a so called “adoptive cell transfer”, or ACT, process \cite{14575691720200915}. The construction of T-cells is dependent on the employment of a gene-editing tool to modify donor-extracted T-cells and prepare them to specifically act against tumor cells with enhanced function and durability with the least amount of side effects \cite{14575691720200915}.

Natural Killer cells are a type of lymphocyte, or a white blood cell, that mediate anti-tumor and anti-viral responses \cite{10.3389/fimmu.2018.01869}. These particular cells have been engineered against Hodgkins’ Lymphoma. Natural killer cells need a shorter time-frame to induce the same effect as T-cells, which is now a niche therapy for Fate Therapeutics’ company \cite{cite_042}.
 
Intellia Therapeutics (NTLA) utilizes CRISPR/Cas9 to edit mutations, particularly cancers via stem/T-cells. The National Cancer Institute defines T-cells as a type of white blood cell developed from stem cells in the bone marrow that help protect the body from infection and may help fight cancer \cite{cite_038}. NTLA utilizes T-cells and/or stem cells to carry out gene editing to cancer-ridden mutations. NTLA’s systemic lipid nanoparticle (LNP)-based delivery system has unlocked treatment of genetic diseases to both selectively knockout disease-causing genes and restore necessary genetic functions by targeted insertion (39).

CAR T therapy is an important advance in how we treat certain forms of cancer. CAR T treatment involves taking naturally occurring infection-fighting cells, called T cells, genetically modifying them, and then putting the altered cells back in the body where they can attack cancer cells \cite{cite-OPtum}. 

\subsubsection{Gene Editing Medicines}
Gene editing companies that are predominantly involved in medical applications such as gene-edited administered drugs include EDIT (editas medicine), Regeneron Pharmaceuticals (REGN), GeneTx Biotherapeutics, Ultragenyx, and Sangamo Therapeutics. Even CRISPR has a platform in medicinal applications explaining that regenerative medicine, or the use of stem cells to repair or replace tissue or organ function lost due to disease, damage or age, holds tremendous potential in both rare and common diseases. Medicinal efforts use unmodified stem cells for gene editing to enable immune evasion, improve existing cell function, and direct cell fate using CRISPR/Cas9 (035). 
	
GeneTx Biotherapeutics developed a drug from gene editing technologies to treat Angelman syndrome. Angelman syndrome (AS) is a rare genetic disorder that affects approximately one in 15,000 live births(040). AS is caused by a loss of function of the maternally inherited UBE3A gene (040). Symptoms of AS include developmental delay, impaired motor function, loss of speech, and epilepsy. GeneTx Biotherapeutics’ GTX-102 medication is designed to compensate for the deficiencies that result from the underlying genetic cause of the disorder (040). GeneTx entered into a worldwide license agreement with The Texas AandM University System and a research collaboration agreement with Texas AandM AgriLife Research, under which GeneTx hopes to further develop and commercialize this novel antisense oligonucleotide that could potentially serve as a targeted therapy for patients with the disorder (040).
	
Ultragenyx is a biopharmaceutical company committed to bringing to market novel products for the treatment of rare and ultra-rare diseases, with a focus on serious, debilitating genetic diseases. Founded in 2010, the company has rapidly built a diverse portfolio of product candidates with the potential to address diseases for which the unmet medical need is high, the biology for treatment is clear, and for which there are no typically no approved therapies treating the underlying disease. Crysvita is the first and only FDA-approved therapy for the treatment of XLH and TIO; Mepsevii is a medication that is used in pediatric and adult patients for the treatment of Mucopolysaccharidosis VII (MPS VII, Sly syndrome), which is the inability to breakdown large sugar molecules in the body. The effect of Mepsevii on the central nervous system manifestations of MPS VII has not been determined, and it has a number of side effects; Dojolvi medication is indicated as a source of calories and fatty acids for the treatment of pediatric and adult patients with molecularly confirmed long-chain fatty acid oxidation disorders (LC-FAOD), with multiple warnings and precautions. All of the medications have warnings regarding side effects which contributes to gene editing technological research rather than consumption of medications.

\subsection{RNA Therapy Applications and Medicines}
RNA therapy applications are best suited for temporary edits and can treat a wider range of genetic disorders and diseases due to its malleability within genes. Fry explains that RNA-targeting mechanisms with engineered deaminase enzymes allows for the therapeutic applications to genetic diseases via its programmable correction of G$>$A and T$>$C mutations in RNA. Base conversions have a greater effect on a variety of mutations, therefore, RNA is better suited to therapies that are typically regarded as temporal applications.
	
Physical pain and inflammation are temporal discomforts that gauge illness or injury severity in an individual. Illness that cause pain and inflammation can be addressed via RNA editing, whereas gene editing therapies can sometimes cause pain. Any temporary condition is easier to treat with RNA editing technologies because it provides short-term alleviation to the physical site. According to Reardon, “RNA editing in select tissues for a limited amount of time could help to alleviate pain without the risk of dependency or addiction associated with conventional painkillers" \cite{reardon2020step}.
 
Rather than separating RNA therapies into the genetic diseases they can solve, there are three broad categories for RNA editing therapeutics. The first category consists of targeting nucleic acids, either DNA or RNA. The two types of nucleic acids that are targeted with this therapy are single-stranded antisense oligonucleotides (ASOs), and double-stranded RNA molecules that operate through a cellular pathway of RNAi \cite{cite_046}. In 2018, the US FDA approved the first therapy using RNAi: “a technique in which a small piece of RNA is inserted into a cell in which it binds to native mRNAs and hastens their degradation” \cite{reardon2020step}. Currently, the clinical applications for antisense RNA therapy contributes to pharmaceutical drug delivery due to the molecule of noncoding mRNA that blocks the translation of a specific proteins \cite{cite_045}. \par
	The second category of RNA therapies involves the targeting of proteins. RNA aptamer is a molecule that is engineered to target proteins. The molecule is designed to bind to a specific site on a specific protein to modulate its function \cite{cite_046}. RNA aptamers have the ability to aid anaesthesia, drug delivery, and modify or control blood clotting in an emergency. Currently, FDA-approved application of targeting proteins with RNA aptamers is used against age-related macular, or eye, degeneration \cite{Germer:2013aa}.
 
The third category of RNA editing for therapeutic applications is the encoding of proteins. These therapies use mRNAs to develop vaccines against infectious diseases and several types of cancers. With chemical modifications, mRNA increases its stability when it’s used in drugs \cite{Kim:2020aa}. Researchers are also exploring whether these types of treatments can be used as protein-replacement therapies for rare conditions such as hemophilia \cite{cite_046}. Though, clinical applications have shown that mRNAs injected into tumor regions cause the antigen-specific immune cells to expand and eliminate the tumor cells \cite{Kim:2020aa}.

%%%%%%%%%%%%%%%%%%%%%%%%%

\section{Evaluation Metrics of Gene and RNA editing} 
\begin{figure*}[h!]
\centerline{
\includegraphics[width=0.99\textwidth]{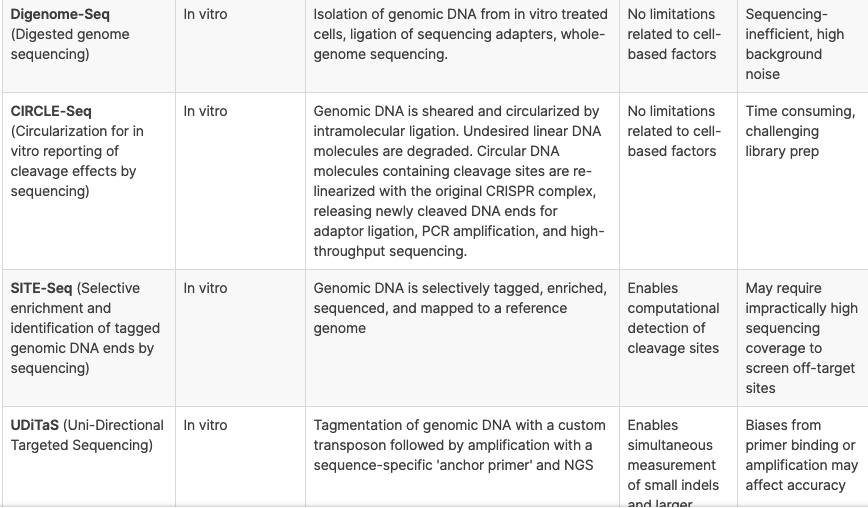}}
\caption{Analysis of different editing systems \cite{cite-Metrics}.}
\label{marker:metrics}
\end{figure*}

Evaluating gene and RNA editing consists of screening and verification to determine the system's functionality, intuitiveness, reliability, and accuracy. The number of methods to assess the efficiency of genome editing is increasing very quickly \cite{germini2018comparison}. Techniques used for laboratory screening and verification include next-generation sequencing (NGS) and computational analysis. These techniques and methods satisfy the ideal CRISPR-Cas system, and thereby screens for both high cleavage efficiency and high target specificity \cite{germini2018comparison}. No matter the technique, each analysis method has its advantages and limitations (see Fig. \ref{marker:metrics}).
 
%\begin{figure}[h!]
%\centerline{
%\includegraphics[width=0.50\textwidth]{Figures/Analysis1.png}}
%\caption{\textcolor{red}{......}}
%\label{fig:analysis1}
%\end{figure}

%\begin{figure}[h!]
%\centerline{
%\includegraphics[width=0.50\textwidth]{Figures/Analysis2.png}}
%\caption{\textcolor{red}{......}}
%\label{fig:analysis2}
%\end{figure}

Understanding the areas of evaluation is essential when screening and validating these editing systems. First, there’s functionality which constitutes the sum performance of a tool that is capable of supporting many different operations on different types of typical data formats. Biomedical data is highly complex, so users may want to filter through several different fields at once or recompute data through the tool itself. An example of this type of functionality includes the filtration and usability of multiple datasets and computational detection of RNA-seq mutation sites. Another pivotal role in functionality is the ability for a editing technology to tune gene expression or modulate the innate immune response, thereby linking it to different human disorders.

\begin{figure}[h!]
\centerline{
\includegraphics[width=0.50\textwidth]{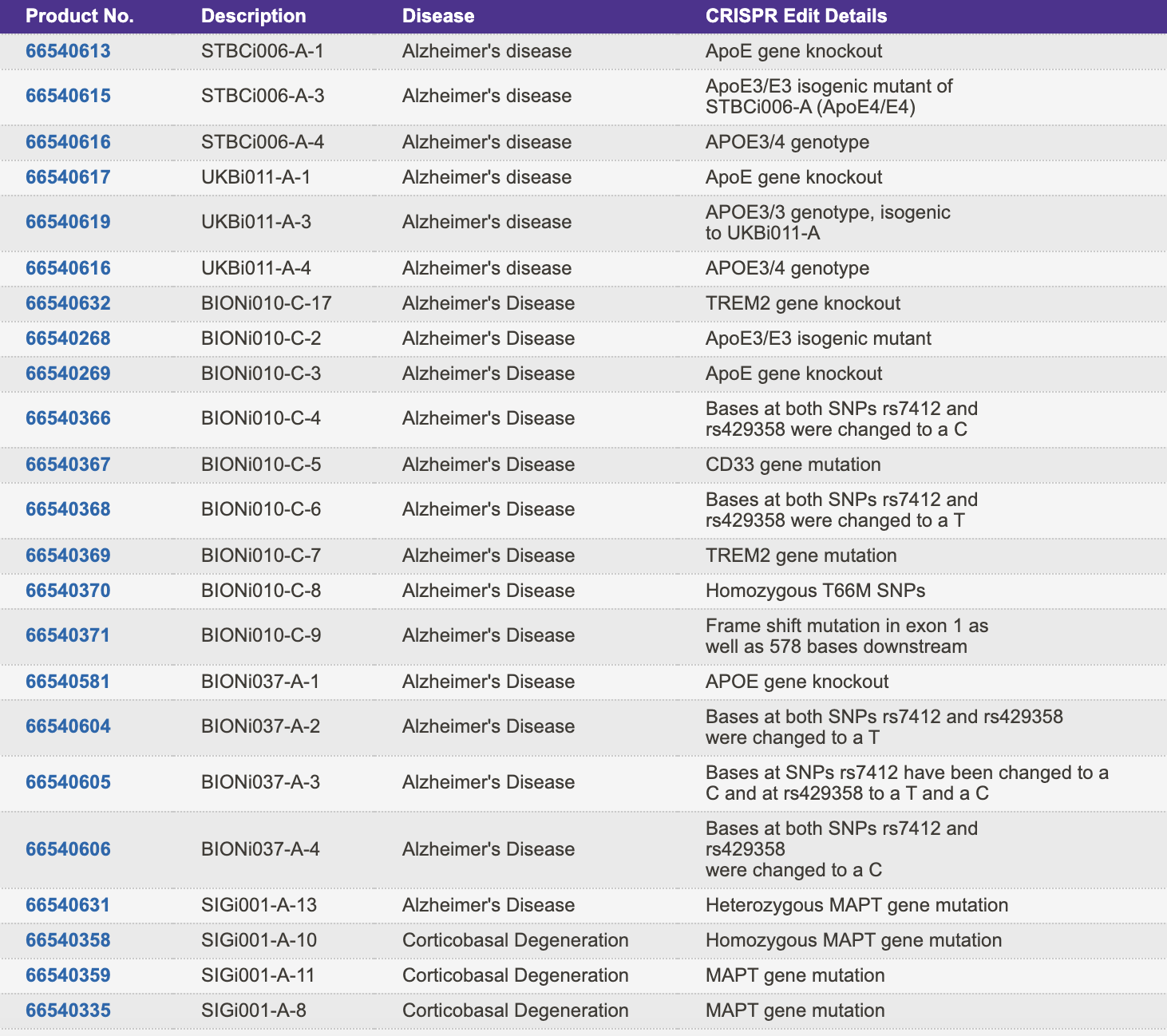}}
\caption{Gene Expression Dataset with CRISPR editing details to measure functionality for applications.}
\label{marker:geneexdataset}
\end{figure}

Functionality of an RNA editing system can be measured by REDItools and the REDIportal database, particularly REDIportal's Gene View. REDIportal stores individual A-to-I positions as well as statistics and relevant metrics per each Genotype-Tissue Expression (GTEx) sample, such as the Alu Editing Index (AEI) or the Recoding Editing Index (REI) or the expression of ADAR genes, that are expected to facilitate the RNA editing investigations \cite{10.1093/nar/gkaa916}. Accumulating gene expression mutations through datasets (Fig. \ref{marker:geneexdataset}) and an editing system's ability to efficiently edit the dataset determines the platform's functionality.

\begin{figure}[h!]
\centerline{
\includegraphics[width=0.50\textwidth]{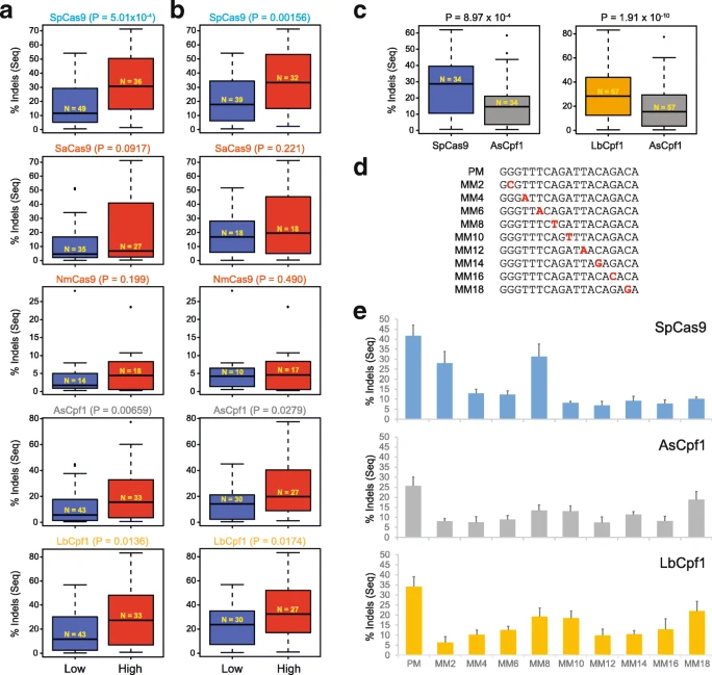}}
\caption{Relationship of DNA cleavage efficiency with gene expression and target specificity.}
\label{marker:relatEval}
\end{figure}

Second is intuitiveness which determines the tool’s usability and informativeness. A platform must maintain a balance between simple visual representation and robust conveyance of information to be considered intuitive. An open question in the research area is how to create visually meaningful solutions that summarize many different dimensions or data. Analyzing sequences and effectively engineering the human genome assists determining intuitiveness of a system.\\
	The amount of targeting sites in genes with varying expression levels is a measurement for intuitiveness. In the CRISPR Cas systems there is a balance of information and usability due to the different enzymes and different PAMs that can be utilized to obtained a variety of editing information. However, there are only two pathways for DSB repair, which can be performed through non-homologous end joining (NHEJ)-mediated and homology-directed repair (HDR)-mediated gene editing. Intuitiveness of these pathways is found in the editing efficiencies of Cas enzymes as affected by the expression of targeted genes as observed in Fig. \ref{marker:relatEval} \cite{wang2018systematic}.  \\
	When selecting an appropriate system based on its intuitiveness, the results rely on the chosen Cas endonuclease. The type of gene modification desired or the type of repair template to be utilized informs a user of the engineered genome's target and non-target efficiencies along with gene expression datasets  \cite{wang2018systematic}. \\

\begin{figure}[h!]
\centerline{
\includegraphics[width=0.50\textwidth]{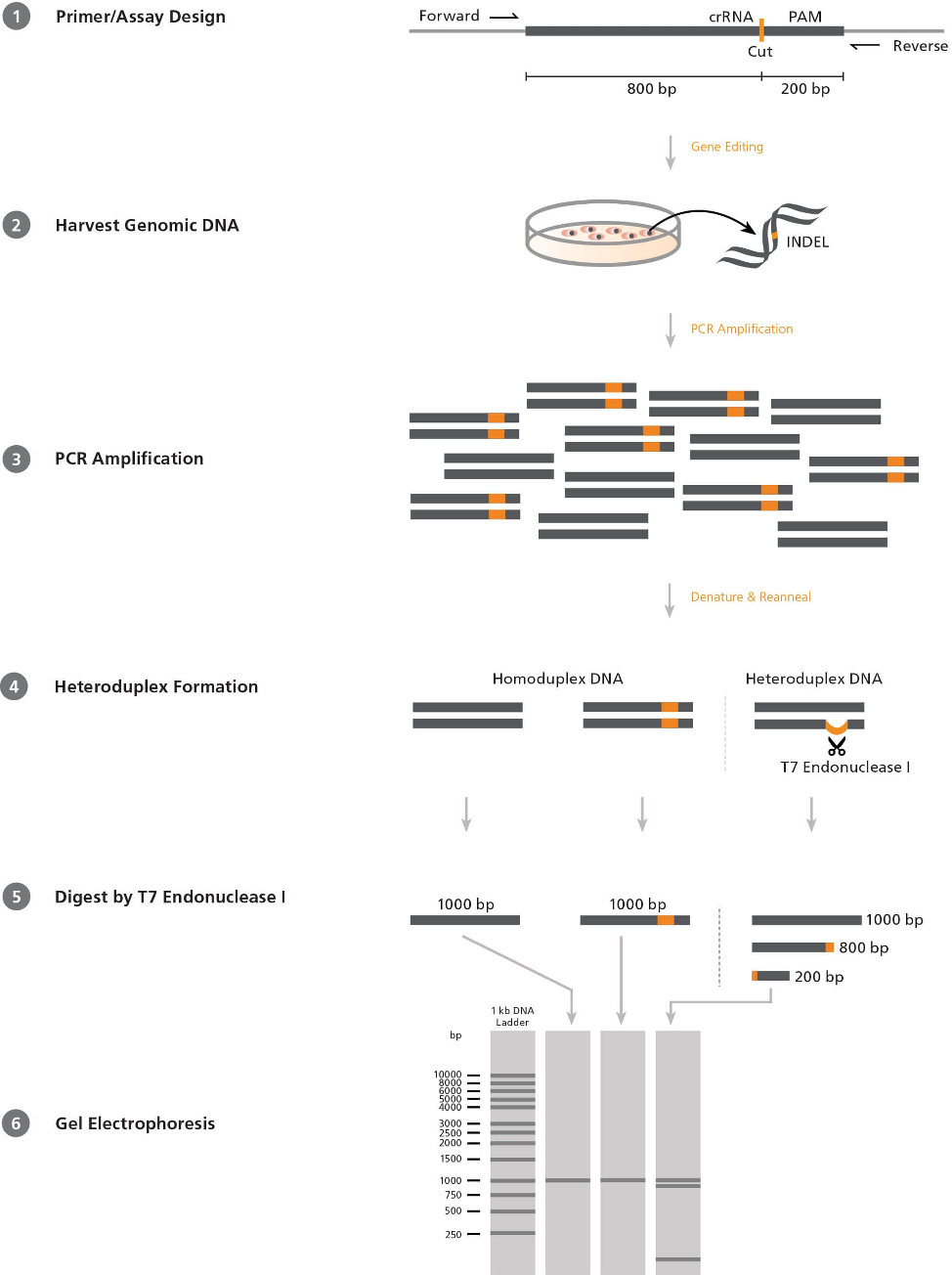}}
\caption{Mismatch Cleavage Assay Using T7 Endonuclease I. 1) Forward and reverse PCR primers are designed to flank the target CRISPR-Cas9 site in an offset manner, e.g. 200 and 800 base pairs (bp) on either side. 2) After CRISPR-Cas9 editing, genomic DNA is extracted from the cells. 3) The target region is amplified by PCR using the above primers. 4) PCR products are denatured and reannealed; DNA from edited cells may reanneal with DNA from non-edited (wild-type) cells to create a heteroduplex. 5) ArciTect™ T7 Endonuclease I will cleave single-strand DNA at heteroduplex structures greater than 2 bp. 6) Due to the offset nature of the primers, the resulting fragments will be of different lengths and can be resolved on an agarose gel. The relative amount of cut fragments detectable on the gel thereby gives an estimate of the mutation frequency within the cell population \cite{cite-EvalMets}.}
\label{marker:EvalMetrics}
\end{figure}

Third is reliability. It identifies a tool that is reliable and failure-free for a user’s operation(s), and this function is relies on the final requirement which is accuracy. Accuracy is based on the specific targeted sequence identified as mutated is being correctly identified and edited. With accurate results, precision can be added to the requirements of an editing system. Precision is the reproducibility of accurate results under the same conditions. In order to accumulate these metrics, screening and validation of the technologies and platforms must be applied. \\
	Screening methods can simultaneously determine the accuracy and precision of genome editing tools by identifying the presence of off-target sites and editing sequences\cite{germini2018comparison}. Genome-wide NGS approaches for detecting CRISPR off-target cleavage sites include cell-based assays (in live or fixed cells) as well in vitro assays\cite{cite-Metrics}. To accomplish this task, screening assay is first performed to detect the presence of a genetic alteration by using a heterogeneous population of cells 48 - 72 hours after CRISPR-Cas9 delivery\cite{cite-EvalMets}. Two systems discovered that satisfy the screening process are AsCpf1 and LbCpf1, model Cas endonucleases, which may assist in pursuing future applications\cite{wang2018systematic}. Altogether, screening is informative regarding relevant proportions of possible editing outcomes\cite{cite-EvalMets}.\\
The current standard screening methods of evaluation are based on DNA sequencing or use mismatch-sensitive endonucleases\cite{germini2018comparison}. Four strategies for screening edited cells are based on mismatch cleavage, Sanger sequencing analysis, polymerase chain reaction (PCR) amplicon length, and restriction endonuclease pattern (Fig. \ref{marker:EvalMetrics})\cite{cite-EvalMets}. There are also differential expression software packages that can be used to screen in RNA-seq datasets such as Cuffdiff, edgeR, DESeq, PoissonSeq, baySeq, and limma\cite{rapaport2013comprehensive}. Regardless of the preliminary tactic used to screen for editing events, researchers found that increasing the number of replicate samples significantly improves detection power over sequencing depth; this confirms the desired mutation is present at the desired allele frequency in potential positive clones\cite{cite-EvalMets}. \\
 
\begin{figure}[h!]
\centerline{
\includegraphics[width=0.50\textwidth]{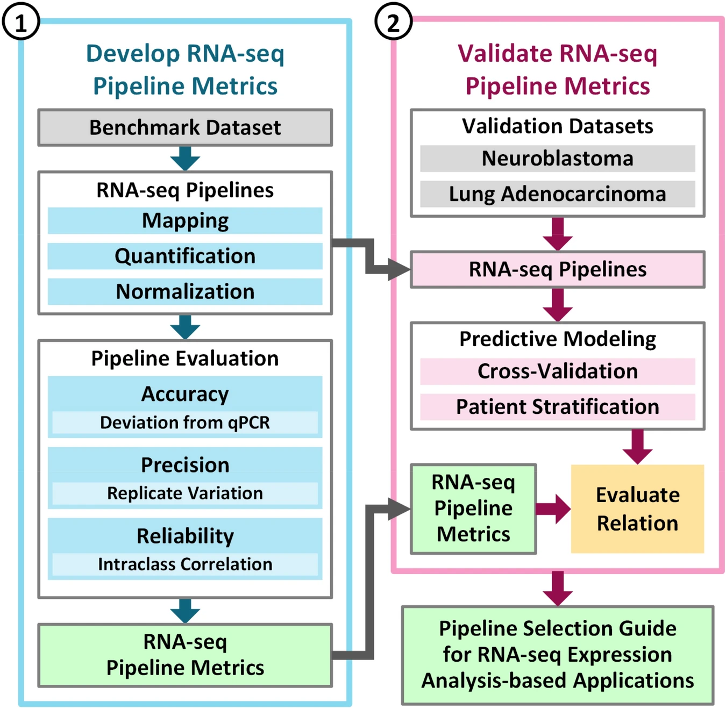}}
\caption{RNA-seq Pipeline Development \cite{tong2020impact}.}
\label{marker:Pipeline}
\end{figure}

Selecting RNA-seq pipelines for gene expression-based predictive modeling assists in developing accuracy and precision evaluation metrics for gene and RNA editing. A metric pipeline is chosen based on the requirements of a clinical application. To create these metrics, one study utilized SEQC-benchmark, SEQC-neuroblastoma, and TCGA-lung-adenocarcinoma datasets to capture the accuracy, precision, and reliability of RNA-seq pipelines (the blue box in Fig. \ref{marker:Pipeline}) \cite{tong2020impact}. By mapping, quantifying, and determining normalization of the datasets, the RNA seq, metrics can be evaluated. Predictive modeling includes cross-validation and patient stratification. However, gene expression variation must be considered as well when determining the most effective evaluation method for editing sites. \\

RNA-seq pipelines that produce accurate, precise, and reliable results dictate the estimation of gene expression which can then predict the outcomes of disease and mutations on the human body \cite{tong2020impact}. Two studies of gene expression variation are probe-based technology and, more recently, high throughput sequencing (HTS). Both of these studies advance scientific knowledge of diseases and classifications. Other sources available to analyze and simplify the application of raw microarray data include Galaxy, GenomicScape, Terra, GenePattern, and Affymetrix. While current platforms are a powerful way to integrate existing programs into pipelines that carry end-to-end data processing, they are limited in their flexibility\cite{Yukselen:2020wd}.
 
\begin{figure}[h!]
\centerline{
\includegraphics[width=0.50\textwidth]{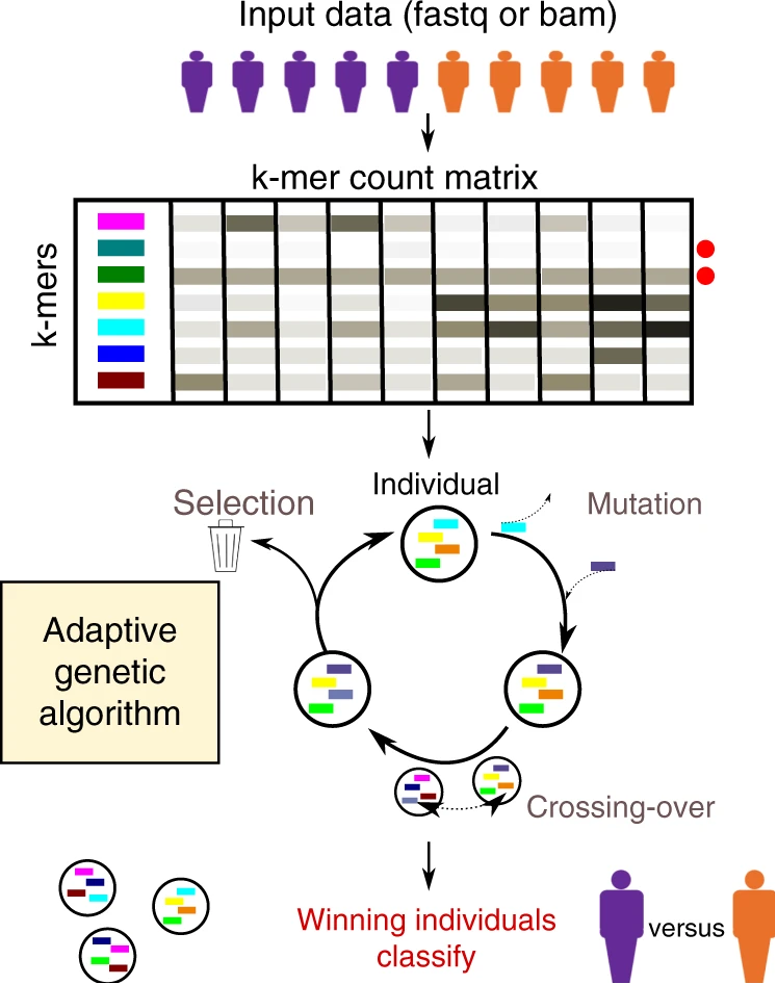}}
\caption{Overview of GECKO algorithm \cite{thomas2019gecko}.}
\label{marker:GECKO}
\end{figure}

One example of HTS data analysis and quantification is known as GECKO, or GEnetic Classification using k-mer Optimization. It is effective at classifying and extracting meaningful sequences from multiple types of sequencing approaches including mRNA, microRNA, and DNA methylome data\cite{thomas2019gecko}. The process includes two steps: k-mer matrix preparation step and an adaptive genetic algorithm (see Fig. \ref{marker:GECKO}).  The adaptive genetic algorithm randomly selects groups of k-mers from the k-mer matrix to form individuals, and these individuals go through rounds of mutation, crossing-over and selection to discover individuals capable of classifying the input samples with high accuracy\cite{thomas2019gecko}.
	
	Considering the current systems' shortcomings for evaluating pipeline metrics, portability and flexibility are essential variables that are not fulfilled in these platforms. Workflow engines that have been developed to address this issue are Nextflow and DolphinNext. The latter integrates the former system in order to improve usability and portability. DolphinNext provides a graphical user interface to create and reproduce pipelines along with addressing many of the needs of HTS data processing \cite{Yukselen:2020wd}. DolphinNext also contributes to RNA-seq pipeline metrics by satisfying the system requirements (e.g., functionality, intuitiveness, reliability, and accuracy) for evaluating gene and RNA editing.

%%%%%%%%%%%%%%%%%%%%%%%%%
\section{Risks and Limitations} 
Based on previously introduced material and application use cases (Table \ref{marker:UCTable}), limitations and risks to gene and RNA editing are surmised into the following categories: off-target risks, delivery limitations, cost limitations, and lack of efficient evidence to support the application. The necessary application for a patient tends to determine the best system and method of delivery. Safety is essential in delivery; however, other issues arise depending on the type of editing introduced and the associated ethical questions.

\begin{figure}[h!]
\centerline{
\includegraphics[width=0.50\textwidth]{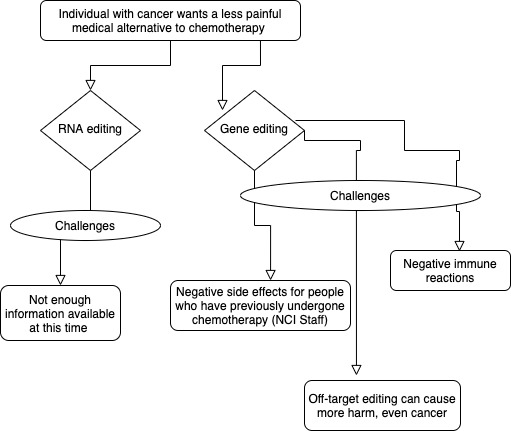}}
\caption{Use case flow chart.}
\label{marker:UseCaseFlow}
\end{figure}

Use cases (Fig. \ref{marker:UseCaseFlow}) are significant in identifying the application area of interest and the challenges/risks that an individual faces when determining the best course of action/treatment. Both gene and RNA editing possesses limitations and risks that, with further research, can be improved and will benefit certain individuals depending on their present disease or genetic disability (see Table \ref{marker:UCTable}). When producing use cases for applications, RNA editing presents less risks; however, the research evidence at this time is limited compared to gene editing that has been utilized for nearly three generations. With more research and clinical trials, scientists might discover more or less risks regarding RNA editing as compared to gene editing.

\begin{table*}[h!]
\centering
\begin{tabular}{|p{7cm}|p{10cm}|}
\hline
\textbf{USE CASE} & \textbf{Challenge(s)} \\
\hline
As an individual with muscular dystrophy, I want to correct my debilitation without medications & 
\begin{itemize}
    \item Efficient delivery of enzymes to site-specific tissues
    \item Validation of system safety
    \item Small packaging size limitations (Chemello)
\end{itemize} \\
\hline
As an individual with chronic pain, I want a therapy that takes away my pain without chemically modified meds & 
\begin{itemize}
    \item Off-target effects can cause more pain and can be cancerous
\end{itemize} \\
\hline
As an individual with genetic blindness, I want to be able to see again & 
\begin{itemize}
    \item Editing is limited in the coding sequence codon preferences of endogenous ADARs, and endogenous ADAR strategies cannot mediate edits other than A-I (Fry)
    \item Difficult to deliver both DNA and RNA base-editing constructs for therapeutic use (Fry)
\end{itemize} \\
\hline
\end{tabular}
\caption{Use cases and challenges.}
\label{marker:UCTable}
\end{table*}

%\begin{figure}[h!]
%\centerline{
%\includegraphics[width=0.50\textwidth]{Figures/USECASEtabledrawio.jpg}}
%\caption{\textcolor{red}{Use case table.}}
%\label{marker:UCTable}
%\end{figure}

Gene and RNA editing comes in many varieties, with many consequences, and any bioethical discussion needs to take into account those distinctions\cite{cite-Perspectives}. One ethical concern is that some people might try to design babies with certain traits, like greater intelligence or athleticism\cite{cite-Yaleinsights}. This leads to the question about where to draw the line between disease treatment and enhancement, and how to regulate the use of editing technologies, considering differing attitudes toward conditions such as blindness\cite{cite-Perspectives}. The scientific community has to ensure that the benefits do not outweigh the costs\cite{cite-Perspectives}.

Despite the following risks and limitations to current editing systems, the need for better solutions to diseases and genetic illnesses is proliferating. Medications and traditional treatments administer temporal solutions and negative side effects that a patient has to endure; these methods cannot target specific cells with mutations, they affect the entire body. Targeted treatments are a new pathway with various limitations and risks depending on the application.

\subsection{Gene Editing Risks and Limitations}
Gene therapy suffered a major setback in 1999 when an 18-year-old man named Jesse Gelsinger with an inherited liver disease died during the first gene-therapy clinical trial at the University of Pennsylvania\cite{cite-RiskStanfordMed}. The procedure involved blood-producing stem cells that are removed from the body, genetically altered — often with the help of a virus — and then reintroduced into the bone marrow \cite{ledford2019gene}. However, during the clinical trial, the virus ran amok, triggering a severe immune response that led to leukemia in the participants\cite{cite-RiskStanfordMed}. Gelsinger died four days later\cite{cite-RiskStanfordMed}. Altogether, the regimen is risky, because the treatment wipes out the participant’s white blood cells, and wreaks havoc on the lining of the gut, potentially leaving them dependent on intravenous nutrition\cite{ledford2019gene}.\\
	In 2016, a different group from the University of Pennsylvania asked a federal panel to green-light the first-ever clinical trial using CRISPR\cite{cite-RiskStanfordMed}. The trial was designed to genetically alter immune cells in cancer patients, then reinject the modified cells to see if they improve the immune system’s ability to fight off the disease. Different from 1999, most CRISPR protocols are ex vivo — they take the cells out of the body, manipulate them and then put them back\cite{cite-RiskStanfordMed}. That, at least, allows for some kind of risk assessment to see if there are any off-target gene modifications, such as the development of acute and severe anemia, or if they’ve accidentally turned the immune cells into cancer cells \cite{cite-RiskStanfordMed}. As a result, participants, particularly those with blood diseases, often need blood transfusions just before harvest of the stem cells to ease cellular stress \cite{ledford2019gene}.

	Risks and limitations of gene editing include bioethical issues, gene drive, off-target effects, delivery limitations, immunogenicity, and unintended consequences for future generations\cite{cite-Speights}. With the manipulation of genes comes the possibility of passing modulated genetic material to other organisms, or even causing hereditary issues with gene alteration; this process of transference is known as gene drive\cite{cite-Yaleinsights}. Whether it’s through genetic drive, off-target side effects, the development of cancer through gene editing, or bioethical consequences, gene editing is a permanent delivery that will affect future generations, because, currently, we don’t have the ability to control and/or reverse runaway evolutionary changes\cite{cite-RiskStanfordMed}. Therefore, long-term danger is unintended changes to an individual’s gene that is carried through to the next generation\cite{cite-Yaleinsights}. The futuristic safety risk is unknown changes in the genome that are transferred to the population which has a possibility of being harmless or harmful\cite{cite-Yaleinsights}. \\

\subsubsection{Bioethical Issues}
\begin{figure*}
\centering
 \includegraphics[width=7in]{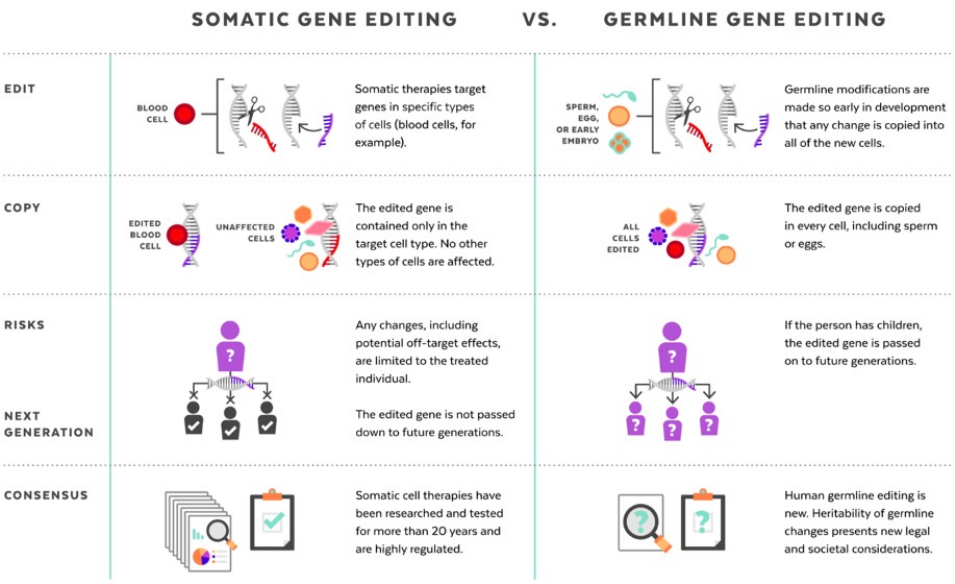}
\caption{Comparison of Somatic and Germline editing.}
\label{marker:SomaticGerm}
\end{figure*}

One of the biggest areas for bioethical debate is between the type of gene editing utilized. For this reason, countries have placed regulations on the type of genetic editing that is permissible. Most countries allow somatic gene editing, while many countries, including the U.S., have regulations to restrict germline editing due to its permanent effects on the genome and the individual\cite{cite-Speights}. Regulations err on the fear of misuse, such as the idea of creating biological weapons or designer babies. \\
Biological warfare is a concept that was a mere idea a generation ago; however, COVID has brought a new age to the science of biological attacks. It is probable that if you want to make a biological weapon, you can use CRISPR to turn ordinary cowpox virus into smallpox \cite{cite-RiskStanfordMed}. In 2016, James Clapper, the then-director of National Intelligence, added gene editing to the WMD [weapon of mass destruction] list, because genetically engineered bacteria or viruses can be used in biological attacks against humans, which we’ve experienced with COVID, or to cause widespread crop damage\cite{cite-Speights}.\\
Another fear is that germline editing could be used to change what it means to be human\cite{cite-Speights}. Creating "designer babies" -- could change society's view of human life. Collins worries that children could be seen as "commodities" instead of as "precious gifts"\cite{cite-Speights}. Germline editing takes away the ability for future generations to choose their own genetic destiny\cite{cite-Speights}, while, on the other hand, enables patients to experience an improved standard of living (see Fig. \ref{marker:SomaticGerm}). \\
Somatic gene editing affects only the patient being treated (and only some of his or her cells), germline editing affects all cells in an organism, including eggs and sperm, and so is passed on to future generations; hence, the possible consequences of that are difficult to predict \cite{cite-Perspectives}. Somatic gene therapies involve modifying a patient’s DNA to treat or cure a disease caused by a genetic mutation\cite{cite-Perspectives}. Germline human genome editing, on the other hand, alters the genome of a human embryo at its earliest stages\cite{cite-Perspectives}. This may affect every cell, which means it has an impact not only on the person who may result, but possibly on his or her descendants\cite{cite-Perspectives}. However, germline editing in a dish can help researchers figure out what the health benefits could be, and how to reduce risks\ cite{cite-Perspectives}.  \\
In 2018, Chinese biophysicist He Jiankui shocked the world by announcing that he had edited implanted human embryos, a germline editing effort to make the resulting children resistant to HIV infection \cite{ledford2020crispr}. The work, which was widely condemned by scientists and yielded prison sentences for He and two of his colleagues, led to the birth of two children with edited genomes \cite{ledford2020crispr}. The problem is that without progressive testing germline editing can generate some unwanted changes to genes and produce a range of different outcomes even among cells in the same embryo\cite{ledford2020crispr}. \\

Although CRISPR–Cas9-induced genome editing is effective in almost all cell types, controlling the exact editing outcome remains a challenge in the field \cite{doudna2020promise}. All of the gene editing types and delivery strategies—albeit elegant in principle—involve large chimeric proteins that pose additional challenges of delivery into primary cells \cite{doudna2020promise}. Therefore, risks remain when considering delivery, potency, and specificity of CRISPR interference, CRISPR activation and CRISPR-mediated base editing and prime editing, which will need to be thoroughly addressed before they are ready for clinical use\cite{doudna2020promise}.\\

\subsubsection{Off-Target Risks}
In previous sections, we’ve learned that off-target effects are a predominant risk in regard to gene editing, because there's a possibility that DNA sequences other than the targeted ones could also be changed\cite{cite-Speights}. Even though the sgRNA is designed to target a specific gene of interest, often a significant number of non-specific genes are targeted by the same Cas9/sgRNA\cite{Mout:2017aa}. In gene-based delivery system, particularly with CRISPR, the long-term constitutive expression of Cas9/sgRNA worsens the problem, because repeated exposure of Cas9/sgRNA to non-specific genes can lead to large off-target effects\cite{Mout:2017aa}.\\

Off-target effects occur with all types of gene editing, and the long-term effects remain a mystery\cite{cite-Speights}. A number of scientists wrote an article in 2017 reporting that the use of CRISPR-Cas9 to edit genes could cause hundreds of unintended mutations\cite{cite-Speights}. Investors upset at the news feared their stakes in gene editing. The fear of unwanted mutations resulted in a retraction, which was followed by a correction to the article. The correction explained that gene editing "can precisely edit the genome at the organismal level and may not introduce numerous, unintended, off-target mutations"\cite{cite-Speights}. However, the question of possible unintended mutations remains. \\
Another targeted therapy with high off-target effects is the human embryo. When it comes to human embryos, gene editing is too dangerous to try, because it places the embryo at impending risk. In more than half of the cases in preclinical trials, gene editing caused unintended changes, such as loss of an entire chromosome or big chunks of it\cite{cite-IntheLab}. Instead of the mutation being fixed, the chromosome carrying the mutation is gone — a profound change that might harm the embryo\cite{cite-IntheLab}. \\

\subsubsection{Cost Limitations}
Though the cost of gene editing has lessened with CRISPR, research funding and the cost of delivery remains expensive. Depending on the type of system and application, costs can range beyond prescription medications and therapies. There’s also a cost to investors that must be considered, because the number of investors limits or expands the research. Articles, like the one mentioned in off-target effects can definitely highlight a danger for investors in such an early-stage technology\cite{cite-Speights}.\\
The cost of medical treatments today is very expensive for individuals with genetic diseases, so introducing another treatment option that includes engineered cells will increase the cost of all products. The treatment for beta-thalassaemia runs to roughly \$1.8 million — not including the hospital stay and other associated costs. \cite{ledford2019gene}. Similarly, cancer immunotherapy costs hundreds of thousands of dollars per year\cite{cite-RiskStanfordMed}. Even if CRISPR proves successful in delivery, many scientists and medical professionals worry that for many patients, the financial cost will be prohibitive\cite{cite-RiskStanfordMed}. \\

\subsubsection{Genetic Delivery Limitations}
There are various delivery systems for gene editing, yet each one has its own limitations and risks attached. Size of the system is a major consideration when choosing a particular delivery system. Delivery options include ex vivo, in vitro, and in vivo; viral and non-viral vectors; and protein-based delivery. Regarding the gene editing of a Cas9 system, there is an immune response that can risk a patient’s overall health and longevity. This is due to sources of Ca9: Staphylococcus aureus (S. aureus; SaCas9) and Streptococcus pyogenes (S. pyogenes; SpCas9), which frequently colonize humans and cause disease (e.g. MRSA and strep throat) \cite{crudele2018cas9}. Introducing this bacteria triggers a destructive immune response to Cas9 as seen in mice where Cas9 was delivered through electroporation of naked DNA\cite{crudele2018cas9}. Though there are more systems beyond Cas9, most gene editing therapeutic applications require systemic delivery, which presents its own limitation\cite{Mout:2017ab}.\\

A delivery system with the least risk is ex vivo gene therapy. Since cells are treated in a dish before transplantation, Cas9 immune responses can potentially be circumvented by using transient Cas9 expression and waiting for the Cas9 protein to clear before administering the corrected cells to patients \cite{crudele2018cas9}. Whereas direct editing of cells in vivo typically utilizes a viral-derived vector to deliver the Cas9 gene, leading to long-term expression in the presence of an intact immune system, which could potentially trigger an anti-Cas9 immune response leading to the killing of Cas9-expressing cells \cite{crudele2018cas9}.\\
The effective delivery of multiple CRISPR components in vivo, with either viral and non-viral vectors, into host cells still remains a major challenge\cite{Mout:2017aa}. Evidence concludes that viral vectors have achieved success in delivery, with the most prominent being adenovirus and adeno-associated virus\cite{gori2015delivery}. However, packaging of CRISPR components into a single vector is a major challenge for therapeutic applications\cite{Mout:2017aa}. For this reason CRISPR/Cas9 in vivo editing efficiency is significantly lower compared to in vitro editing\cite{Mout:2017aa}. Researchers discovered the limitations of gene editing in vivo delivery with an injection of CRISPR components which resulted only 1 in 250 edited cells\cite{yin2014genome}. Such a low editing percentage may be enough for alleviating certain diseases (e.g. muscular dystrophy, liver tyrosinemia), however, other diseases such as cancer require 100\% editing on-target efficiency\cite{Mout:2017aa}. \\

Another challenge is presented in protein-based delivery. While SpCas9 protein is a large protein (160 kDa, approx. 7.5 nm hydrodynamic diameter) with a net positive surface charge, sgRNA (approx. 31 kDa, 5.5 nm hydrodynamic diameter) is negatively charged\cite{Mout:2017ab}. Thereby, packaging these elements through supramolecular chemistry may be a major limitation for designing delivery vehicles\cite{Mout:2017aa}. Targeting is particularly difficult to achieve, as incorporation of additional biomolecules to a delivery vector alongside the CRISPR components complicates the packaging\cite{Mout:2017aa}.

\subsubsection{Immunogenicity}
In 2018, articles were published that raised cancerous concerns about gene editing systems, specifically CRISPR-Cas9\cite{cite-Speights}. Studies have suggested that CRISPR may cause cells to lose their cancer-fighting ability, and that it may do more damage to the genome than initially understood\cite{cite-Yaleinsights}. Scientists at Cambridge University and the Karolinska Institute in Sweden discovered that CRISPR-Cas9 gene editing can induce a response in cells that attempts to protect against DNA damage\cite{cite-Speights}. This response involves p53 gene activation, which tries to repair the DNA break or cause the cell to self-destruct causing cancer\cite{cite-Speights}.

	Since Cas9 or other CRISPR-based genome editing proteins are derived from bacteria, the systems thereby elicit host immune response\cite{Mout:2017aa}. This causes immunogenicity against the very system that is trying to help and cure the individual. Another reaction to gene-based delivery of CRISPR elements is due to its integration of Cas9 into host cells. The constitutive expression of foreign Cas9 protein in the host cell will engage a specific immune response that might result in the elimination of Cas9 genetic cells in the host \cite{neefjes2011towards}. In fact, there is evidence that shows AAV-based CRISPR delivery in vivo elicits a strong immune response against Cas9 protein\cite{Mout:2017aa}. As a result, the potential for pre-existing antibodies against CRISPR components can cause inflammation, and leave researchers questioning the long-term safety and stability of genome-editing outcomes\cite{doudna2020promise}.

\subsection{RNA Editing Risks and Limitations}
Most of the RNA editing risks is surrounded by the limited knowledge and research in the field of study. With the current applications available, the two predominant challenges are separating true editing sites from false discoveries and accurate estimation of editing levels \cite{bahn2012accurate}. Concerning software of RNA editing technologies, there remain challenges in terms of the efficiency, accuracy, and usability of data analysis, and the necessity to overcome these issues are associated with Web‐based tools (i.e., lack of flexibility and reliability), especially since most tools can process only one sample at a time or have a data upload limit or require preprocessed input \cite{veneziano2016noncoding}.

Current RNA sequencing methods limit editing due to the precision and amounts of readings available. One reason is due to sequence reads being blind to DNA and RNA modifications (with the exception of A-to-I editing, which causes an A-to-G mismatch), because they are based on sequence-by-synthesis (SBS) technologies \cite{jonkhout2017rna}. The yield of these sequencing technologies is also a limitation for RNA editing compared to that of gene editing. First, only one genetic read (maximum approx. 1kb) can be taken at a time, thus presenting substantial limitations in terms of speed of execution and costs \cite{veneziano2016noncoding}. The number of reads from direct RNA sequencing is only 1 million while DNA sequencing is over 6 million reads \cite{jonkhout2017rna}. Another challenge in this context lies in the interpretation of read alignments in detecting fragment boundaries \cite{veneziano2016noncoding}. Genes of tRNA are distributed across multiple locations throughout the human genome which leads to ambiguity of the genomic origin due to multimapping\cite{veneziano2016noncoding}.

Despite the limited number of reads, RNA editing has better accuracy when compared to gene editing. Nonetheless, RNA editing methods, such as HTS and A-I editing which are designed to identify RNA editing events in protein‐coding RNA, have limitations in terms of false positives produced \cite{kleinman2012rna}. One of the most important challenges is to increase the sensitivity with which the RNA fragments are detected and annotated, reducing false positives, and increase the comprehensive computational pipelines to handle and analyze the huge amount of data in an automated manner \cite{veneziano2016noncoding}.

the sensitivity and specificity of the detection methods as well as the elucidation of the biological function of lncRNAs, considering their limited annotation status and low expression levels miRNA sequencing (miRNA‐seq) has produced many analytical challenges, 
Challenges still remain the ability to appropriately detect their low level of expression, with its temporal and spatial patterns, and the challenge posed by the relatively low sequence and transcriptional conservation that would otherwise aid in a more precise identification and functional characterization.
\cite{veneziano2016noncoding}

%%%%%%%%%%%%%%%%%%%%%%%%%
%\newpage
\section{Future Directions of RNA and Gene Editing}
Since the body is similar to a computer–particularly cellular behavior which is like that of computer bytes that parallel biological codons–future biotechnologies can incorporate processors to edit genes and RNA, rather than remaining dependent solely on enzymes/proteins. Recently, researchers at ETH Zurich have used CRISPR to build functional biocomputers inside human cells and slot dual-core processors into human cells by first modifying the CRISPR gene-editing tool \cite{cite_Irving}. The team created a special version of the Cas9 enzyme that can act like a digital half adders processor that reads gRNA as inputs, expresses genes, and outputs proteins \cite{cite_Irving}. Potentially, RNA and gene editing can boost their diagnostic ability and treatment of diseases by implementing computing power to cells. 

Based on research, future directions of editing focus on RNA approaches due to its safety to a patient. The expansion of gene editing can be discovered with further research into RNA, particularly as a delivery system rather than using RNA as a mere guide for gene editing. Lessening the debilitating effects of gene editing is a goal, while the need for company and research investment in RNA editing will expand beneficial applications.

%\newpage
%%%%%%%%%%%%%%%%%%%%%%%%%
\section{Conclusions}
Evidence of the results involving gene and RNA editing supports the basis that RNA editing is a better, more desirable approach as compared to gene editing. Gene editing is a permanent system that can cause harm to a patient who is seeking treatment for a disease, a genome mutation. This article provides the methods, the architectures of the editing systems so as to compare and contrast the technologies, and the applications. Methods are applied from different architectures and technologies based on effectiveness and accuracy of target site sequencing. Efficient delivery of enzymes determines the applicability of each device. Applications in their first stages, in pre-clinical trials, and in clinical trials are established in this article, along with the companies that are investing in each type of integration. RNA and gene editing therapies and medications are currently available for application toward different diseases and genetic disabilities. Altogether, start-up companies are growing to improve the initial methods, architectures, and technologies to expand applications. 

%\newpage
%%%%%%%%%%%%%%%%%%%%%%%%%
\bibliographystyle{ieeetr}
\bibliography{References}

\end{document}